\documentclass[useAMS]{mn2e}

\usepackage{mathptmx}
\usepackage{lscape}
\usepackage{rotating}
\usepackage{xspace}
\usepackage{amssymb}
\usepackage{color}
\usepackage{subfigure}
\usepackage{deluxetable}
\usepackage{graphicx}
\usepackage[T1]{fontenc}

\def\simgt{\lower 2pt \hbox{$\, \buildrel {\scriptstyle >}\over {\scriptstyle \sim}\,$}}
\def\simlt{\lower 2pt \hbox{$\, \buildrel {\scriptstyle <}\over {\scriptstyle \sim}\,$}}
\def\chandra{{\it Chandra}}

\def\rosat{{\it ROSAT}}

\def\ha{H$\alpha$}

\def\xray{\hbox{X-ray}}

\begin{document}

% -----------------------------------------------------------------------------
% Title
% -----------------------------------------------------------------------------

\title[Gemini spectroscopy of Galactic Bulge Sources]
 {Gemini spectroscopy of Galactic Bulge Sources: a
  population of hidden accreting binaries revealed?\thanks{Based on observations
  obtained at 
  the Gemini Observatory, which is operated by the 
    Association of Universities for Research in Astronomy, Inc., under a cooperative agreement 
    with the NSF on behalf of the Gemini partnership: the National Science Foundation 
    (United States), the National Research Council (Canada), CONICYT (Chile), the Australian 
    Research Council (Australia), Minist\'{e}rio da Ci\^{e}ncia, Tecnologia e Inova\c{c}\~{a}o 
    (Brazil) and Ministerio de Ciencia, Tecnolog\'{i}a e Innovaci\'{o}n Productiva (Argentina).}}

% -----------------------------------------------------------------------------
% Authors and their affiliations
% -----------------------------------------------------------------------------

\author[Jianfeng Wu et al.]
{Jianfeng~Wu,$^1$\thanks{E-mail: jianfeng.wu@cfa.harvard.edu} 
P.~G.~Jonker,$^{1,2,3}$ 
M.~A.~P.~Torres,$^{2,3}$ 
C.~T.~Britt,$^{4,5}$\thanks{Visiting astronomer, Cerro Tololo
  Inter-American Observatory, National Optical Astronomy Observatory,
  which are operated by the Association of Universities for Research
  in Astronomy, under contract with the National Science Foundation.} 
C.~B.~Johnson,$^{4}$\footnotemark[3]
\newauthor R.~I.~Hynes,$^{4}$\footnotemark[3]
S.~Greiss,$^{6}$ 
D.~T.~H.~Steeghs,$^{1,6}$ 
T.~J.~Maccarone,$^{5}$
C.~O.~Heinke,$^{7,8}$
\newauthor T.~Wevers$^{3}$\\
$^1$Harvard-Smithsonian Center for Astrophysics, 60 Garden Street,
     Cambridge, MA 02138, USA\\
$^2$SRON, Netherlands Institute for Space Research, Sorbonnelaan 2,
     3584 CA, Utrecht, The Netherlands\\
$^3$Department of Astrophysics/IMAPP, Radboud University,
     Heyendaalseweg 135, 6525 AJ, Nijmegen, The Netherlands\\
$^4$Department of Physics and Astronomy, Louisiana State University,
     Baton Rouge, LA 70803-4001, USA\\
%$^{6}
%    {School of Physics and Astronomy, University of Southampton,
%     Highfield, Southampton SO17 1BJ, UK} 
$^5$Department of Physics, Texas Tech University, Box 41051, Lubbock TX, 
     79409-1051, USA\\
$^6$Department of Physics, University of Warwick, Coventry, CV4~7AL,
     UK\\
$^7$Department of Physics, University of Alberta, CCIS 4-183,
     Edmonton, AB T6G 2E1, Canada\\ 
$^8$Max Planck Institute for Radio Astronomy, Auf dem Hugel 69, 
    53121 Bonn, Germany\\
%$^9$Visiting astronomer, Cerro Tololo Inter-American Observatory,
%    National Optical Astronomy Observatory, which are operated \\ by the
%    Association of Universities for Research in Astronomy, under
%    contract with the National Science Foundation.
}

% -----------------------------------------------------------------------------
% Abstract
% -----------------------------------------------------------------------------

\maketitle
\begin{abstract}

We present Gemini spectroscopy for 21 candidate optical counterparts
to X-ray sources discovered in the Galactic Bulge Survey (GBS). 
%Goals of the GBS include determining accurate masses of neutron stars and
%black holes using eclipsing quiescent low-mass X-ray binaries. 
For the majority of the 21 sources, the optical spectroscopy establishes that they
are indeed the likely counterparts. One of the criteria we used for the
identification was the presence of an H$\alpha$ emission line. The
spectra of several sources revealed an H$\alpha$ emission line only
after careful subtraction of the F or G stellar spectral absorption
lines. In a sub-class of three of these sources the residual H$\alpha$
emission line is broad ($\simgt400$~km~s$^{-1}$) which suggests that it is
formed in an accretion disk, whereas in other cases the line width is
such that we currently cannot determine whether the line emission is
formed in an active star/binary or in an accretion disk. GBS source
CX377 shows this hidden-accretion behaviour most 
dramatically. The previously-identified broad H$\alpha$ emission of
this source is not present in its Gemini spectra taken $\sim$1 year
later. However, broad emission is revealed after subtracting an F6
template star spectrum.  The Gemini spectra of three sources (CX446,
CX1004, and CXB2) as well as the presence of possible eclipses in 
light curves of these sources  suggest that these sources are accreting
binaries viewed under a high inclination. 

\end{abstract}

\begin{keywords}
binaries: close --- stars: emission line, Be --- Galaxy: Bulge
  --- X-rays: binaries
\end{keywords}

% -----------------------------------------------------------------------------
% Introduction
% -----------------------------------------------------------------------------

\section{Introduction}\label{intro}

  %Comprehensive surveys of Galactic \xray\ sources provide new impetus
  %to the studies of accreting binaries. Multiwavelength follow-up campaigns
  %are necessary for classifying faint \xray\ sources. 
  Previous surveys of 
  faint \xray\ sources have been focused on the Galactic Center or
  globular clusters. While the Galactic Center Survey (e.g., Muno et~al. 
  2003) benefits from a high source density, the crowding
  and significant extinction make the optical/infrared follow-up
  necessary for classification difficult (e.g., Mauerhan et~al. 2009). 

  The Galactic Bulge Survey (GBS; Jonker et~al. 2011; Jonker
  et~al. 2014; Paper I \& II hereafter) is a 
  multiwavelength project that is designed to allow optical/infrared
  classification of X-ray sources detected in the Galactic Bulge. The
  GBS consists of {\it Chandra} and multiwavelength observations of
  two $6^\circ\times1^\circ$ strips centered $1.5^\circ$ above
  and below the Galactic plane (see Fig.~1 of Paper~I),
  thus avoiding the $|b|<1^\circ$ regions with serious crowding
  and extinction problems, while still maintaining a relatively high
  source density. The GBS  
  utilizes \chandra\ observations with an exposure of 2~ks for each
  pointing; the exposure time is chosen to maximize the relative
  numbers of low-mass \xray\ binaries (LMXBs) to cataclysmic variables
  (CVs). The completed \chandra\ observations have detected 1640 unique 
  \xray\ sources (Paper~II), agreeing well with
  the estimation in Paper~I which also gave a break-down
  of the expected numbers of various kinds of objects based on source
  density, 
  expected \chandra\ flux limit, etc. Among the 1640 \xray\ sources,
  $\sim600$ are expected to be CVs, including both intermediate polars
  (IPs) and non-magnetic CVs, while the number of LMXBs is expected to be
  $\sim250$ (see Table~2 of Paper~I). We also expect
  $\sim$600 chromospherically active stars or binaries, e.g., RS Canum 
  Venaticorum variables (RS CVn systems; Walter et~al. 1980). 

  The GBS
  combines a large sky coverage with the good sensitivity to faint
  X-ray sources and the excellent positional accuracy possible with
  {\it Chandra}.  
  There are two main science goals of the GBS (see \S1 of Paper~I
  for more details): 1) constraining the nature of the
  common-envelope phase in binary evolution by comparing the
  observed number of sources with model predictions in each class,
  e.g., CVs and LMXBs; 2) measuring the mass of the compact objects in
  X-ray binaries, e.g., eclipsing quiescent black hole (BH) and
  neutron star (NS) LMXBs, to investigate the Galactic BH mass
  distribution (e.g., \"{O}zel et al. 2010) and to constrain the
  NS equation of state (EoS).  

  Both of the science goals rely on the multiwavelength identification
  and classification of this large sample of faint \xray\ sources. A
  variety of optical/infrared follow-up campaigns have been 
  conducted. Hynes et~al. (2012) identified 69 \xray\ sources in the
  GBS using the {\it Tycho}-2 catalogue. These sources are coincident
  with or very close to the bright stars in that catalogue, most of
  which are likely to be the real optical counterparts to the \xray\
  sources. This sample is a mix of objects with a broad range of
  spectral types, including both late-type stars with coronal \xray\
  emission and early-type stars with wind \xray\ emission. Many
  sources are foreground objects instead of residing in the
  Galactic Bulge. Britt et~al. (2014) reported on an optical photometric
  survey of three quarters of the sky area covered by the \chandra\
  GBS, and presented the light curves of variable objects consistent
  with the \xray\ positions of GBS sources catalogued in Paper~I. About
  a quarter of the optical counterparts are 
  variable, and they are expected to be a mix of IPs, 
  non-magnetic CVs, LMXBs, and RS CVns. Greiss et~al. (2014) 
  provided likely near-infrared identification of GBS \xray\ sources
  using current near-infrared sky surveys.
  %In many cases, there are multiple near-infrared
  %sources within the \xray\ positional uncertainty
  %($2.8^{\prime\prime}$ median). Greiss et~al. (2014) developed a
  %method to calculate the likelihood of being the true counterpart for
  %each near-infrared source. 
  Maccarone et~al. (2012) found 12
  candidate radio matches to the GBS \xray\ sources using the archival
  NRAO VLA Sky Survey (NVSS; Condon et~al. 1998). The 
  majority of them appear to be background active galactic nuclei.   

  Optical/infrared spectroscopy of the detected \xray\ sources is an
  essential tool to investigate their nature. Accreting binaries can be 
  identified by the emission features in their optical spectra. The
  only firm way to distinguish white dwarf (WD), NS, and BH as the
  primaries of the X-ray 
  binaries is via measurements of the accretor masses, which requires
  high-quality optical/infrared spectroscopy. Britt et~al. (2013)
  presented five accreting binaries identified in the GBS
  based on the strong emission lines in their spectra, including three likely IPs, 
  one CV undergoing a dwarf nova outburst, and one likely quiescent LMXB (qLMXB). 
  Torres et~al. (2014) identified 22 new accreting binaries via the
  H$\alpha$ emission lines in their optical
  spectra. They developed criteria of accreting binaries based on
  the equivalent width (EW) of H$\alpha$ emission line
  (EW$>18$~\AA), the breadth of the H$\alpha$ emission line (FWHM $>400$~km~s$^{-1}$), 
  or the strength of He~{\sc i} $\lambda$5876,6678 in case of narrow and
  weak H$\alpha$ lines. There are also several extensive spectroscopic
  studies on individual GBS sources. Ratti et~al. (2013) presented a
  dynamical analysis of GBS source CX93 in which they measured the mass
  of the compact primary and the companion star, and concluded that the
  source is a long orbital period CV. Hynes
  et~al. (2014) identified a symbiotic \xray\ binary with a carbon
  star companion in the Galactic Bulge based on the spectra of its
  optical counterpart. 

  In this work, we present Gemini spectroscopy of
  21 GBS \xray\ sources with a better \xray\ positional accuracy than
  the median (due to
  low off-axis angles). These 21 sources are listed in
  Table~\ref{log_table}. We will refer to these sources with their
  labels, i.e. CX (or CXB) IDs, introduced in the GBS source catalogue
  (Paper~I and II). GBS sources listed in Paper~I have the prefix of
  ``CX'', while the remaining GBS sources, detected in the last
  quarter of the \chandra\ coverage published in Paper~II, have
  the prefix of ``CXB''. The sources in each catalogue were ranked by
  their \chandra\ counts, where CX1 has the most counts among the CX
  sources. The majority of the sources in
  this work are CXB sources, while previous works were focused on CX
  sources. 
  %\footnote{At the time of the Gemini observations,
  %the list of sources with the ``CXB'' prefix had not been finalized,
  %and thus most of our targets have different CXB IDs in the headers
  %of the FITS files from those in the final GBS catalogue. For reference,
  %Table~\ref{log_table} lists both the
  %finalized CXB IDs (Column 1) and the previously used CXB IDs (Column
  %2) that are present in the FITS headers. All the CXB IDs referred in this paper are the
  %finalized version unless otherwise noted.}\label{cxb_note} 
  This paper is 
  structured as follows. In \S\ref{obs} we describe the Gemini
  observations and data reduction. In \S\ref{ana} we present the
  analysis of the Gemini spectroscopy, including spectral
  classification and radial velocity analysis. In \S\ref{discuss} we give
  results and discuss each interesting object. Overall conclusions are
  summarized in \S\ref{conclusion}. 

% -----------------------------------------------------------------------------
% Sample Selection
% -----------------------------------------------------------------------------

\section{Observation \& Data Reduction}\label{obs}

\subsection{Gemini Spectroscopy}\label{obs:spec}

The list of 21 GBS sources of which we obtained Gemini spectroscopy
consists of six CX sources and 15 CXB sources. The six CX sources were
proven to be interesting on the basis of earlier spectra and/or
photometric variability. For example, three of the CX sources (CX377,
CX446, and CX1004) have shown H$\alpha$ emission lines in their
previous spectra (Torres et~al. 2014) obtained by the VIsible
Multi-Object Spectrograph (VIMOS) mounted on the Very Large Telescope
(VLT). Based on our follow-up strategy (i.e., prioritizing sources
with higher positional accuracy and brighter in optical/infrared), the
15 CXB sources are chosen to have off-axis angles less than 5 arcmin
in their \chandra\ observations and also have sufficient counts to
allow for an accurate \xray\ position
($<1^{\prime\prime}$). Optical/infrared brightness, colour and
photometric variability are also among the factors of sample selection.  

The finding charts of our sources are shown in Appendix~A (see
Fig.~\ref{fc_fig}--\ref{fc_fig2}). The coordinates listed in
Table~\ref{log_table} are for the candidate optical counterparts, i.e., the
objects for which we took Gemini spectroscopy. The astrometry was
performed on images from VLT/VIMOS (CX377, CX446, and CX1004; Torres
et~al. 2014), Gemini (CXB117), and Mosaic-II/Dark Energy Camera (DECam; 
other sources; see \S\ref{obs:photo}). All of
them have a RMS accuracy of $<0.2^{\prime\prime}$, while some sources
(CX377, CX1004, and CXB117) have $<0.1^{\prime\prime}$ positional
accuracy (Britt et~al. 2014; Torres et~al. 2014). 

%\begin{figure*}
%    \centering
%    \includegraphics[width=3in]{CX84lc.ps}
%    \includegraphics[width=3in]{CX377lc.ps}\\
%    \includegraphics[width=3in]{CX446lc.ps}
%    \includegraphics[width=3in]{CX1004lc.ps}
%    \caption{The Mosaic-II $r^\prime$-band light curves for CX sources
%    (CX84, CX377, CX446, and CX1004). CX84 shows periodic behaviour
%    with a period of 4.67 days in its light curve, while the other
%    objects only have random 0.05--0.1 magnitude flickering. CX446 has a
%    possible eclipsing event in its light curve.}
%             \label{lc_fig}
%\end{figure*}% 

Optical spectroscopy of the 21 GBS sources was obtained with
GMOS (Gemini Multi-Object Spectrograph; Davies et~al. 1997) mounted on the
Gemini-South Telescope in Chile. All the observations were taken
between 2012 Apr 20 and 2013 May 4 under programmes GS-2012A-Q-44 and
GS-2012A-Q-67 (see Table~\ref{log_table} for an observation
log). Nine objects have multi-epoch spectroscopy, while other objects
have one epoch; each epoch has 900--3600~s integration time. The seeing of
each spectroscopic observation was measured from the corresponding
acquisition image (see the last column of
Table~\ref{log_table}). Typical seeing was around $0.7^{\prime\prime}$
(with a range of 0.5--1.3$^{\prime\prime}$). GMOS was
operated in long-slit mode. We used the R400\_G5325 grating
(400~line~mm$^{-1}$) and a $0.75^{\prime\prime}$ slit. The
two-dimensional spectra were binned by a factor of two in both spatial
and spectral dimensions, resulting in a spatial dispersion of
$0.15^{\prime\prime}$/pixel and a spectral dispersion of
$1.36$~\AA/pixel. The spectral resolution is estimated to be
$\approx5$~\AA\ FWHM 
%by measuring the width of the strong, unblended emission
%line at $\lambda$7272.9359 in the arc spectra. This spectral
%resolution applies 
for the sources that had filled the whole slit
during the observation (i.e., the seeing was greater than the slit
width 0.75$^{\prime\prime}$), while spectral resolution should be
better than $\approx5$~\AA\ FWHM for the sources with seeing less than
0.75$^{\prime\prime}$. All the Gemini/GMOS spectra were  
split into three equal parts in the wavelength dimension by
detector gaps. The bluest part of the 
spectra was ignored in the analysis because of the low signal-to-noise
ratio ($S/N$) in the spectra due to the extinction towards our
sources and the lack of arc lines in this part of the spectra. The
middle part (wavelength range $\sim4800$--6100~\AA) and 
the red part (wavelength range $\sim6200$--7600~\AA) of the spectra
were reduced and analyzed separately. The results presented in this
work are mainly based on the analyses of the red part of the spectra
because it is least affected by extinction.

The Gemini/GMOS data were reduced using the {\sc figaro} package implemented in the
{\sc starlink} software suite and the packages of {\sc pamela} and
{\sc molly} developed by T. Marsh.\
%\footnote{See
%  http://www2.warwick.ac.uk/fac/sci/physics/research/astro/people/ marsh/software/.} 
The two-dimensional spectra were bias-corrected and flat-fielded.
The bias was corrected using the overscan areas of
the detectors. We utilized the flat fields taken directly following each
target observation for flat fielding. We fit the background on both
sides of the spectra with a second-order polynomial and determined the
background at the position of the spectra. The object spectra were 
optimally extracted using the \verb+optexp+ procedure in the {\sc pamela} 
package (Marsh 1989). The spectra were
wavelength-calibrated in {\sc molly} using the helium-argon arc
spectra which were taken either right after observing the target or at the
end of the night. The average arc spectrum was used in case of
multiple arc spectra in one night. 
The resulting root mean square (RMS) scatter on the wavelength
calibration is $\simlt0.3$~\AA. The wavelength calibration was
examined using the skyline O~{\sc i}~$\lambda 6300.303$. Offsets
relative to the wavelength of this line have been corrected. 
For sources with seeing less than the slit width ($0.75^{\prime\prime}$), 
the centroiding uncertainty (i.e., if the source is not placed 
in the middle of the slit) may introduce a small wavelength shift
(Bassa et~al. 2006). This 
wavelength shift cannot be corrected by examining the sky-line
wavelength as those fill the whole slit. 
However, it can potentially be assessed by checking the wavelengths of 
diffuse interstellar bands (DIBs; Herbig 1995). Three such sources (CX84, 
CXB149, and CXB174) have strong DIBs at $\lambda5780$ for which the 
line profiles do not deviate from Gaussians. We checked 
these features and found they have minor shifts relative to the rest-frame 
wavelength (65~km~s$^{-1}$ for CX84, $-30$~km~s$^{-1}$ for CXB149, and 
$-75$~km~s$^{-1}$ for CXB174 in heliocentric frame). We corrected the 
wavelength scale for these small shifts. Each spectrum
was normalized by dividing it by the result of fitting a 5th-order
polynomial fit to the continuum. 

\subsection{Optical Photometry from Mosaic-II \& DECam}\label{obs:photo}

\begin{figure*}
    \centering
    \includegraphics[width=3in]{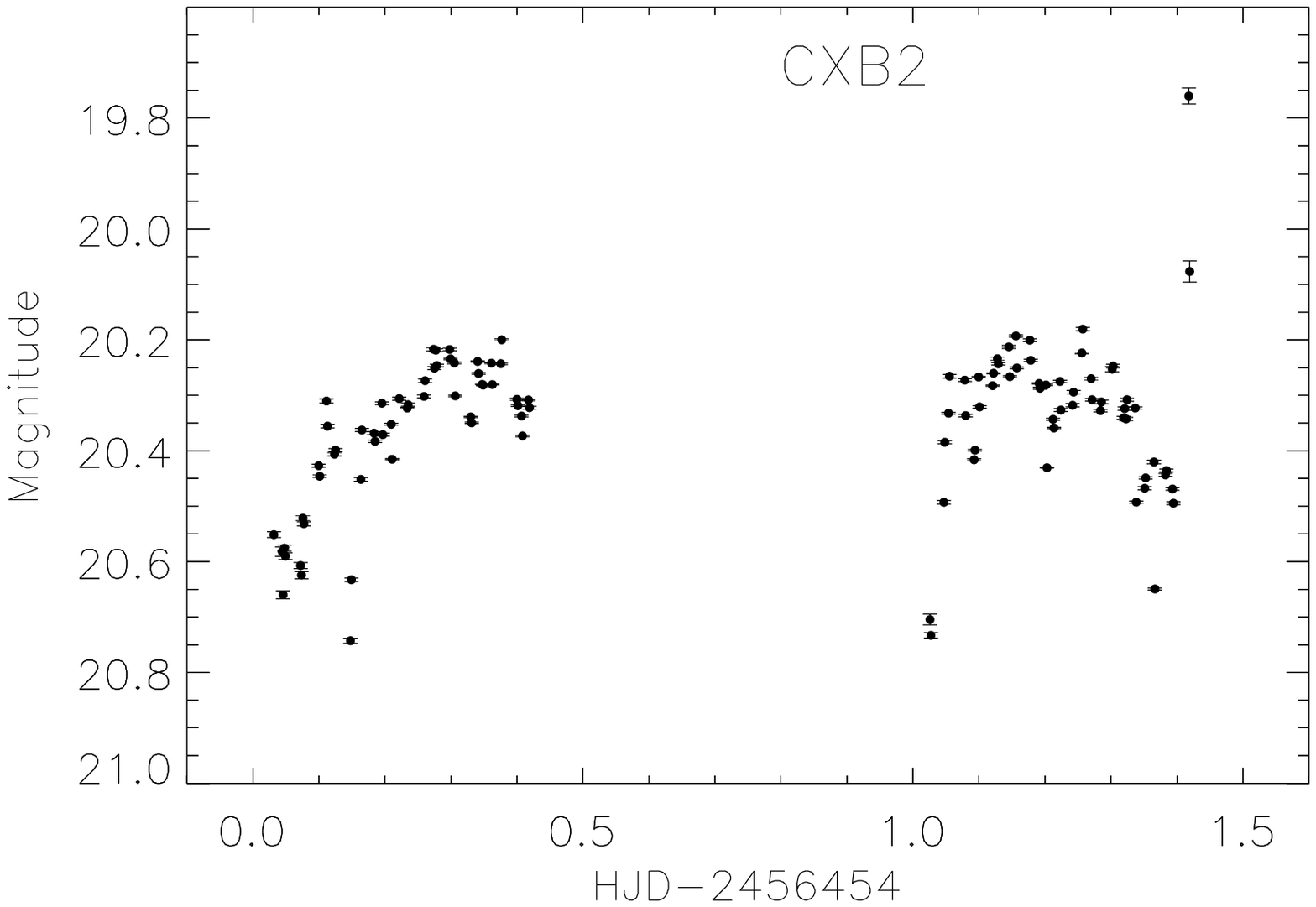}
    \includegraphics[width=3in]{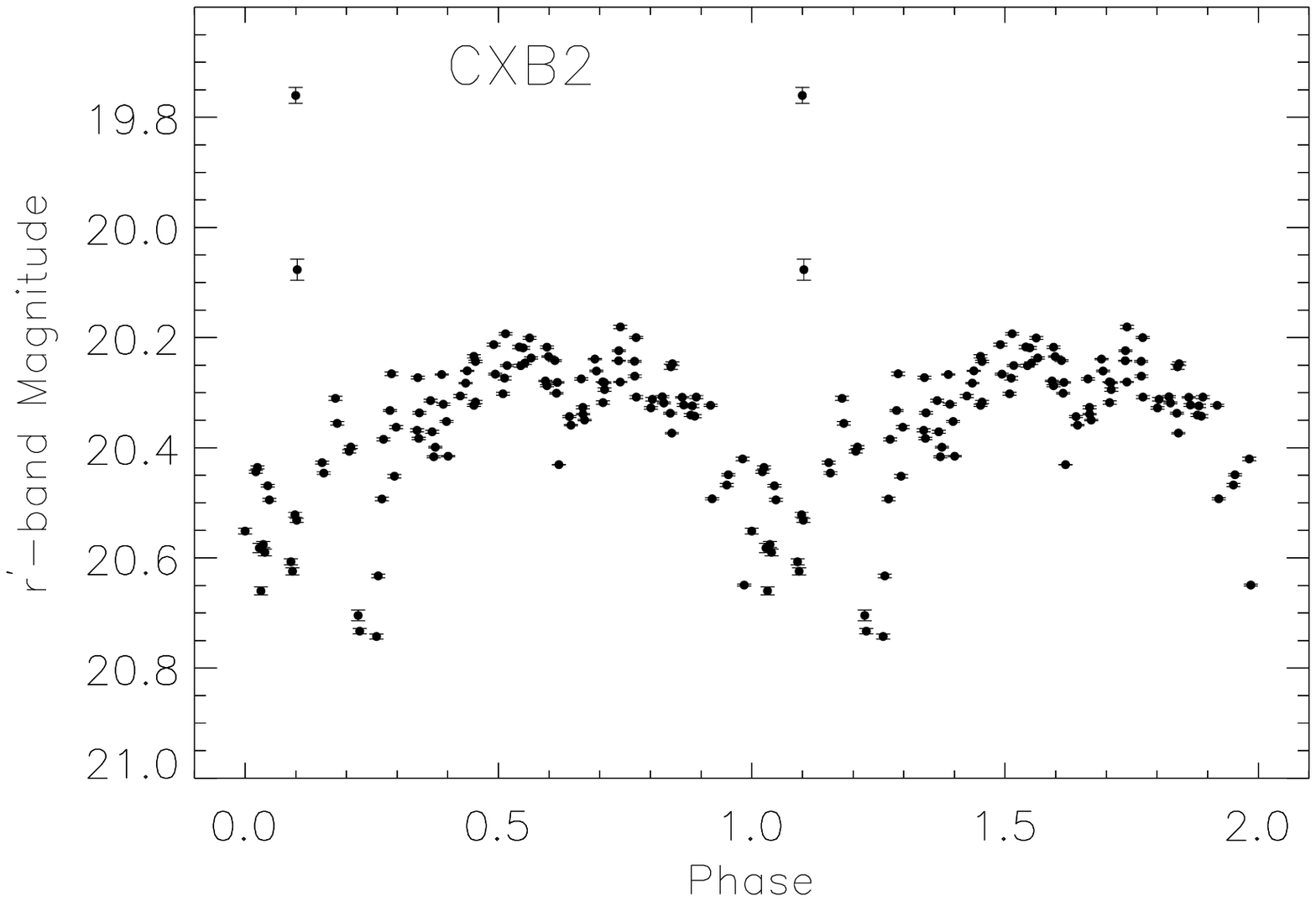}\\
    \includegraphics[width=3in]{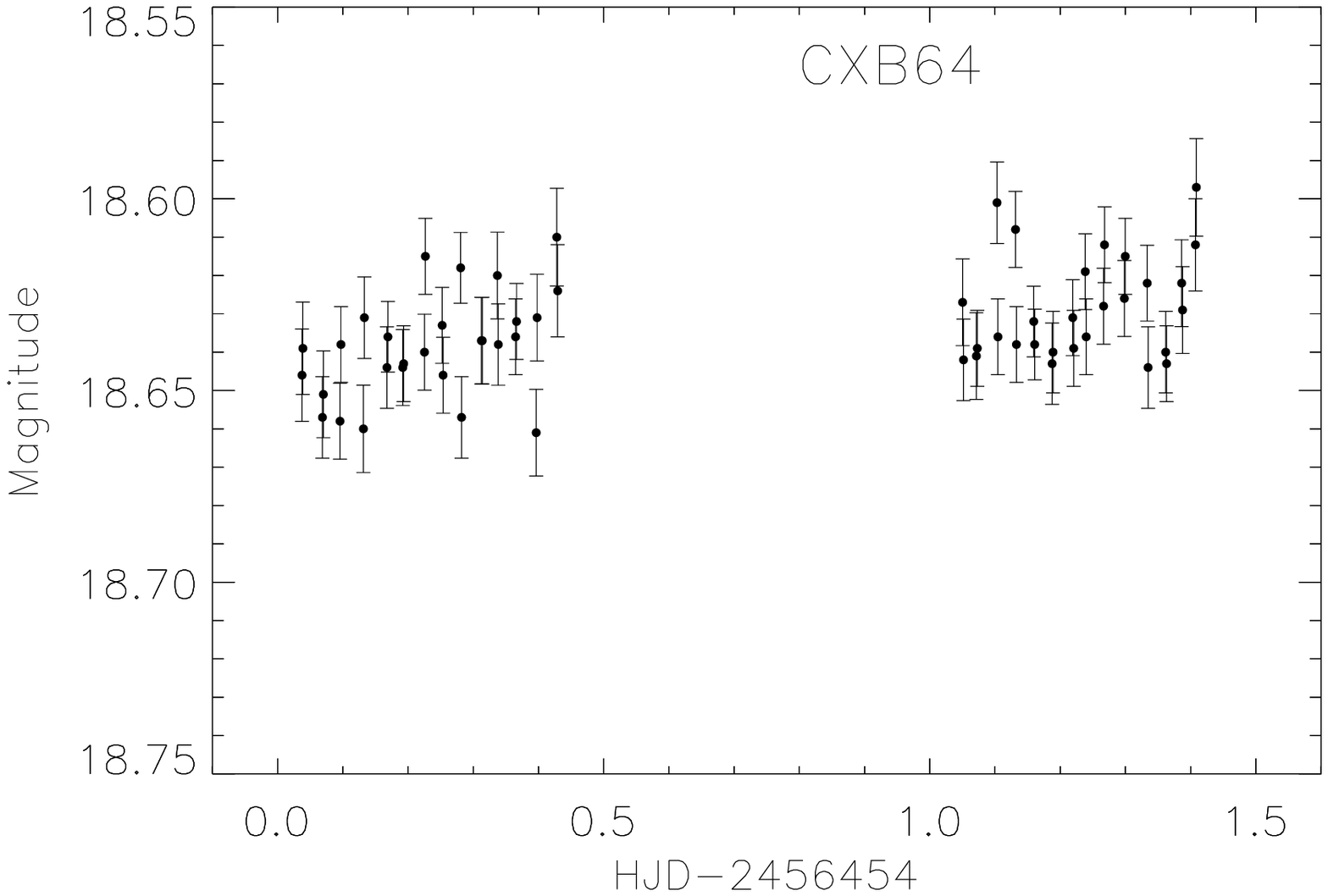}
    \includegraphics[width=3in]{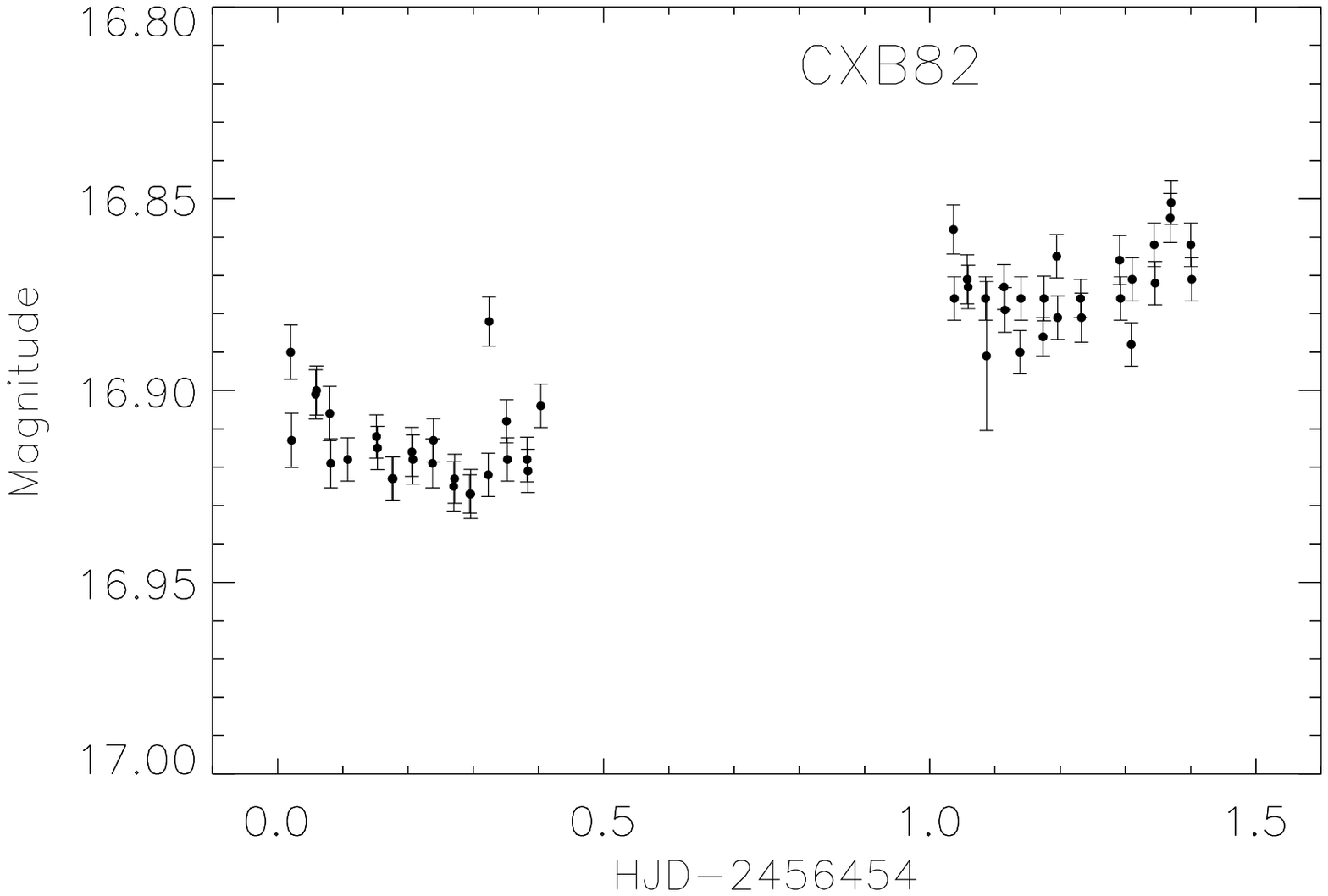}\\
    \includegraphics[width=3in]{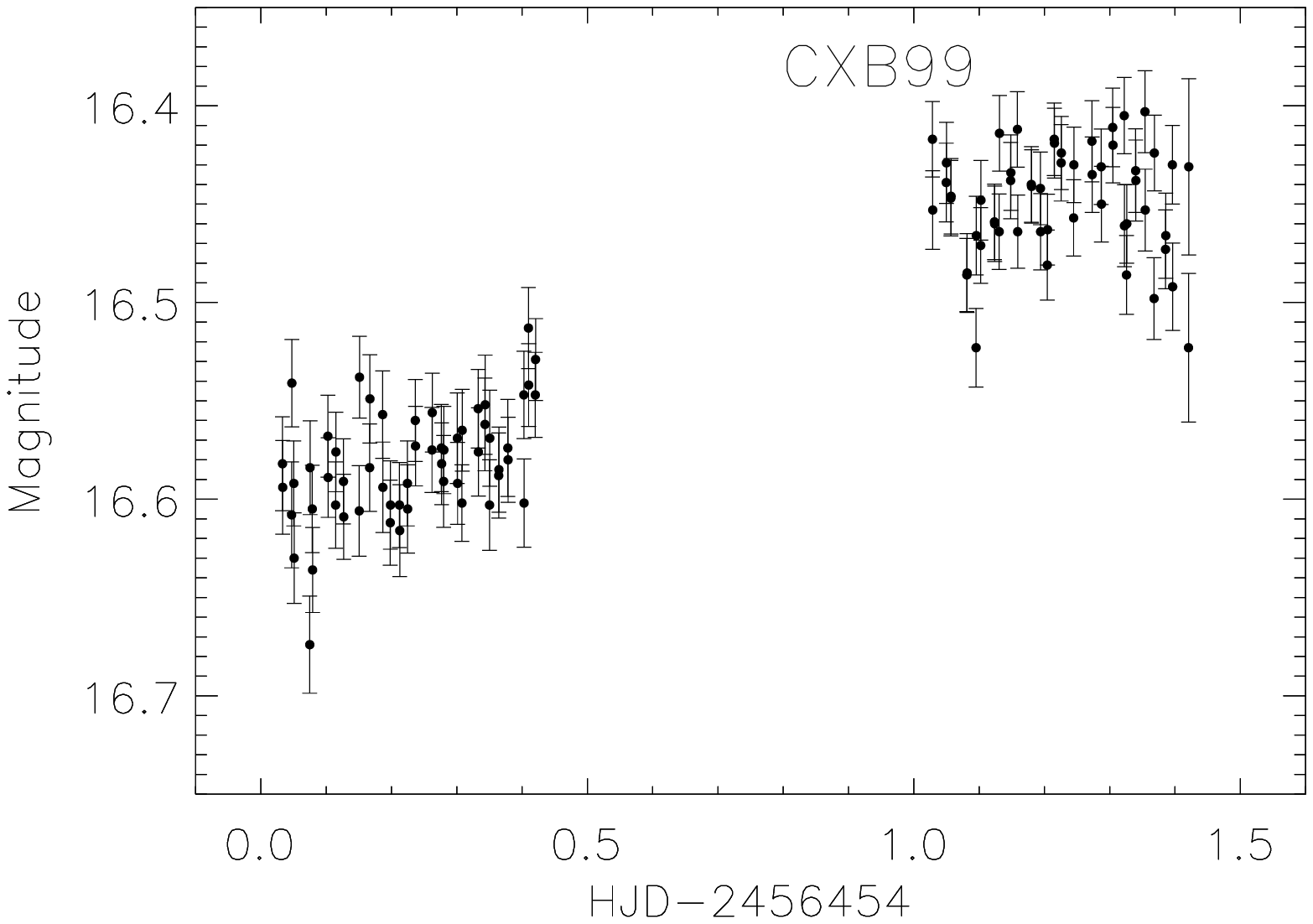}\\
    \includegraphics[width=3in]{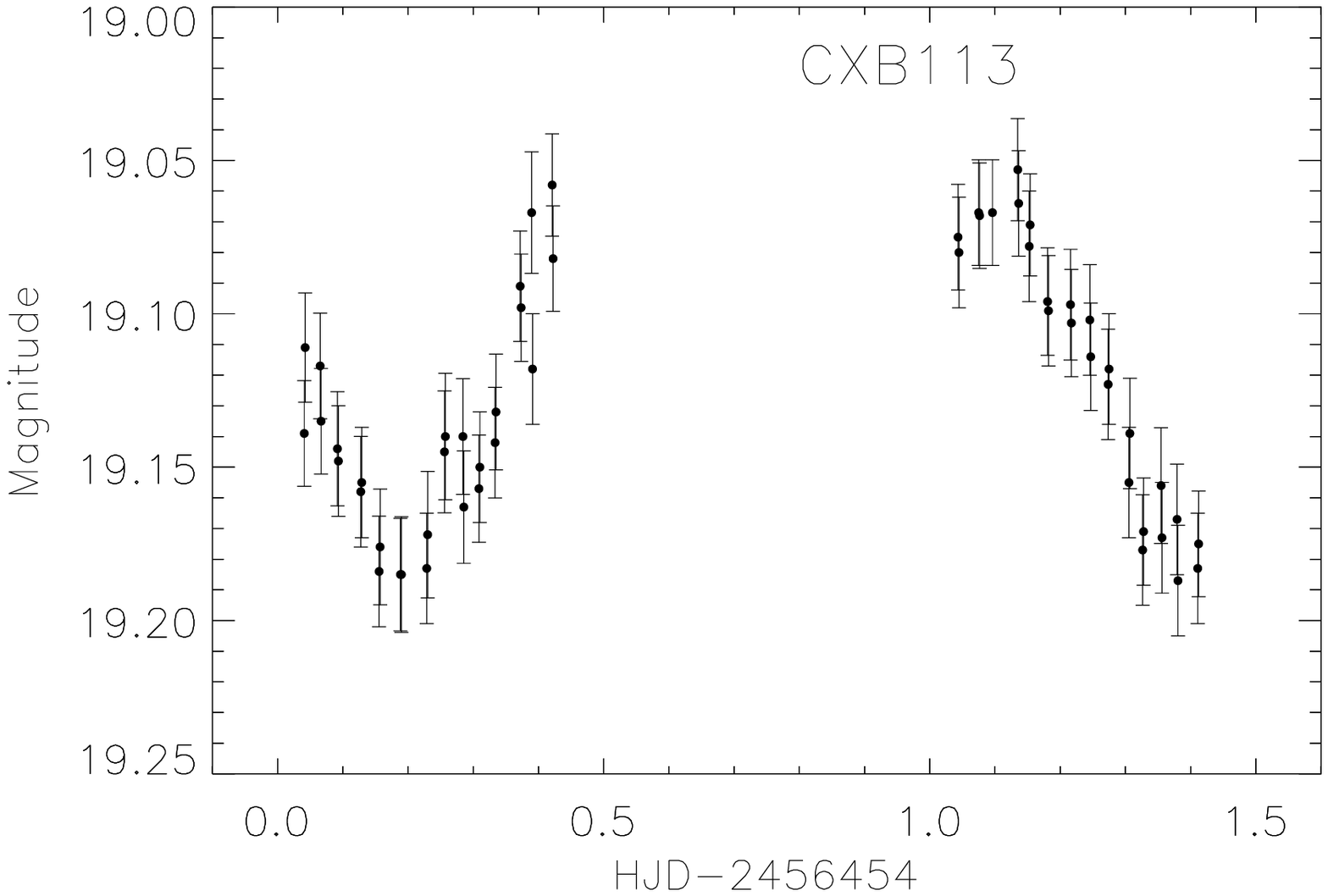}
    \includegraphics[width=3in]{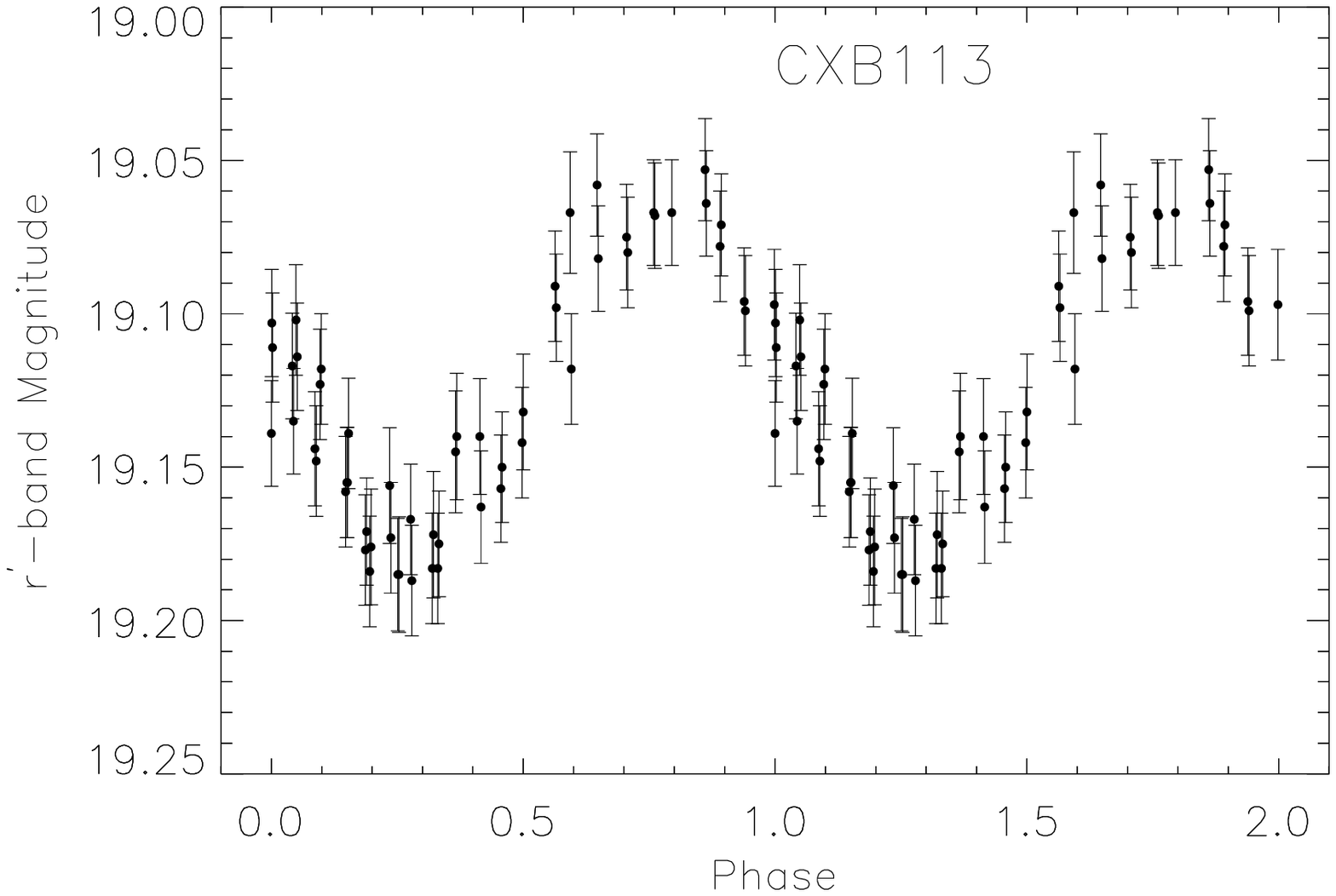}
    \caption{The DECam $r^\prime$-band light curves for CXB sources
    (CXB2, CXB64, CXB82, CXB99, and CXB113). CXB2 shows a periodic 
    modulation of 0.447 days. CXB113 has a
    sinusoidal modulation of 0.588 day in its light curve.
    The phase-folded light curves of these two objects are also included.}
             \label{lcb_fig}
\end{figure*}% 

Time-resolved optical photometry for all six CX sources was obtained
with the Mosaic-II imager mounted on the Blanco 4-meter telescope at
the Cerro Tololo Inter-American Observatory (CTIO) in 2010 July
8--15. Nineteen exposures in the Sloan $r^\prime$-band with an
integration time of 120~s were taken on 45 overlapping fields to cover
a nine square degree area, which contain all but seven of the X-ray
sources identified in Paper~I (see Britt et~al. 2014 for the full
description and results of this variability campaign). Typical seeing
for these imaging observations was around $1^{\prime\prime}$ (with a 
range of 0.8--3.0$^{\prime\prime}$). 
%During each imaging cycle, the
%order in which the 45 fields were taken was randomized to minimize the
%aliasing caused by regular sampling. 
The data were reduced  
using the NOAO Mosaic Pipeline (Shaw 2009). 
%The pipeline 
%searches for instrumental artifacts, and performs a variety of
%corrections on CCD cross talk, pupil ghost, bias and flat fields, 
%and adds a world coordinate system (WCS) for each image based 
%on USNO-B1 stars in the field (see \S2.3 of Shaw 2009). 
We performed differential photometry using Alard's image subtraction
ISIS (Alard \& Lupton 1998; Alard 2000) to obtain the changes in flux
with respect to reference images. The zero-point flux in the reference images was
measured with either aperture photometry or DAOPHOT-II (Stetson
1987). 
%The surface density of variables in the whole covered sky area
%is 0.00128 variable per arcsec$^2$. For the 1216 GBS \xray\ sources
%covered by the Mosaic-II frames, there are $\sim20$ chance interlopes within
%$2^{\prime\prime}$ of their \xray\ positions. 
The number of variable
interlopers within the 95\% error circles of the \xray\ position is
$\sim40$ ($\sim3\%; $see \S3.1 of Britt et~al. 2014 for details). 

The optical photometry information 
of four CX sources (CX84, CX377, CX446, CX1004) in this paper is shown in
Table~\ref{rmag_table}, while their light curves have been presented in 
literature (CX84 in Fig.~6 of Britt et~al. 2014; CX377, CX446 and CX1004 
in Fig.~A2 of Torres et~al. 2014). 
For CX138, there are two blended sources in the Mosaic-II image at
its \xray\ position; neither of the possible counterparts is 
variable. The counterpart to CX139 is saturated in the Mosaic-II imaging. 
Among the four CX sources with light curves, only CX84 shows possibly periodic
behaviour with a period of 4.67~days. However, it is worth noting that it is 
currently not possible to confirm the periodic nature of this
variability because the baseline of  
our monitoring was only 8 days, which is less than two full cycles.
The light curves of the other three
sources only have random flickering with an RMS$\sim0.05$--0.1 magnitude,
although CX446 possibly experiences eclipse events. 

Optical photometry for five of the CXB sources in our sample (CXB2,
CXB64, CXB82, CXB99, and CXB113) was obtained 
using the Dark Energy Camera (DECam) instrument mounted on the 
Blanco 4-meter telescope at CTIO on the nights of June 10 and 11 of
2013. The average seeing for both nights was $1.3^{\prime\prime}$ with
a range of 0.9--1.9$^{\prime\prime}$. DECam provides a $2.2\times2.2$
square degree field of view combining 62 
science CCDs, 8 focus CCDs, and 4 guiding CCDs with a scale of
0.27$^{\prime\prime}$ per pixel. For all of our images, the SDSS
$r^\prime$ filter was used with 
exposure times of either $2\times90$~s or $2\times1$~s for the faint
and bright sources, respectively. The DECam pipeline reduction provided the
resampled images with cross-talk corrections, overscan, trimmed
sections, bias subtraction, flat-fielding and saturation
masks.\footnote{See DECam Data Handbook at
  http://www.noao.edu/meetings/decam/ media/DECam\_Data\_Handbook.pdf.}  
We then used standard IRAF tasks (including
\verb+wcsctran+, \verb+digiphot+, and \verb+apphot+) to extract magnitudes and
fluxes and to generate light curves through the use of differential
photometry. Calibration of the target stars was achieved by using
reference stars in the field of view that were contained in the
Carlsberg Meridian 14 catalogue (Evans et al. 2002) and VizieR catalogue:
I/304.
%\footnote{See http://vizier.u-strasbg.fr/.} 

The photometry of
these five sources are included in Table~\ref{rmag_table} while their
light curves are shown in Fig.~\ref{lcb_fig}. Among the five GBS 
sources, CXB2 shows possible eclipsing/dipping events and an ``outburst''. 
Although the outburst was towards the end of the night with higher airmass, 
visual inspection of the images confirms that the brightening is
real. This outburst is possibly the reprocessed \xray\ emission 
from a Type~I \xray\ burst, which is similar to the case of NS LMXB 
EXO~0748$-$676 (Hynes et~al. 2006). The light curve shows a likely  
periodic modulation of $P\approx0.447$~day (see the phase-folded light curve). 
CXB82 and CXB99 both appear to be long period variables
with periods longer than our 2-night observing run. For both sources, 
there is a steady increase in brightness by about $\sim0.1$ and
$\sim0.25$ magnitude, respectively. Due to the lack of $r^\prime$
standard stars in their fields, CXB64 and CXB113 only have approximate 
magnitudes without absolute calibration. CXB64 
has a counterpart USNO B1.0 star 0578$-$0732346 with the
optical magnitude of $R\sim18.7$ and $I=16.9$. CXB113 shows a
sinusoidal modulation with period of $P = 0.58791(12)$ day. The
phase-folded light curve for this source is also shown in
Fig.~\ref{lcb_fig}. The optical counterpart to CXB113 was also 
identified by OGLE (Optical Gravitational Lensing Experiment; 
Field: BUL\_SC37, StarID: 9614; Udalski et~al. 1992). 
The OGLE source has optical magnitudes of $V =
19.135(154)$ and $I=15.847(49)$. It also shows a periodic modulation
of $P\approx0.588$~day.
%exactly the same as the period obtained from our DECam data, 
%which means the OGLE source is exactly the same object as the
%optical source that we ascribe as the counterpart to CXB113. 

\subsection{Near-Infrared Photometry}\label{obs:irphoto}

%Figure 3: source spectra
\begin{figure}
    \centering
    \includegraphics[width=3in]{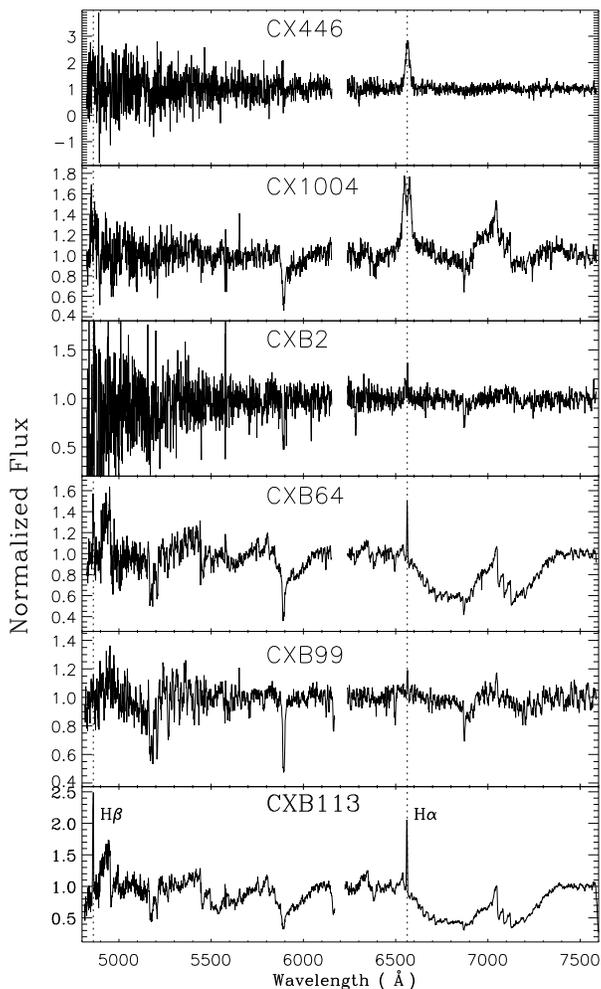}
    \caption{Gemini/GMOS spectra for six sources with
    H$\alpha$ emission lines (CX446, CX1004, CXB2, CXB64, CXB99, and
    CXB113). The broad H$\alpha$ emission of CX446, CX1004, and CXB2
    likely originate in the accretion disk, while the narrow
    H$\alpha$ and H$\beta$ emission of CXB64 and CXB113 are an
    indication of chromospherically active binaries. The H$\alpha$ emission 
    of CXB99 is narrow and weak. All the spectra
    are normalized to unity. The positions of the H$\alpha$ and the H$\beta$
    emission lines are labeled by the dotted lines.}   
             \label{spec_fig1}
\end{figure}

\begin{figure*}
    \centering
    \includegraphics[width=5.5in]{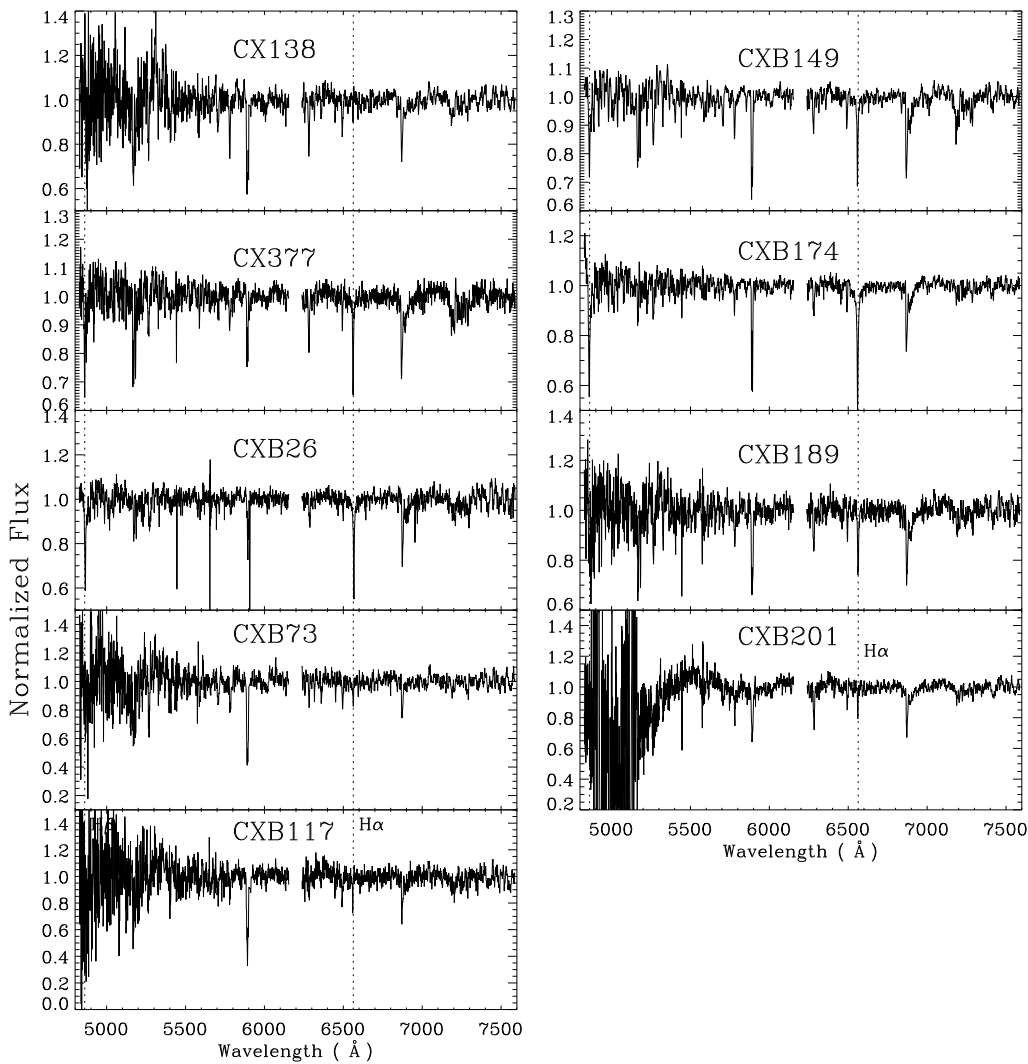}
    \caption{Gemini/GMOS spectra for nine sources with
    H$\alpha$ absorption features (CX138, CX377, CXB26, CXB73, CXB117, CXB149,
    CXB174, CXB189, and CXB201), many of which also have
    H$\beta$ absorption. All the spectra are normalized to
    unity. The positions of the H$\alpha$ and the H$\beta$ emission lines are
    labeled by the dotted lines.} 
             \label{spec_fig2}
\end{figure*}

\begin{figure}
    \centering
    \includegraphics[width=3in]{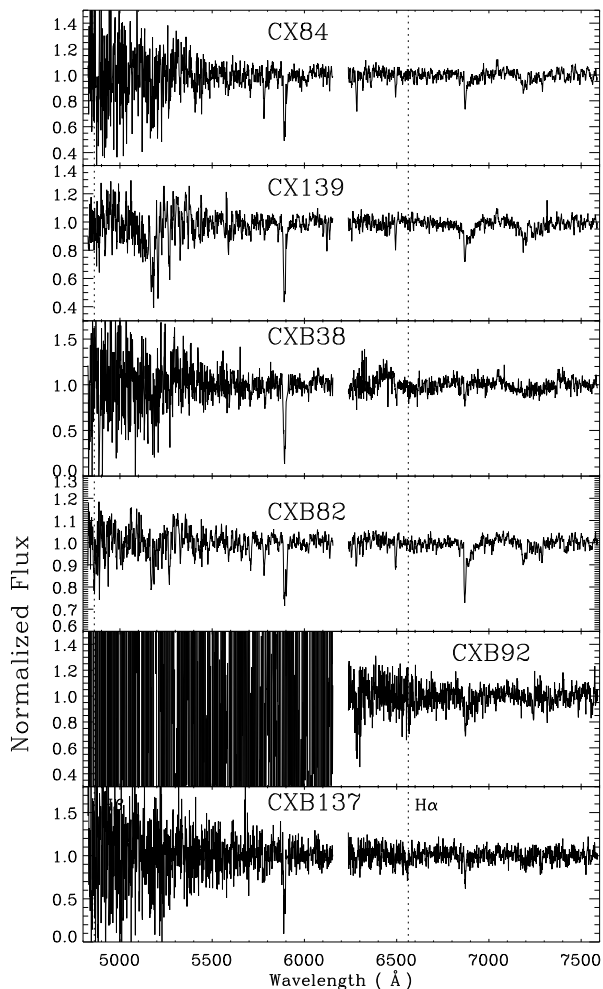}
    \caption{Gemini/GMOS spectra for the remaining 6 sources
    presented in this paper. All the spectra are normalized to 
    unity. The positions of the H$\alpha$ and the H$\beta$ emission lines are
    labeled by the dotted lines.} 
             \label{spec_fig3}
\end{figure}

The likely near-infrared matches to the GBS \xray\ sources detected by \chandra\ 
(Greiss et~al. 2014) were mainly obtained from the  
public variability survey VISTA Variables in the V\'{i}a L\'{a}ctea
(VVV; Minniti et~al. 2010). The VVV survey provides the most 
complete near-infrared coverage of the GBS area with consistent depth. 
Complementary coverage from the Two-Micron All Sky Survey 
(2MASS; Skrutskie et~al. 2006) and UKIDSS Galactic Plane Survey (GPS; Lucas 
et~al. 2008) were also utilized for bright ($K_s < 12.5$) and faint ($K_s > 16$) sources, 
respectively. Greiss et~al. (2014) developed a method to estimate the likelihood 
(by calculating the false alarm probability) of a near-infrared source to be the 
 counterpart of the X-ray source. However, for the optical sources where we have 
obtained Gemini/GMOS spectra, we are able to search for the near-infrared 
counterparts that matches our optical sources by comparing the optical and
near-infrared images. We set a $0.2^{\prime\prime}$ astrometric error
circle, and searched for any matches within that  
error circle between the optical images and VVV images. Then we visually examined both 
images and selected the true near-infrared counterparts. A visual inspection is crucial 
given that the GBS fields are crowded.

Table~\ref{vvv_table} lists the near-infrared counterparts to 13 of the 21 optical 
sources with Gemini/GMOS spectroscopy. The other eight sources do not have 
reliable near-infrared counterparts. Three of them (CX446, CXB26, and CXB137) are 
too faint in the VVV frames and there are no near-infrared sources at the 
optical position. For the other five sources (CX1004, CXB64, CXB117, CXB189, 
and CXB201), there could be near-infrared matches but they are blended with 
nearby sources, for which the current VVV data release does not provide photometry. 

The infrared colours can be 
used to estimate the distance of the source by calculating the needed
extinction to match the source infrared colours to those of a standard
star with the same spectral type (e.g., Ratti et~al. 2013). Some of the GBS
sources have an infrared excess, i.e., their infrared colours are
redder than expected based on optical colours (see \S4 of Hynes
et~al. 2012), which is possibly for cases where the optical classification
may represent the hotter component of a binary (or of a blend, if another 
star is serendipitously along the same line of sight) while the infrared
colours are from the cooler component. Accreting systems (CVs and
LMXBs) and Be stars with a circumstellar disk
may also have colours that do not match those of single stars. 
Be stars could also have significant colour variations due to the 
formation/dissolution of the circumstellar disks.

%Table 2: Optical photometry for Gemini Sources

%Figure 2: Light curves (only for interesting objects)

% -----------------------------------------------------------------------------
% Results
% -----------------------------------------------------------------------------

\section{Data Analysis}\label{ana}

%\subsection{Spectral Features}\label{ana:features}

The Gemini/GMOS spectra of the 21 GBS sources are shown in 
Fig.~\ref{spec_fig1}--\ref{spec_fig3}. Three sources (CX446, CX1004, 
and CXB2) show apparent broad H$\alpha$ emission lines, while another three 
(CXB64, CXB99, and CXB113) have narrow H$\alpha$ emission (see
Fig.~\ref{spec_fig1}). We fit Gaussian profiles to these broad H$\alpha$
  emission lines to measure their width and velocity separations (if
  double-peaked) using the \verb+mgfit+ procedure in the {\sc molly}
  package. The equivalent widths (EWs) of these H$\alpha$ lines were measured 
  using the \verb+light+ procedure in {\sc molly}. Nine other objects
  have H$\alpha$ absorption features (Fig.~\ref{spec_fig2}). The stellar 
  absorption features 
  shown in the Gemini/GMOS spectra can be
  utilized to perform spectral classification, and radial velocity
  analysis. The spectra of the remaining six sources 
  are shown in Fig.~\ref{spec_fig3}; they appear at first sight to 
  have neither H$\alpha$ 
  emission features nor H$\alpha$ absorption features. 

Some of the objects in our sample have strong DIBs 
in their spectra. We estimated the reddening
for these sources via the equivalent width (EW) of the DIB at
$\lambda5780$ with the calibration in Table~3 of  
Herbig (1993). We also compared the measured reddening to the Bulge
reddening along the line of sight provided by Gonzalez
et~al. (2011,2012), which utilized the Red Clump stars in the
Bulge.\footnote{See the Bulge Extinction And Metalicity (BEAM)
  calculator at http://mill.astro.puc.cl/BEAM/calculator.php.}   

\subsection{Spectral Classification}\label{ana:class}

%The width of the absorption features, presumably originated from the 
%companion star, is expected to be dominated by the rotational broadening. 
We utilize the optimal subtraction technique following previous works 
(e.g., Marsh et~al. 1994; Ratti et~al. 2013) to classify the spectra. 
A set of template star spectra was chosen from the library of the 
Ultraviolet and Visual Echelle Spectrograph Paranal Observatory Project 
(UVES POP; Jehin et~al. 2005),
%\footnote{http://www.eso.org/sci/observing/tools/uvespop.html},
covering spectral types from A0 to M6 with luminosity class of V 
(see Table~\ref{uves_table}). The UVES POP template spectra provide
coverage over the wavelength range of 3000--10000~\AA\ with
a spectral resolution of $\sim80,000$. The templates were re-sampled and 
Gaussian-smoothed to match the spectral resolution of the object spectra. 
The object spectra were Doppler-shifted into the same rest frame and averaged. 
Each stellar template was optimally subtracted from the object 
spectrum, while a $\chi^2$ test was performed on the residuals. 
%A scaling factor representing the spectral contribution
%from the donor star could also be determined in the procedure.
%However, since we are not controlling luminosity class in the spectral 
%classification, we do not have constraints on this scaling factor.  
All the emission lines, DIBs, 
and telluric features (e.g., Kurucz 2006; Wallace et~al. 2011) were 
masked during the procedure. The resulting $\chi^2$ values for each template 
were compared with each other. The template
with the minimum  $\chi^2$ value provides our best estimate of the
spectral classification of each GBS source in
our sample. These spectral classification procedures were first performed 
on the red part of the Gemini spectra. We verified our results by performing 
the same procedures to the middle part of the Gemini spectra and the 
results are consistent with each other. The results of spectral
classifications are listed in Table~\ref{vsini_table}. Note we are 
not controlling luminosity class in the spectral classification. Nine of 
the 21 GBS sources in our sample were spectrally classified. 
The uncertainty of spectral classification is estimated to be 
one or two spectral sub-classes.
%\footnote{The uncertainty can be derived
%  in principle by searching for the spectral types resulting
%  $\Delta\chi^2=1$. However, for all of our sources, the adjacent spectral
%  types in our set of templates (e.g., G6 and K2 are adjacent to G9)
%  generate $\Delta\chi^2>1$. Therefore, the uncertainty range of the
%  spectral classification should be within the adjacent spectral
%  types.} 
The other three sources listed in Table~\ref{vsini_table} are
classified as early--mid M-type based on their spectral features. 
%For other sources, we were not able to obtain reliable minimum
%$\chi^2$ values from the $\chi^2$--spectral type curve. 

%For GBS sources with more than one Gemini/GMOS spectrum, 
%the normalized spectra were
%Doppler-shifted into the same rest frame (the frame of the
%first spectrum of this source) and then averaged. All the stellar 
%templates were Gaussian-smoothed using the {\sc molly} procedure \verb+gsm+ 
%to match the spectral resolution of the source spectra, and then 
%shifted to the rest frame of the averaged source
%spectrum. If the seeing $d$ (measured from the acquisition image) for
%a Gemini/GMOS spectrum is greater than the slit width
%($0.75^{\prime\prime}$), the spectral resolution should be the same as
%that measured from the arc spectra ($\approx5$~\AA; see
%\S\ref{obs:spec}) because the source light fills the slit. If the
%seeing $d$ is less than the slit width, the spectral resolution is
%estimated to be $\approx(d/0.75^{\prime\prime})\times5$~\AA. 
%The stellar templates were re-sampled to match the velocity dispersion of the
%source spectrum using the \verb+vbin+ procedure in {\sc molly}. All
%the rebinned template spectra were normalized using a 5th-order
%polynomial continuum fit. The velocity dispersion per pixel in the 
%red part of the source spectra is $\approx60$~km~s$^{-1}$; the velocity 
%resolution (corresponding to the $\approx5$~\AA\ spectral resolution) 
%is $\approx230$~km~s$^{-1}$. 

We have also tried to measure the rotational broadening of the
absorption features. However, our limited spectral resolution of
$\sim200$~km~s$^{-1}$ precluded the determination of rotational
velocities.

\subsection{Radial Velocity Analysis}\label{ana:vel}

The radial velocities of the optical counterpart of each GBS source are
measured by cross-correlating the object spectra using the \verb+xcor+
procedure in {\sc molly}. For each source, all the object spectra and
the template
spectrum were rebinned to the same velocity dispersion with the
\verb+vbin+ procedure. We also masked all the emission lines,
interstellar/telluric features  
during the cross-correlation procedure. We performed two sets 
of the cross-correlation analysis. The first set uses the first source 
spectrum as the cross-correlation template, i.e., we derive the radial 
velocity (RV1) relative to the first source spectrum. For the GBS sources that 
were spectrally classified using the procedure in \S\ref{ana:class}, 
we also performed a second set of cross-correlation analyses, taking the 
UVES POP standard star with the best-fit spectral type as the template. 
Thus the second set of radial velocity (RV2) is relative to the template 
star. All spectra had been shifted to a heliocentric frame.

Radial velocity values were derived by fitting a Gaussian profile to the 
cross-correlation function. The results are listed in
Table~\ref{rvc_table}. Seven sources have non-zero radial velocities
relative to the standard star template. Three sources (CX84, CX138,
CX139) have radial velocity variations of $\sim100$~km~s$^{-1}$ on
timescales of days.   
%One third (7/21) of the sources do not 
%show significant cross-correlations between the source spectra, and 
%thus there is no radial velocity measurement. 
%The other 14 sources show significant 
%cross-correlations among source spectra and/or between source spectra and standard 
%star templates, the majority of which, however, have relative radial
%velocities that are consistent or close to zero. Three 
%sources (CX84, CX138, CX139) have radial velocity variations of $\sim100$~km~s$^{-1}$ 
%on timescale of days. The results of the radial velocity analysis are
%listed in Table~\ref{rvc_table}. 

%Table 4: radial velocity values for interesting sources

%Figure 3: radial velocity curves for interesting sources

% -----------------------------------------------------------------------------
% Results \& Discussion
% -----------------------------------------------------------------------------

\section{Results \& Discussion}\label{discuss}

%\subsection{Identification of Accreting Binaries}\label{discuss:hidden}

%In this paper we report on the Gemini/GMOS spectra of a sample of
%candidate counterparts to \xray\ sources discovered in the GBS (Jonker
%et al. 2011; 2014). We have performed the data analysis procedures
%outlined in \S\ref{ana} for all these Gemini spectra. We are able 
%to classify the spectral types for nine of the 21 sources that have Gemini/GMOS 
%spectroscopy (see Table~\ref{vsini_table}). For these sources, we
%generated the residual spectrum after optimally subtracting the
%best-fit stellar template from the object spectra (see Fig.~\ref{resid_fig}). 
%In the following subsections, we will discuss the identification of
%accreting binaries based on their H$\alpha$ features. 

\subsection{Objects with H$\alpha$ Emission Lines}\label{discuss:emission}

There are six sources with apparent H$\alpha$ emission in their
Gemini/GMOS spectra (Fig.~\ref{spec_fig1}). Three of them (CX446,
CX1004, and CXB2) have broad H$\alpha$ emission lines
(FWHM$\simgt800$~km~s$^{-1}$), while those of the other three sources
(CXB64, CXB99, and CXB113) are narrow (FWHM$\simlt200$~km~s$^{-1}$). The
strong, broad H$\alpha$ emission likely originates from an accretion
disk, indicating CX446, CX1004, and CXB2 are likely accreting
binaries (\S\ref{discuss:cx446}). 

CXB64 and CXB113 are possibly 
chromospherically active stars or binaries because of their narrow 
and weak H$\alpha$ emission lines (Torres et~al. 2014; also see Fig.~\ref{spec_fig1}). 
%and molecular absorption features from late-type stars
The $H-K$ colour ($0.300\pm0.033$; see
Table~\ref{vvv_table}) of CXB113 agrees well with that 
of a M4V--M5V star ($H-K\approx0.30$; see Table~5 of Pecaut \& Mamajek
2013), while its $J-H$ colour ($0.509\pm0.030$) appears 
slightly bluer than that of a M4V--M5V type star ($J-H\approx0.57$). 

CXB99 has a weak, narrow H$\alpha$ emission. After the optimal subtraction 
with the best-fit K2V template, the H$\alpha$ emission appears to be 
stronger, indicating that it partially fills in the H$\alpha$ absorption 
line in the spectrum.

\subsubsection{Potential Quiescent Accreting Binaries: CX446,
  CX1004, and CXB2}\label{discuss:cx446} 
 
%\subsection{CXOGBS~J174627.1$-$254952 (CX446)}\label{discuss:cx446}

\begin{figure}
    \centering
    \includegraphics[width=3.4in]{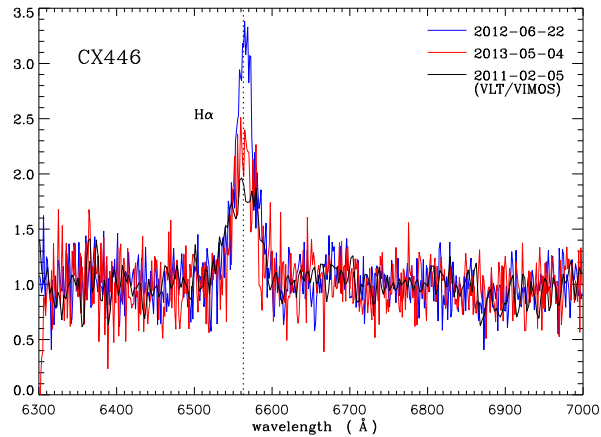}
    \caption{The H$\alpha$ region of both epochs of 
    Gemini/GMOS spectra for CX446, acquired on 2012 Jun 22 (blue line) and
    2013 May 4 (red line), respectively, showing evidence for strong
    variability in the EW of the H$\alpha$ between the two epochs. 
    The VLT/VIMOS spectrum of CX446 is also overlaid (black line).}
             \label{cx446_fig}
\end{figure}% 

\begin{figure}
    \centering
    \includegraphics[width=3.4in]{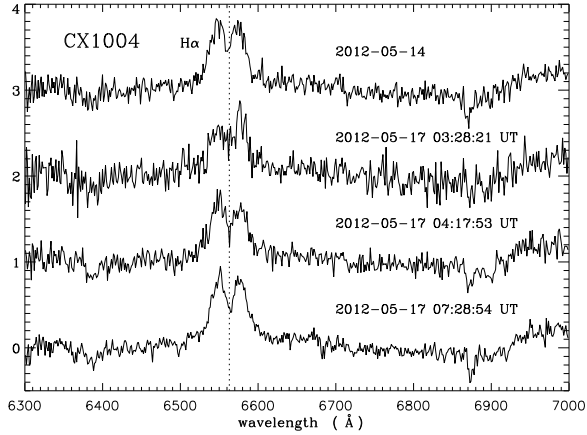}
    \caption{The H$\alpha$ region of the four epochs of 
    Gemini/GMOS spectra for CX1004 (from top to bottom in
    chronological order; see Table~\ref{log_table}). All four spectra
    show a broad, double-peaked H$\alpha$ profile which varied in 
    strength between the observations.}
             \label{cx1004_fig}
\end{figure}

\begin{figure}
    \centering
    \includegraphics[width=3.4in]{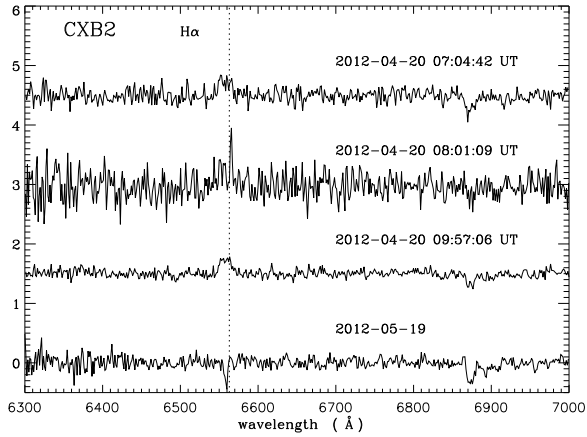}
    \caption{The H$\alpha$ region of the four epochs of 
    Gemini/GMOS spectra for CXB2 (from top to bottom in
    chronological order; see Table~\ref{log_table}). The spectra of
    the first and third epochs show broad, weak H$\alpha$ emission,
    whereas it is absent in the bottom spectrum. Both the emission and
    absorption features have substantial negative velocity offset
    which may indicate the high space velocity of this object.}
             \label{cxb2_fig}
\end{figure}

Both epochs of the Gemini/GMOS spectra of CX446 show broad H$\alpha$
emission (see Fig.~\ref{cx446_fig}). The strength of the H$\alpha$
line is weaker in the spectra taken on 2013 May 4 than that on 2012
Jun 22. The line width ($1250\pm50$~km~s$^{-1}$) is smaller than that 
of the H$\alpha$ emission lines 
in the VLT/VIMOS spectra of CX446 ($2200\pm50$~km~s$^{-1}$; Torres
et~al. 2014), while the line EW is bigger. No absorption lines
from the companion star are visible in the Gemini spectra. We also see
no evidence of He~{\sc i}~$\lambda$~6678. The light curve of CX446 does  
show a possible eclipse event with a depth of 0.4 magnitude (see
Fig.~A2 of Torres et~al. 2014; on HJD$=2455387.82$). No significant periodicity is
found in the light curve. CX446 is a candidate eclipsing CV or 
qLMXB.

%\subsection{CXOGBS~J174623.5$-$310550 (CX1004)}\label{discuss:cx1004}

The Gemini/GMOS spectra of CX1004 show double-peaked H$\alpha$
emission in all four epochs (see Fig.~\ref{cx1004_fig}). We measure
the H$\alpha$ emission line properties using the spectra of the last
epoch because of its best $S/N$. The line width is FWHM =
$2500\pm100$~km~s$^{-1}$, which is broader than that in the VLT/VIMOS
spectra of CX1004 (FWHM = $2100\pm20$~km~s$^{-1}$; see Torres
et~al. 2014). The line strength (EW$=38.0\pm0.6$~\AA) has also slightly
increased compared to that in the VLT/VIMOS spectra
(EW$=32.9\pm0.4$~\AA). The velocity separation between the red and
blue peaks 
is $\Delta v = 1160\pm30$~km~s$^{-1}$, which is consistent with the
result in Torres et~al. (2014). The centroid of the H$\alpha$ line does 
not have significant radial velocity (velocity offset $-15\pm20$~km~s$^{-1}$), 
while that of the VLT/VIMOS spectra shows substantial radial velocity 
($-170\pm20$~km~s$^{-1}$; Torres et~al. 2014). One possible
scenario for this is that the accretion disk of CX1004 is
precessing. The double-peaked profile of the H$\alpha$ emission line
is asymmetric in three of the four epochs; the relative strength of
the  
blue peak and the red peak varies with time (see Fig.~\ref{cx1004_fig}). 
%A similar H$\alpha$ profile variation is also found for \xray\ binaries, 
%e.g., A0620$-$00 (see Johnston et~al. 1989), for which the disk-originated 
%H$\alpha$ emission profile evolves with orbital phase. 
This behaviour can be
explained by the presence of an S-wave originating in a hot-spot or
the donor star (see \S IV of Johnston et~al. 1989). 
%The deep absorption features in
%between the double peaks of the H$\alpha$ emission is an indication of
%the high system inclination for CX1004.  

\begin{figure*}
    \centering
    \includegraphics[width=2.1in]{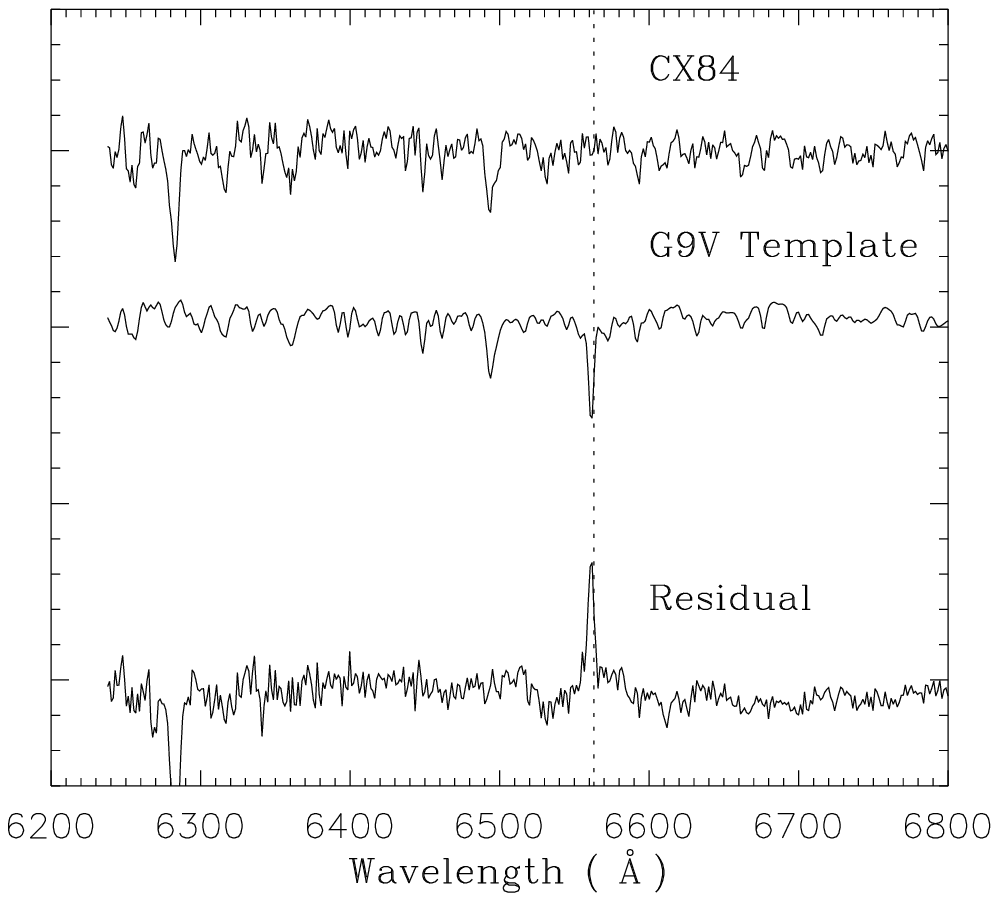}
    \includegraphics[width=2.1in]{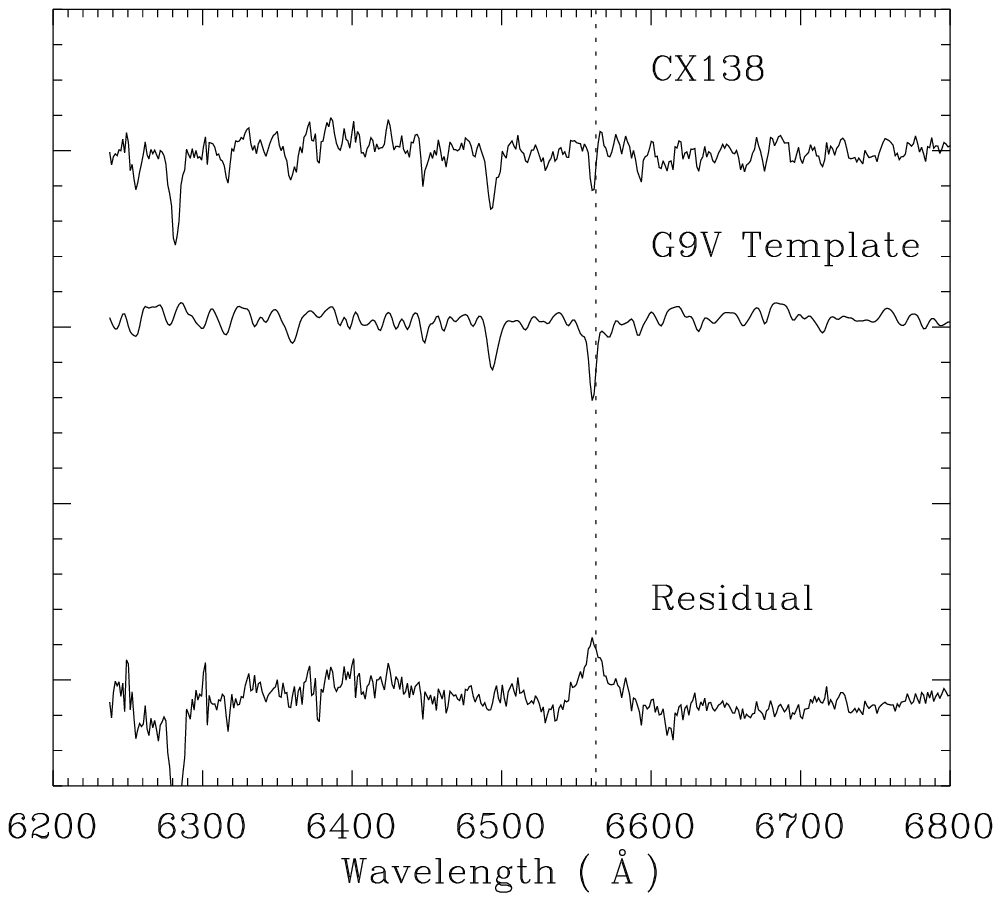}
    \includegraphics[width=2.1in]{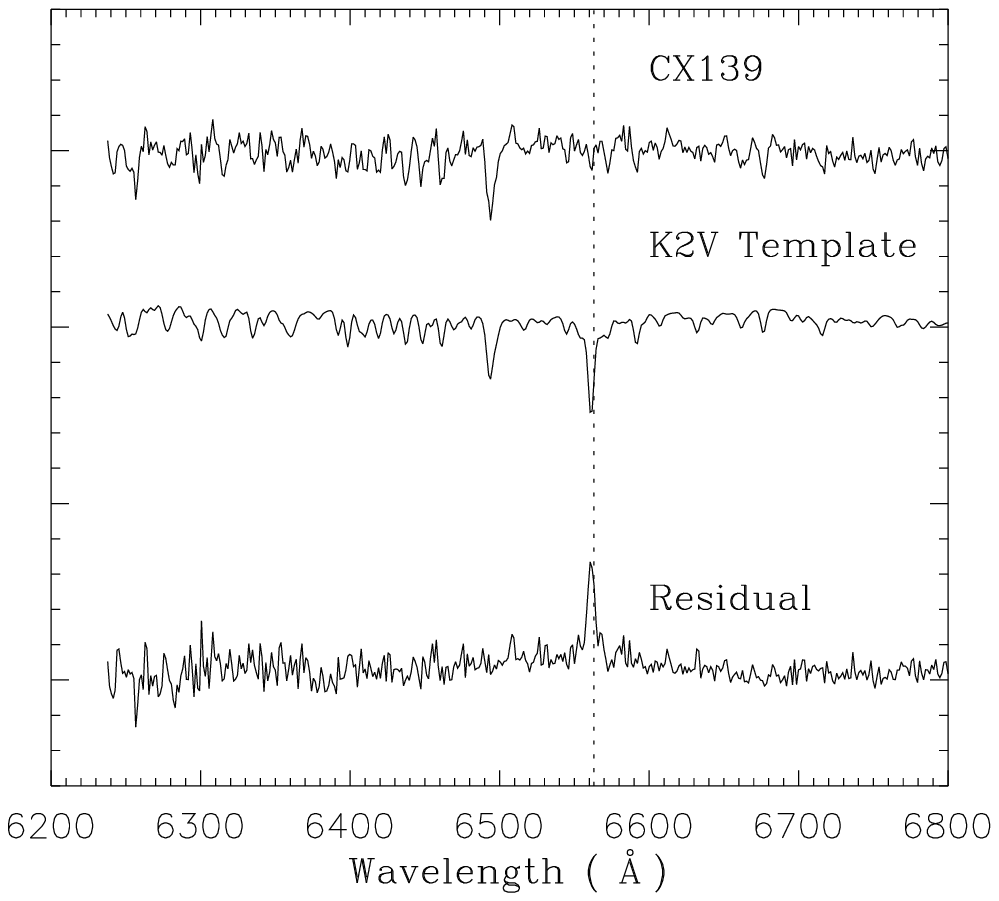}\\
    \includegraphics[width=2.1in]{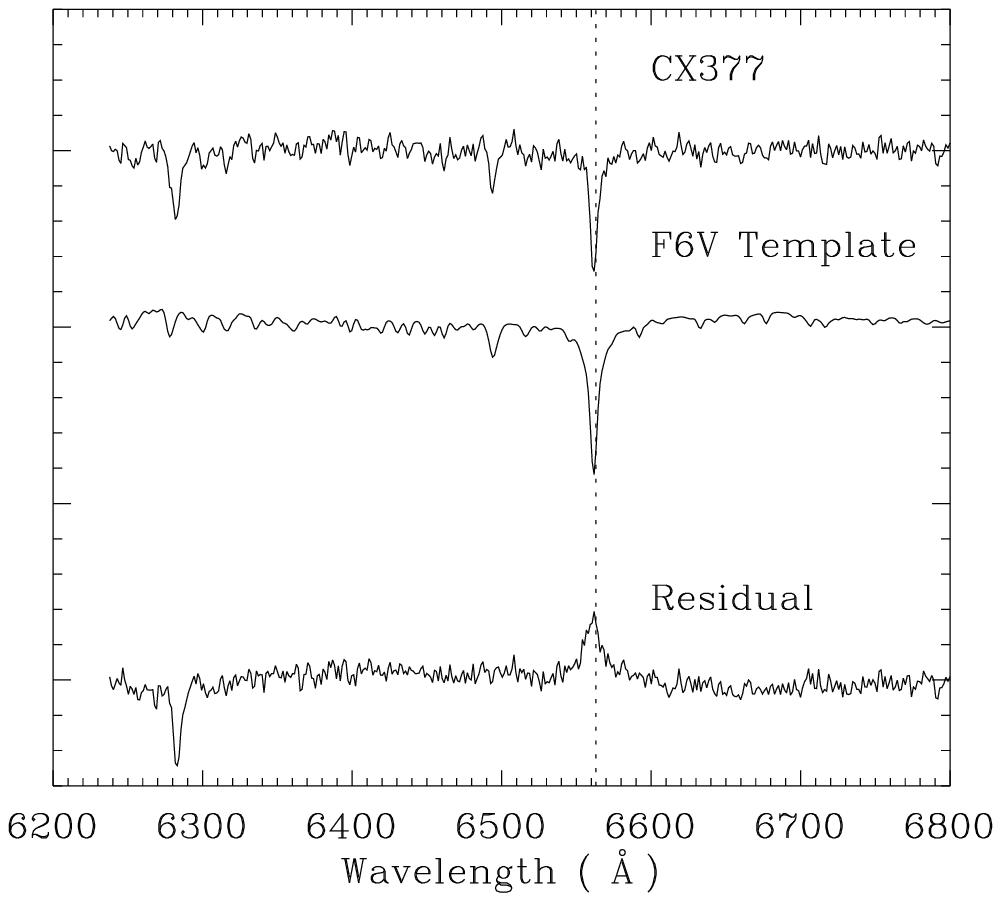}
    \includegraphics[width=2.1in]{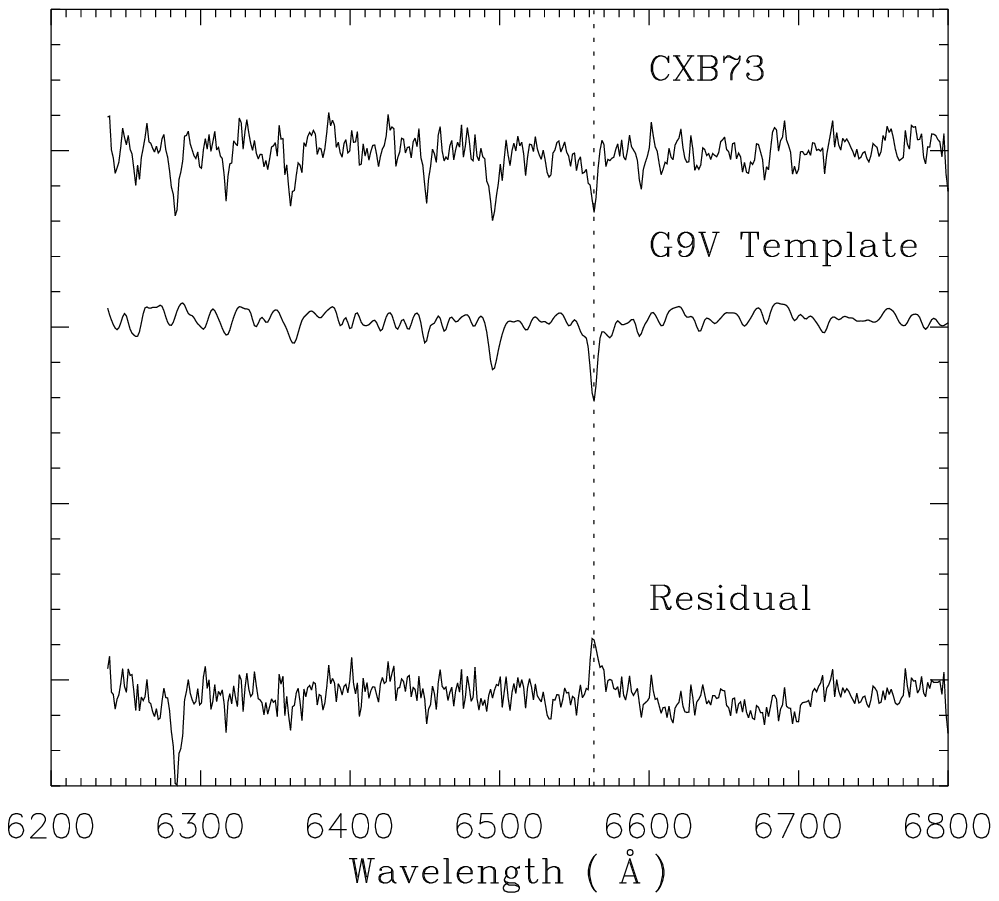}
    \includegraphics[width=2.1in]{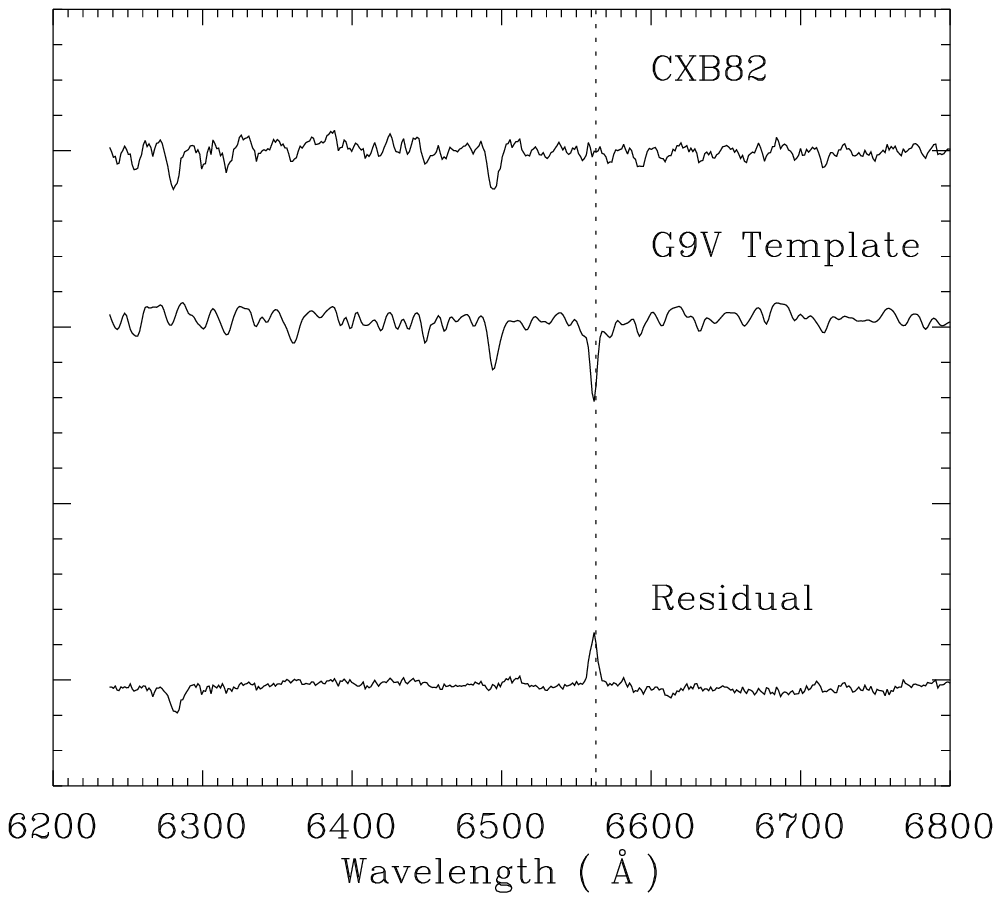}\\
    \includegraphics[width=2.1in]{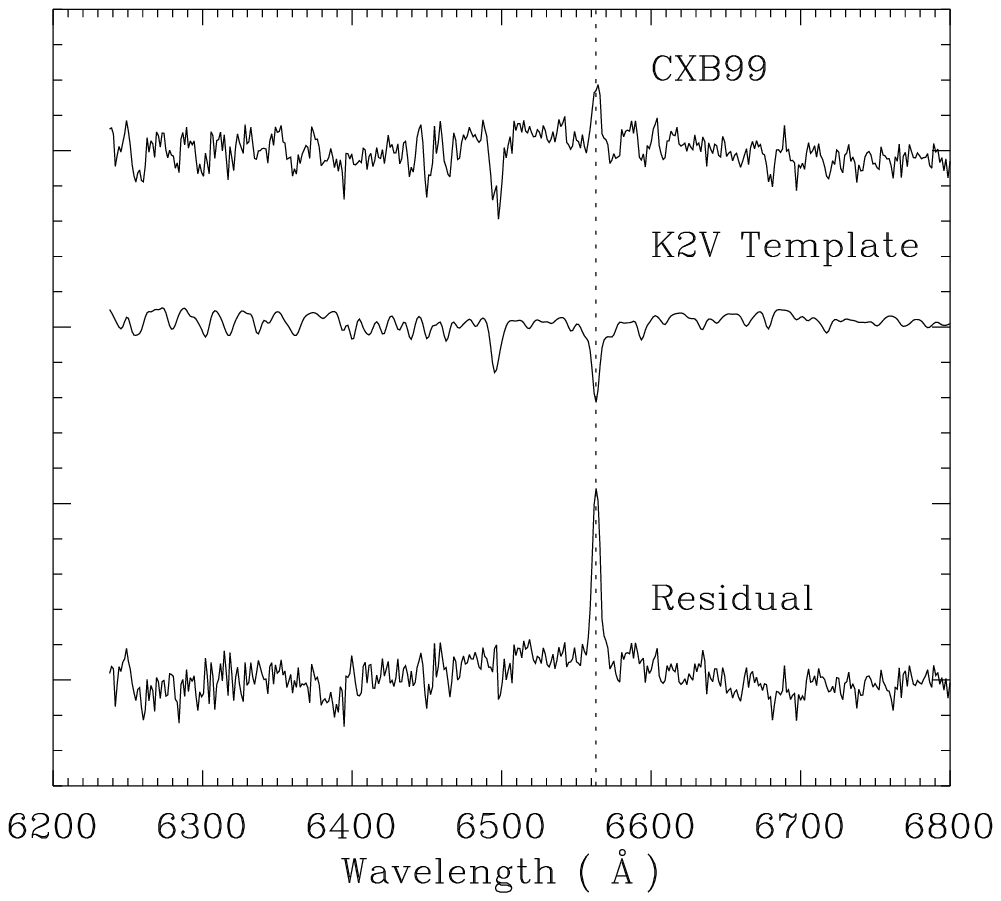}
    \includegraphics[width=2.1in]{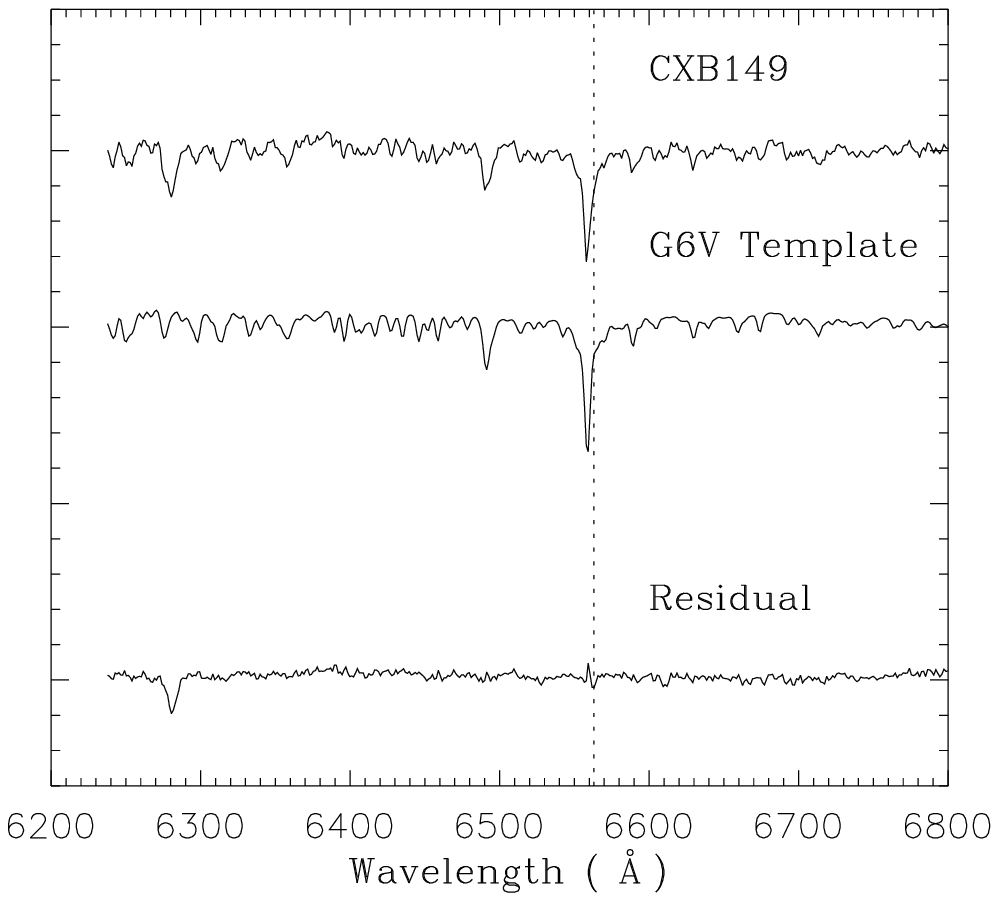}
    \includegraphics[width=2.1in]{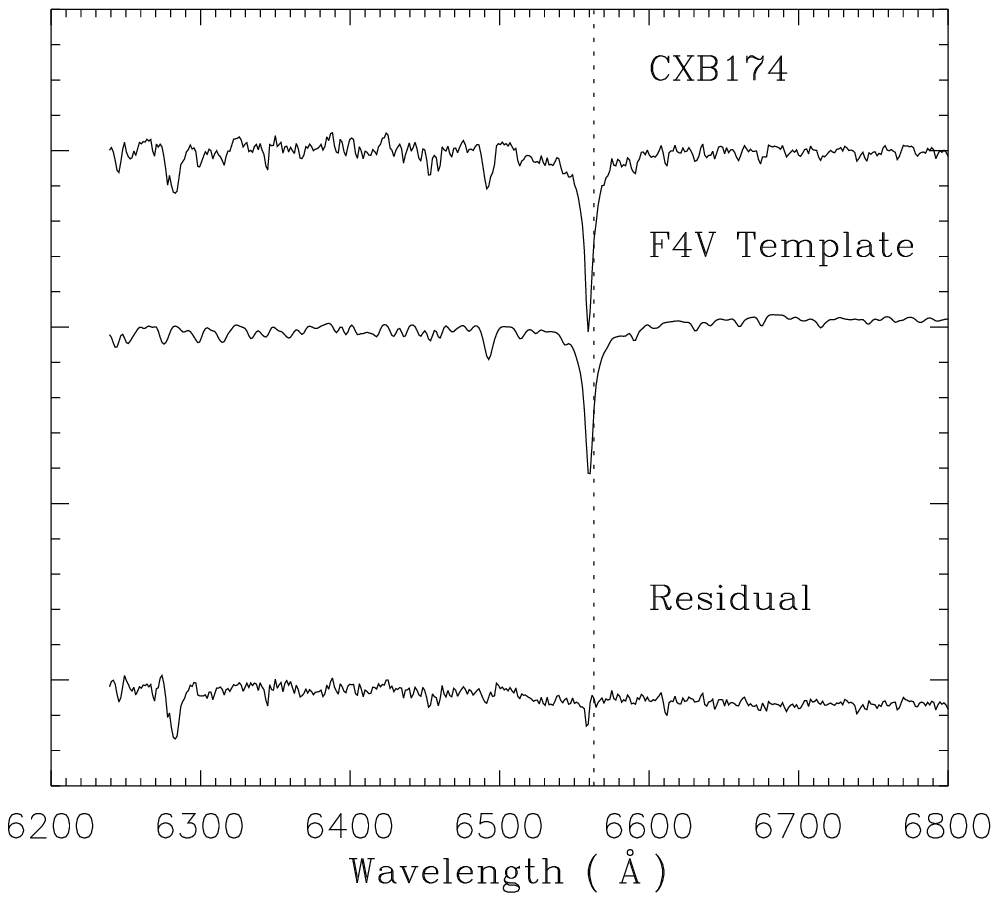}
    \caption{The residual spectra for the nine spectrally classified sources 
    (CX84, CX138, CX139, CX377, CXB73, CXB82, CXB99, CXB149, and
    CXB174) after optimally subtracting the 
    standard star template. The dotted line in each panel labels the position of H$\alpha$. 
    All the residual spectra show H$\alpha$ emission features, except
    CXB149 and CXB174. For all panels we plot the observed spectrum on top, the best-fit 
    template star spectrum in the middle and the residual spectrum, obtained after 
    subtracting the best-fit template from the observed spectrum, at the bottom. The offset 
    between the three spectra per panel is chosen so that the spectral
    features are discernible. The interstellar absorption feature at
    $\sim6280$~\AA\ is masked out when optimal subtraction was performed.}
             \label{resid_fig}
\end{figure*}% 

The prominent molecular absorptions in the wavelength range of
6300--7300~\AA~ are signatures of an M-type companion star. 
However, the detailed rotational broadening analysis and spectral 
classification are not feasible for the Gemini/GMOS spectra of CX1004. 
%due to the complex combination of the accretion-disk emission 
%(contributing both continuum and emission lines) and the companion star emission. 
As discussed in Torres et~al. (2014),
the line width and the double-peak velocity separation are consistent
with that of an eclipsing quiescent CV or qLMXB. However, possibly owing 
to the faintness of the source, the light curve 
of CX1004 (see Fig.~A2 of Torres et~al. 2014) does not show significant periodic
behaviour. Torres et~al. (2014) suggested CX1004 to be a nearby source
due to the lack of diffuse interstellar bands. 

%\subsection{CXOGBS~J175359.9$-$292907 (CXB2)}\label{discuss:cxb2}

%The light curve of CXB2 shows a periodicity of $\approx0.447$~days,
%with signatures of a possible eclipse and flaring. 
The four epochs of Gemini/GMOS spectra of CXB2 are shown in  
Fig.~\ref{cxb2_fig}. The spectra of the first and third epochs appear 
to have weak, broad H$\alpha$ emission. Both of the lines have large 
velocity offsets (which are also visible in the figure) with 
$v_{\rm peak} = -260\pm60$~km~s$^{-1}$ and $-200\pm30$~km~s$^{-1}$, respectively. 
The widths of the two lines are FWHM = $825\pm175$~km~s$^{-1}$ and 
$725\pm100$~km~s$^{-1}$, respectively. The spectrum of the fourth epoch 
shows a narrow \ha\ absorption feature (FWHM = $180\pm40$~km~s$^{-1}$), 
which also has a significant velocity offset ($v_{\rm peak} =
-175\pm15$~km~s$^{-1}$). These large negative 
offsets could originate in a high space velocity of this source, which may 
imply a NS or a BH as the primary star. CXB2 also shows the H$\alpha$ 
emission variation as we have seen in CX377. No photospheric line from 
the companion star is detected from the spectra of CXB2. 
%The H$\alpha$ emission of CXB2 occasionally vanishes, revealing a weak
%H$\alpha$ absorption line (see Fig.~\ref{cxb2_fig}). The vanishing of
%the Halpha emission has two plausible causes.  The central part of the
%disk may be eclipsed during one of our observations.  Alternatively,
%the system may be a transitional millisecond pulsar, where a radio
%pulsar occasionally turns on and sweeps away the disk, and the disk
%later re-forms and accretes (see Burderi et al. 2001, Archibald et
%al. 2009, Papitto et al. 2013).  Further spectroscopy can decide
%between these alternatives.   
%although the spectra of 
%CXB2 have much lower $S/N$. 
%Further optical spectroscopy with more complete coverage on the orbital 
%phase is needed to investigate the cause of the velocity offset 
%and the H$\alpha$ emission variability of CXB2. 

CXB2 is brighter in \hbox{X-rays} than most of the GBS sources. It has
147 \chandra\ ACIS-I counts in 0.3--8.0~keV band in a 2~ks
exposure. CXB2 was serendipitously detected by an archival Suzaku observation 
with 53~ks exposure time (ObsID: 507031010). Full details of the \xray\ analysis 
will be presented elsewhere, but we briefly summarize relevant results here. The 
Suzaku XIS0 and XIS1 data were reduced and analysed using standard \verb+FTOOLS+. 
A good spectral fit ($\chi^2/\nu$=1.0) was found for CXB2 with an absorbed power-law 
model (photon index $\Gamma=0.96\pm0.11$, 
$N_{\rm H}=1.5^{+1.4}_{-1.0}\times10^{21}$~cm$^{-2}$). 
The \xray\ flux is $1.5\times10^{-12}$~erg~cm$^{-2}$~s$^{-1}$, which is 
consistent with that from our Chandra observation 
($1.3\times10^{-12}$~erg~cm$^{-2}$~s$^{-1}$, assuming the same spectral model). 
This source is probably associated with the {\it ASCA}
source AX~J1754.0$-$2929 which is 7.2$^{\prime\prime}$ away (see
catalogue in Sakano et~al. 2002). The \xray\ flux in the {\it ASCA}
observation is $1.4\times10^{-12}$~cm$^{-2}$~s$^{-1}$ in 0.7--10~keV
band, which is similar to the \xray\ flux in the \chandra\
observation. This source was, however, not detected in \rosat\ and is
classified as a transient (Paper~II).
The $N_{\rm H}$ value is smaller than that predicted to the Galactic
Bulge ($\sim10^{22}$~cm$^{-2}$), 
hence we infer that CXB2 probably lies in the 1--4~kpc distance range. Its 
near-infrared colours also indicate low reddening. The \xray\ luminosity  
would be 1--40$\times10^{32}$~erg~s$^{-1}$ (assuming the above distance 
range), which is consistent with either a high-inclination CV or a qLMXB. 

\subsection{Objects with H$\alpha$ in Absorption}\label{discuss:cx377} 

\begin{figure}
    \centering
    \includegraphics[width=3.4in]{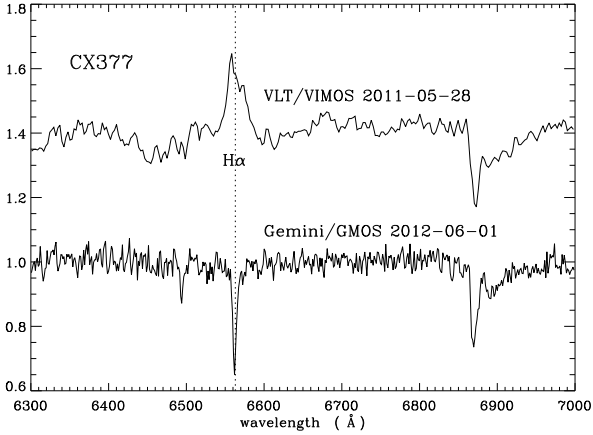}
    \caption{The comparison of the H$\alpha$ region of
    CX377 optical spectra, acquired by VLT/VIMOS (upper line) and
    Gemini/GMOS (lower line), respectively. The broad H$\alpha$
    emission line in the VLT/VIMOS spectrum is absent in the
    Gemini/GMOS spectrum, indicating strong spectral variability even though
    the source remains in quiescence.}
             \label{cx377_fig}
\end{figure}

Nine sources (CX138, CX377, CXB26, CXB73, CXB117, CXB149, CXB174, 
CXB189, and CXB201) in our sample
only show H$\alpha$ absorption lines in their Gemini/GMOS
spectra (Fig.~\ref{spec_fig2}). We are able to obtain a spectral
classification for five of them (CX138, CX377, CXB73, CXB149, and CXB174). 
The residual spectra of these sources after
optimally subtracting their best-fit stellar templates are shown in
Fig.~\ref{resid_fig}. Three of them (CX138, CX377, and CXB73) have
unambiguous H$\alpha$ emission features, which partially fill
in the stellar H$\alpha$  absorption features in our Gemini/GMOS
spectra. We measured the strength (EW) and width (FWHM) of the
H$\alpha$ emission shown in these residual 
spectra (see Table~\ref{line_table}). 
The broad H$\alpha$ emission (FWHM$\simgt400$~km~s$^{-1}$; see
Table~\ref{line_table} and Fig.~\ref{resid_fig}) is most likely 
due to the accretion disk. Therefore, for the first time, we may have 
discovered a hidden population of accreting binaries where the 
accretion disk contribution to the optical light is small, and the
H$\alpha$ emission lines become apparent only after subtracting the
stellar contribution to the spectra. In what follows,
we provide further details on the spectroscopic and/or photometric 
properties of several individual objects.\\

%{\it \centering (a) CX138}
\noindent {\it (a)} CX138 has a broad H$\alpha$ emission line 
(FWHM$=1350\pm75$~km~s$^{-1}$) in its residual spectrum. It also has
a radial velocity variation of $\sim100$~km~s$^{-1}$ between its Gemini
spectra. Therefore, CX138 is a likely candidate of the hidden
accreting binaries. The reddening measured via the strength of
DIB $\lambda5780$ (EW = 1971$\pm$131~m\AA) is $A_K
= 1.29 \pm 0.08$, which also puts the object in the Galactic Bulge, if
not farther. For the Galactic Bulge distance ($\sim8$~kpc), the absolute 
$V$-band magnitude of CXB138 would be $M_V=-2$, suggesting a giant 
companion star.   \\

%\#\#\#\#ToDo:
%Done 1) Section 4.2 with CX377, CX138, also wavelength shift of CXB149/174;
%Done 2) Section 4.3;
%Done 3) Conclusion;
%4) Table 5;
%Done 5) write something about DIB measure reddening in Section 3. 

%\subsection{CXOGBS~J174316.5$-$274537 (CX377)}\label{discuss:cx377}

\begin{figure}
    \centering
    \includegraphics[width=3.4in]{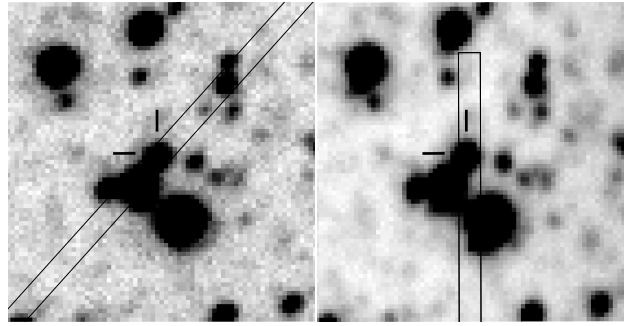}
    \caption{The finding charts of CX377 for our Gemini/GMOS 
    spectroscopy (left panel) and the VLT/VIMOS spectroscopy in Torres et
    al. (2014; right panel). The two short thick bars in each panel
    indicate the position of CX377, while the thin lines show the slit
    position. The sky area shown in both panels is
    $15^{\prime\prime}\times15^{\prime\prime}$. North is up and east is left in both panels.}
             \label{fc_cx377_fig}
\end{figure}

%Two epochs of Gemini/GMOS spectra were obtained for CX377 separated by 
%eight days. The spectra of the second epoch (2012 Jun 9) were contaminated 
%by an adjacent brighter source, and thus we do not use these spectra in 
%our analysis (see Fig.~\ref{fc_cx377_fig}). 
%{\it \centering (b) CX377}
\noindent {\it (b)} CX377 is also a candidate hidden accreting binary. 
The Gemini/GMOS spectra of
CX377 do not show any notable emission  
features (see Fig.~\ref{spec_fig2} and Fig.~\ref{cx377_fig}). 
However, the optical spectra of CX377 taken $\sim1$ 
year before (2011 May 28 and 2011 July 23) by VLT/VIMOS show strong broad H$\alpha$ 
emission with an asymmetric double-peaked profile (Torres et~al. 2014; 
see Fig.~\ref{cx377_fig}). The intrinsic width of the H$\alpha$ emission is 
$\sim1200$~km~s$^{-1}$, while the double-peak separation is $\sim700$~km~s$^{-1}$. 
This apparent dramatic change of the spectrum is not caused 
by incorrect targeting. The finding charts of CX377 for our 
Gemini/GMOS spectroscopy and for the VLT/VIMOS spectroscopy
(Fig.~\ref{fc_cx377_fig}; also see Fig.~9 in Torres et~al. 2014) show
the exact same targets. The slit was placed  
during our Gemini/GMOS observations (see Fig.~\ref{fc_cx377_fig}) to contain both the target 
and the adjacent object so that we are able to distinguish their spectra. 
The seeing during the Gemini/GMOS observations
($\sim0.6^{\prime\prime}$) was better than that during the VLT/VIMOS
observations ($\sim1^{\prime\prime}$). Thus, the contamination of the
nearby bright sources is less of a  
problem for the Gemini/GMOS spectra. Therefore, the H$\alpha$ emission feature of CX377 indeed 
has significant variation within one year. Although the Gemini/GMOS spectra of CX377 only 
show H$\alpha$ absorption lines, its residual spectrum after optimally 
subtracting the F6V standard star template shows broad H$\alpha$
emission (see Fig.~\ref{resid_fig}). The H$\alpha$ 
emission has not completely disappeared, but it is significantly weaker
during our Gemini/GMOS observation than during the VLT/VIMOS
observations of Torres et~al. (2014). It has also become narrower (FWHM =
$660\pm30$~km~s$^{-1}$ in the Gemini/GMOS residual spectrum). 
This rare behaviour 
of H$\alpha$ emission line variations 
indicates significant accretion disk variations for CX377. The soft \xray\ 
transient GRO~J1655$-$40 has shown similar strong variability of its H$\alpha$ emission line 
(e.g., Soria et~al. 2000). GRO~J1655$-40$ has a period of 2.62 days;
the spectral type of its companion star had been classified as F3--F6
(Orosz \& Bailyn 1997),
similar to that of CX377 (see below). However, the H$\alpha$ emission line
variability of GRO~J1655$-$40 occurred when this object was going
through an outburst cycle, while CX377 has remained in
quiescence. It is possible 
that CX377 experienced a faint \xray\ outburst ending
before June 1, 2012 when our Gemini spectra were taken, as long as the
\xray\ flux remained below the detection threshold of all sky monitors
such as Monitor of All-sky \xray\ Image (MAXI; Matsuoka et al. 2009). Another 
possibility is that CX377 has variable accretion rate during quiescence.
%The optical spectra of the BH binary
%GRO~J1655$-$40 did not show  
%H$\alpha$ emission when it was in the quiescent state, while the H$\alpha$ emission 
%began to emerge when the object entered high-soft state. The strength of H$\alpha$ 
%emission is correlated with the hard \xray\ flux of this object. 
%This may provide 
%insights into the mechanism of H$\alpha$ emission variation for CX377. 
%Further optical spectroscopy with higher spectral resolution is
%needed to measure the mass of the primary and to determine
%whether it is a WD, a NS or a BH. 
%, despite 
%CX377 is likely to be a CV ({\bf[why? besides its mass should be $<1.25$~M$_\odot$ if 
%we assume $q < 1$]}) rather than a BH system as GRO~J1655$-$40.

CX377 is spectrally classified as F6 type. 
%which may be a main sequence star (with typical mass of
%1.25~M$_{\odot}$) or a giant/subgiant star (see Torres et~al. 2014 and
%the references therein). 
The relative radial velocities between the Gemini spectra of CX377 (see
Table~\ref{rvc_table}) are consistent with zero or less than
10~km~s$^{-1}$, which is not 
surprising given that the three object spectra were only 15 minutes
apart, and the orbital period of CX377 is expected to be much longer
than 15 minutes for an F6 main sequence star filling its Roche lobe. 
Then we cross-correlated the object spectra to the F6V star
template (HD~16673). The radial velocities relative to the star
template range from $-50$ to $-20$~km~s$^{-1}$. The radial velocity 
of the star template itself is consistent with zero ($-4\pm5$~km~s$^{-1}$).
The light
curve of CX377 obtained in July 2010 only shows non-periodic flickering 
with a RMS scatter of 0.06 magnitude (see Fig.~A2 of Torres et~al. 2014).
The strength of DIB $\lambda5780$ (EW $= 1093\pm86$~m\AA) in the
spectra of CX377 
corresponds to $A_K=0.73\pm0.06$. Extinction map from Gonzalez
et~al. (2012) gives $A_K=0.72\pm0.15$ for CX377, which is
consistent with the extinction derived above. Therefore,
it is likely that CX377 resides in the Galactic Bulge.

The neutral hydrogen column density,
$N_H=(1.44\pm0.14)\times10^{22}$~cm$^{-2}$, was estimated from the 
extinction following the relation in G{\"u}ver \& 
{\"O}zel (2009). CX377 has 7 counts in the \hbox{ACIS-I} 0.3--8~keV band in
the 2~ks exposure. Assuming an absorbed power-law spectrum with
photon index of 1.6, the unabsorbed \xray\ flux of CX377 is
$\approx1\times10^{-13}$~erg~cm$^{-2}$~s$^{-1}$. At a distance of 
$\sim8$~kpc, the \xray\ luminosity would be $L_{\rm X}=8\times10^{32}$~erg~s$^{-1}$.
With the estimated extinction and the apparent $r^\prime$-band
magnitude (see Table~\ref{rmag_table}), we calculate the absolute
magnitude of $M_{r^\prime}=-1.3$
%,which is consistent with that of an F-giant companion. 
and the \xray\ to optical flux ratio
$f_{\rm X}/f_{r^\prime}\sim0.005$; both values are consistent with those
for a giant companion star. However, 
the disk luminosity at the time of the VIMOS observation should 
have been larger than that of the F6 giant ($M_{r^{\prime}}=-1.3$), implying 
that the source might have experienced a faint \xray\ outburst
(cf. Wijnands \& Degenaar 2013). \\

%{\it \centering (c) CXB149 \& CXB174}
\noindent {\it (c)} CXB149 and CXB174 were spectrally
classified as G6 and F4, respectively. Their residual spectra after
optimal subtraction do not
show unambiguous H$\alpha$ emission or absorption features. The
Gemini/GMOS spectra could possibly be from interlopers instead of the true 
optical counterparts of the \xray\ sources. The H$\alpha$ absorption
features of these two sources are significantly blueshifted from the
laboratory wavelength ($-109\pm7$~km~s$^{-1}$ for CXB149 and
$-179\pm8$~km~s$^{-1}$ for CXB174). This shift is not introduced by
the wavelength calibration process since offsets to the sky lines were 
small and they have been corrected. It cannot be explained by the
wavelength shifts caused by the centroiding uncertainty within the slit
(which are $-30$~km~s$^{-1}$ and $-75$~km~s$^{-1}$, respectively; see
\S\ref{obs:spec}). These velocities are consistent with that for stars
residing in the Galactic 
Bulge (e.g., Zoccali et~al. 2014). 
%We also estimated the reddening for
%these two sources via the equivalent width (EW) of the interstellar 
%bands at $\lambda5780$~\AA\ with the calibration in Table~3 of 
%Herbig (1993), and compared it to the Bulge reddening along the 
%line of sight provided by Gonzalez et~al. (2011,2012), which utilized 
%the Red Clump stars in the Bulge.\footnote{See the Bulge Extinction And
%Metalicity (BEAM) calculator at http://mill.astro.puc.cl/BEAM/calculator.php.}  
Regarding the reddening measurements, CXB149 has EW(DIB $\lambda5780$)
= 1099$\pm$28~m\AA; the corresponding reddening  
is E(B-V) $=$ 2.11$\pm$0.05 and $A_K=0.73\pm0.02$, which put this 
object in the Galactic Bulge or farther ($A_K=0.30\pm0.07$ from the map of 
Gonzalez et~al. 2012). For CXB174, EW(DIB $\lambda5780$) = 616$\pm$34~m\AA; 
the corresponding reddening is E(B-V) $=$ 1.21$\pm$0.06 and $A_K=0.42\pm0.02$, 
which is consistent with the Bulge reddening ($A_K=0.35\pm0.08$) from 
Gonzalez et~al. (2012). The absolute magnitudes for CXB149 and CXB174 residing 
in the Bulge would be $M_V=0$ and $M_V=-0.7$, respectively, which are 
consistent with those of the giant stars with their spectral types. \\

%{\it \centering (d) CXB26 \& CXB189}
\noindent {\it (d)} CXB26 and CXB189 were not spectrally classified using our
method. CXB26 is considered to be likely associated with an OGLE
source $1^{\prime\prime}$ away from the \chandra\ position (OGLE
BUL-SC3 6033) which is identified 
as a CV (see Paper~II). However, the Gemini/GMOS spectra of
CXB26 do not show apparent features of an accretion disk. It is
possibly a hidden accreting binary, which could be verified after it
is spectrally classified with future spectroscopy. 
%\footnote{Note CXB26 is not the same object as the source 
%`GBS 26' classified as a G3V star in Hynes et~al. (2012; see their Table~4), 
%which would be labeled as `CX26'.} 
The physical nature of CXB189 remains unclear. 

\subsection{Objects without Apparent H$\alpha$ Emission/Absorption Features} 

\begin{figure}
    \centering
    \includegraphics[width=3.4in]{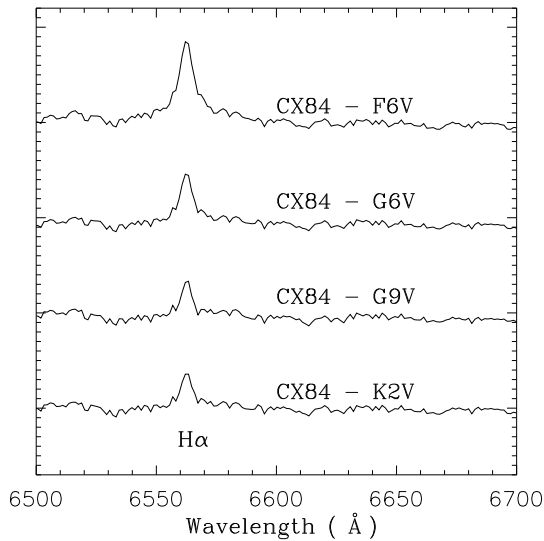}
    \caption{The residual spectra of CX84 after
    subtracting the best-fit star template (G9), and other templates
    with spectra types (F6, G6, K2) bracketing that of the best-fit
    template. The H$\alpha$ emission feature exists in all four
    residual spectra.}
             \label{cx84test_fig}
\end{figure}

The remaining six sources show neither clear H$\alpha$ emission nor
absorption features in their Gemini/GMOS spectra (see
Fig.~\ref{spec_fig3}). Three of them (CX84, CX139, 
 CXB82) were spectrally classified (see Table~\ref{vsini_table}). The
 residual spectra of these sources after optimally subtracting their
 best-fit stellar templates are shown in Fig.~\ref{resid_fig}. Similar
 to CX138, CX377, and CXB73, they all have H$\alpha$  emission
 features, which fill in the stellar H$\alpha$
 absorption features in our Gemini/GMOS spectra. To test the
 robustness of our results, 
we also subtracted other non-best-fit stellar 
template from the object spectra to see how the residual H$\alpha$
emission would vary. Fig.~\ref{cx84test_fig} shows one example of this
test. The H$\alpha$ emission feature remains in the residual spectra
of CX84 after subtracting several stellar templates with 
spectral types bracketing that of the best-fit template, which supports
the detection of a residual accretion disk. Note that the choice of 
templates with non-best-fit spectral types could affect the profile of 
the residual H$\alpha$ emission. For example, the mis-matched templates 
generate broader wings of the H$\alpha$ emission which may result in a higher 
FWHM value (e.g., FWHM increases from $225$~km~s$^{-1}$ to $320$~km~s$^{-1}$ 
if using an F6 template spectrum). 

It is worth noting that even in the case 
that the residual H$\alpha$ emission features are real, 
there could be 
alternative explanations. For example, it is possible
that the observed optical light is dominated by a foreground,
physically unrelated star (instead of the companion star in the 
above scenario) so that we can only see the accretion after subtracting the
foreground stellar emission. However, based on the analysis in Britt
et~al. (2014), we expect $\sim0.7$ interloper in our sample of 21 GBS
sources (also see \S\ref{obs:photo}). In \S\ref{discuss:cx377}, we
already discussed two possible interlopers (for CXB149 and CXB174) for
which the residual spectra do not have clear H$\alpha$
emission. Therefore, it is not likely that the six optical sources
that only have H$\alpha$ emission in their residual spectra are all
chance interlopers. Another possibility is that these sources are
active stars/binaries, where the H$\alpha$ emission fills in the
stellar absorption features. Sources with narrower H$\alpha$ emission
(e.g., $\simlt200$~km~s$^{-1}$) in their residual spectra could fit into
this scenario. However, it is difficult to explain the broad residual
H$\alpha$ emission lines of CX138 and CX377. The periodic modulation in 
the light curves of CX84, CXB82, and CXB99 
could originate from either binary motion or stellar activity
(hotspots). Alternative explanations for the narrowness of their
H$\alpha$ emission lines are that the binaries are relatively face-on,
or that the disks are highly recessed. 

%Sources with narrow H$\alpha$ emission
%(FWHM$\sim200$--300~km~s$^{-1}$) in their residual spectra are
%possibly chromospherically active stars/binaries. However, 
%For CX84, its high \xray\ luminosity may support a 
%quiescent LMXB interpretation. 
CX84 is spectrally 
classified as G9. The reddening estimated from the strength of 
interstellar band $\lambda5780$ (EW = 2499$\pm$218~m\AA) is 
$A_K = 1.62 \pm 0.14$, which means the object is in the Bulge or 
farther. Assuming 
CX84 resides in the Galactic Bulge, the absolute $V$-band 
magnitude would be $M_V=-1$, which indicates a G9 giant star. 
The \xray\ luminosity is $\sim2.6\times10^{33}$~erg~s$^{-1}$, which  
appears higher than that for G9 giants (see Fig.~2 of G{\"u}del 2004).  
However, it is typical of long orbital period qLMXBs
(e.g., Jonker et~al. 2012). The optical light curve of CX84 indeed shows a
modulation with a tentative period of 4.67 days. If it is a qLMXB, the
orbital period should be twice that value ($\approx9.3$ days)  
because of the ellipsoidal modulation. In summary, the \xray\ luminosity 
of CX84 suggests the qLMXB nature, while the narrow residual H$\alpha$ 
emission line favors the chromospherically active star/binary scenario. 
Similarly, the reddening of CXB82 derived from its near-infrared colour
($A_K\approx0.32$) is also consistent with the reddening expected if
it resides in the Galactic Bulge. The absolute $V$-band 
magnitude ($M_V=-0.3$) suggests a G9 giant star, while its \xray\ luminosity 
($\sim1.1\times10^{33}$~erg~s$^{-1}$) indicates that it is more likely to
be a qLMXB. The H$\alpha$ emission line in the residual spectrum of CXB82 is 
also narrow. Based on current data, we are not able to determine whether 
CX84 and CXB82 are chromospherically active stars/binaries or qLMXBs. \\

%In summary, the Gemini/GMOS spectra of the 21 GBS sources have shown the effectiveness of 
% our optical spectroscopic follow-up campaign on the \xray\ sources detected in 
%the GBS because we have found the likely optical counterparts for most 
%of our sources. Twelve of the 21 sources have H$\alpha$ emission
% features in their Gemini/GMOS spectra or the residual spectra after
% subtracting their best-fit stellar templates. Five of them (CX138,
% CX377, CX446, CX1004, and CXB2) with broad H$\alpha$ emission lines
% are likely to be accreting binaries. Another two  
%objects (CX84 and CXB82) are also possibly accreting binaries although their H$\alpha$ 
%emission lines are relatively narrow. The other five sources (CX139,
% CXB64, CXB73, CXB99, and CXB113) are more likely to be
% chromospherically active stars/binaries.   

Among the seven potential accreting binaries identified in our sample
with a spectral classification, one source (CX377) has an F-type  
companion star, another four (CX84, CX138, CXB73, CXB82) have G-type 
companion stars, while the other two (CX139 and CXB99) have K-type companion stars. 
Given the mass of F/G-type companion stars, the mass ratios
($q=M_2/M_1$ here) for the sources 
with F/G-type companion stars should be higher than
those of the more typical \xray\ binaries with later spectral types (K, M). 
In contrast to LMXBs, very
 few of these ``intermediate'' mass \xray\ binaries (IMXBs) have been
 found. Existing examples include Cyg X-2 (e.g., Casares et~al. 2010)
 and Her X-1 (e.g., Reynolds et~al. 1997). Our 
Gemini/GMOS spectroscopy indicates that the GBS 
may be able to provide an 
example of this type of X-ray binary.
%Those sources with low values
%of the EW of the H$\alpha$ emission line that is apparent after
%optimally subtracting the template star spectrum
%could also be RS Canum Venaticorum variables (RS CVn). Further variability 
%and/or spectroscopic data would be necessary to distinguish between these scenarios. 

%\subsection{Individual Objects}\label{discuss:individual}

%In the following subsections, we will discuss further the five likely accreting 
%binaries (CX138, CX377, CX446, CX1004, and CXB2) 
%in detail. The H$\alpha$ line properties of these four sources are listed in 
%Table~\ref{line_table}.

%\subsection{CXOGBS~J174623.1$-$254930 (CX138)}\label{discuss:cx138}

%The optical counterpart of CX138 is spectrally classified as G9 type. After the 
%corresponding star template is subtracted, the residual spectrum reveals a broad 
%H$\alpha$ emission line (FWHM = $1350\pm75$~km~s$^{-1}$), which may indicate the 
%existence of an accretion disk. CX138 

%[Tables: optical magnitude, vsini, radial velocity]

% -----------------------------------------------------------------------------
% Summary
% -----------------------------------------------------------------------------

\section{Conclusions}\label{conclusion}

The GBS is a multiwavelength survey project where one of the goals is
to identify \xray\ binaries in the Galactic Bulge area. 
%The identification and classification of the \xray\ sources via optical 
%spectroscopy is a crucial step to fulfill the main science goals of the GBS, which 
%includes measuring the compact object mass, constraining the binary formation \& 
%evolution models, and investigating the Galactic structure. 
In this work, we 
present optical and infrared photometry, and Gemini/GMOS spectroscopy of 
21 GBS \xray\ sources detected in the GBS. 
%We also performed line
%measurements, spectral classification, and radial velocity analysis for
%these sources. The majority of the 21 optical sources with Gemini/GMOS
%spectroscopy are likely to be the real counterparts of the
%\xray\ sources. The optical sources at the 
%position of the \xray\ sources CXB149 and CXB174 
%are probably interlopers (see \S\ref{discuss:hidden}); they only show
%H$\alpha$ absorption features in their Gemini/GMOS
%spectra and optimally subtracting their best-fit template spectra does
%not reveal H$\alpha$ emission from a disk or the companion star. 
%Three sources (CX446, CX1004, and CXB2) have broad H$\alpha$ 
%emission lines in their Gemini/GMOS spectra which are likely the
%signature of accretion disks, while another two (CXB64 and CXB113)
%have narrow H$\alpha$ emission lines and molecular absorption features
%from late-type companion stars indicating that they are likely
%chromospherically active stars or binaries. 
One prime goal of the GBS is to identify eclipsing qLMXBs. CX446,
CX1004 and CXB2 are promising candidates to be eclipsing 
qLMXBs. CX446 has broad H$\alpha$ emission, and its optical light curve shows 
likely eclipsing events. CX1004 has broad, double-peaked H$\alpha$
emission. The light 
curve of CXB2 also contains eclipsing events suggesting a high
systemic inclination. 

We may have discovered a population of hidden accreting
binaries. After optimally subtracting the stellar templates with
matched spectral type, the residual spectra of seven sources show
H$\alpha$ emission. Three of them (CX138, CX377, and 
CXB73) have broad ($\simgt400$~km~s$^{-1}$) H$\alpha$ lines, which are 
likely produced by an accretion disk. CX84 and CXB82 are also 
possibly hidden accreting binaries based on their \xray\ luminosity, 
while their residual H$\alpha$ emission lines are narrow ($\simlt200$~km~s$^{-1}$).
%These hidden accreting sources may constitute a substantial
%portion of the full accreting binary population, for which the
%physical and demographic properties have not been
%investigated. Therefore, the hidden population of accreting source is
%one of the most important discoveries of this work. 
Previous VLT/VIMOS spectra of CX377 showed a strong, double-peaked
H$\alpha$ emission line, while our Gemini/GMOS spectra of CX377 only 
contains an H$\alpha$ absorption line. This may indicate the strong 
variability of the accretion disk
of this source. The residual spectrum of CX377 after optimal
subtraction supports this scenario by showing weak, broad H$\alpha$
emission. 
%This is the first case to our knowledge that a quiescent accreting 
%binary shows this kind of H$\alpha$ variability.
%CXB2 also shows similar H$\alpha$ variability.

In summary, based on the emission features, 
the spectral classification, \xray\ luminosity, and the residual 
spectra after optimal subtraction, we
are able to constrain the likely nature for eight sources in our sample:
CX446, CX1004, and CXB2 are accreting binaries (CVs or qLMXBs)
while CX446 and CXB2 are likely also eclipsing; CXB64 and CXB113 are
chromospherically active stars or binaries; CX138, CX377, and CXB73 could be 
hidden accreting binaries.  
%The seven sources (CX84, CX138, CX139, CX377, CXB73, CXB82, and CXB99) that we 
%have obtained spectral classification for have F-, G-, or K-type companion 
%stars.  Three sources (CX84, CX138, and CX139) have relative radial velocity variation
%of $\sim100$~km~s$^{-1}$, while the majority of others do not show significant 
%radial velocity variations. 

%This work, together with Torres et~al. (2014) and Britt et~al. (2013), illustrates 
%the optical follow-up procedures for the GBS \xray\ sources. Identification of 
%accreting binaries is mainly based on the broad Balmer emission lines; while 
%the absorption features can be used for rotational broadening measurements, 
%spectral classification, and radial velocity
%measurements. Further investigations on the residual spectra after subtracting
%stellar templates with matched spectral classification may be able to
%reveal accreting binaries that do not have apparent H$\alpha$
%emission at first sight. Radial
%velocity variations are an integral part of 
%identifying the counterparts to the \xray\ sources. Investigations on 
%more GBS \xray\ sources will be continued with the ongoing optical spectroscopic 
%campaigns. Meanwhile, further spectroscopy with higher resolution 
%and more complete photometry/variability data are needed for confirming the
%nature of the promising objects 
%identified in the GBS (e.g., CX138, CX377, CX446, CX1004, and CXB2).

% -----------------------------------------------------------------------------
% Appendix
% -----------------------------------------------------------------------------

\section*{Appendix A: Finding Charts}

\begin{figure*}
%    \figurenum{A1}
    \centering
    \includegraphics[width=1.5in]{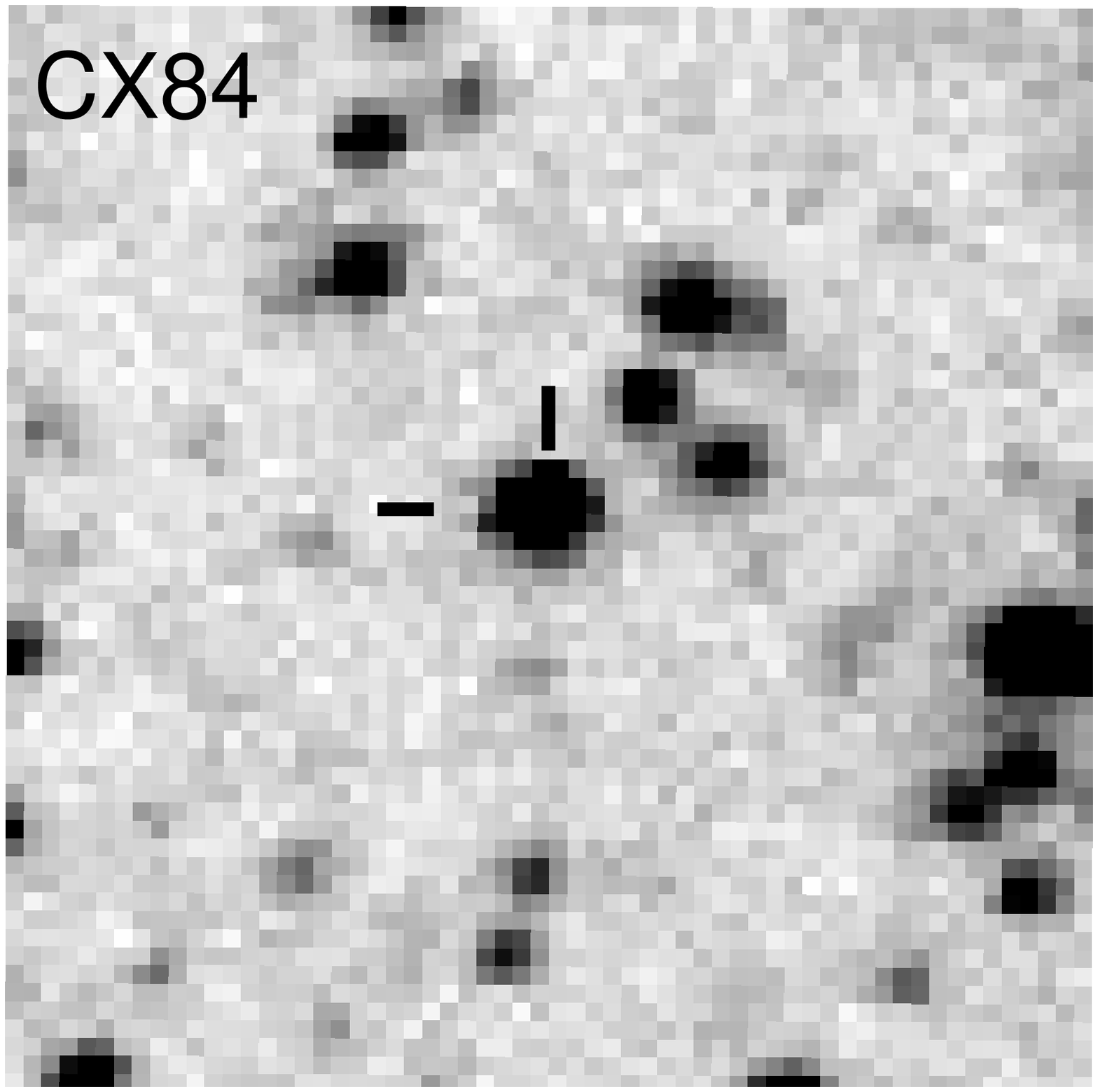}
    \includegraphics[width=1.5in]{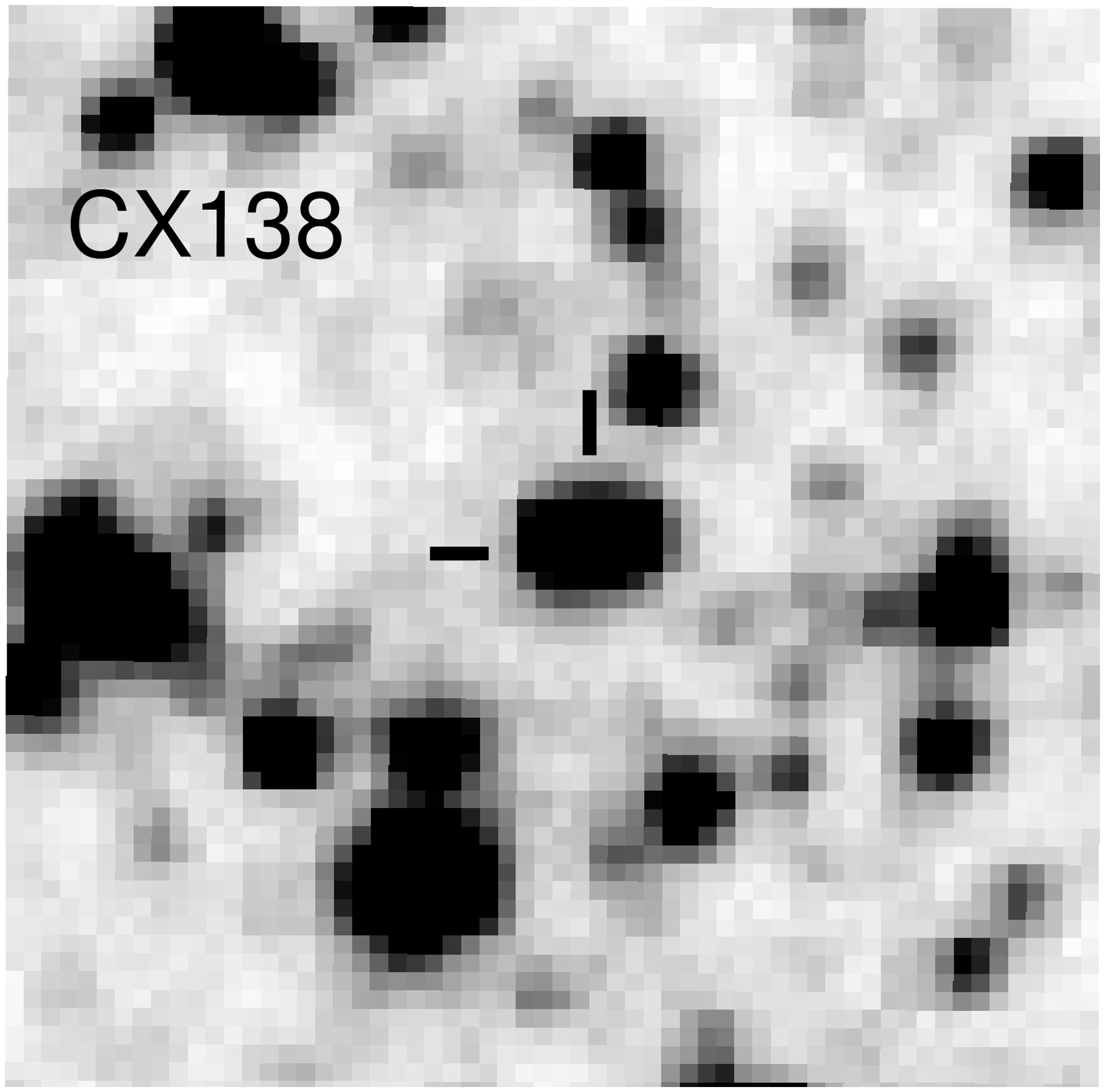}
    \includegraphics[width=1.5in]{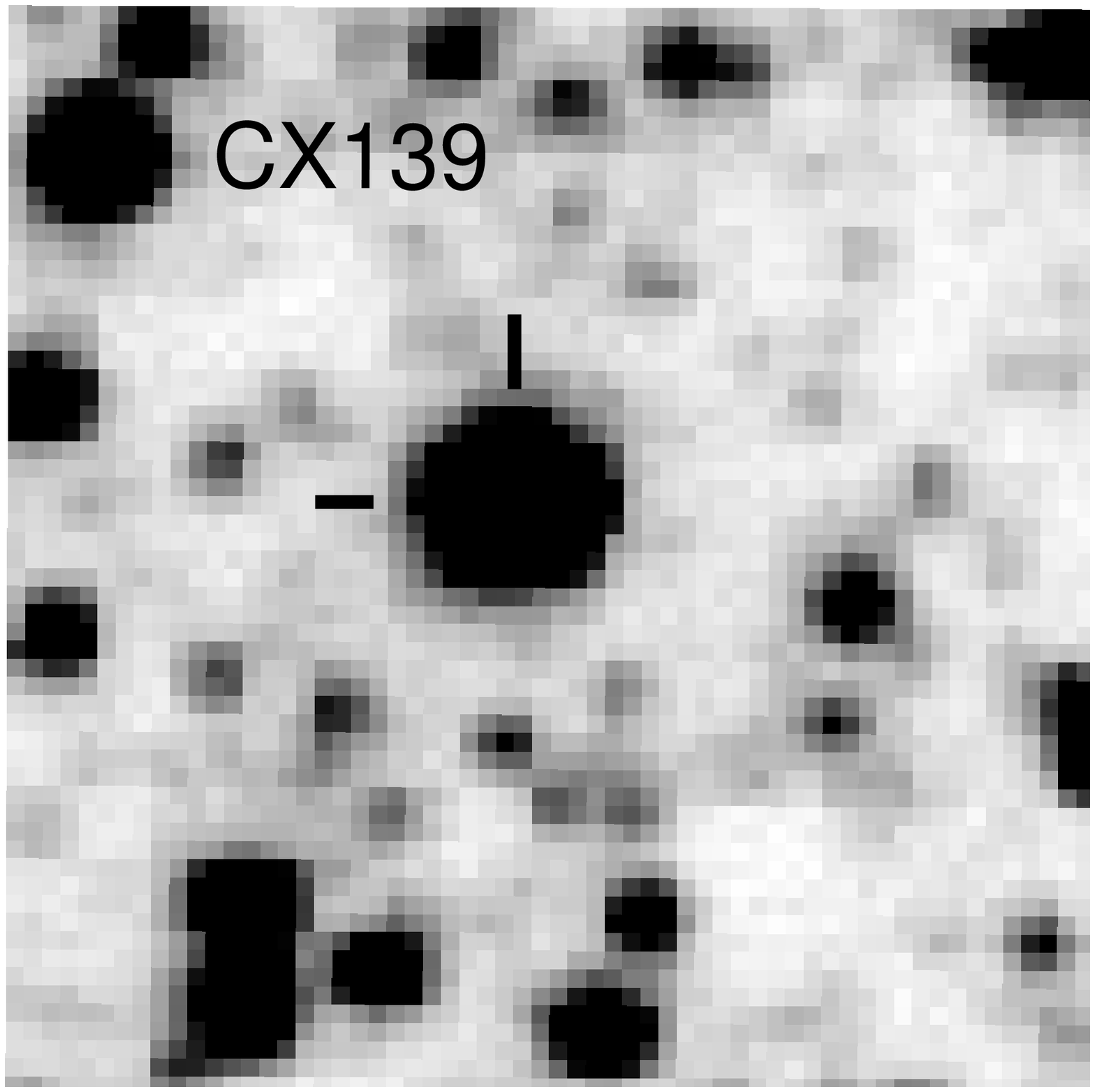}\\
    \includegraphics[width=1.5in]{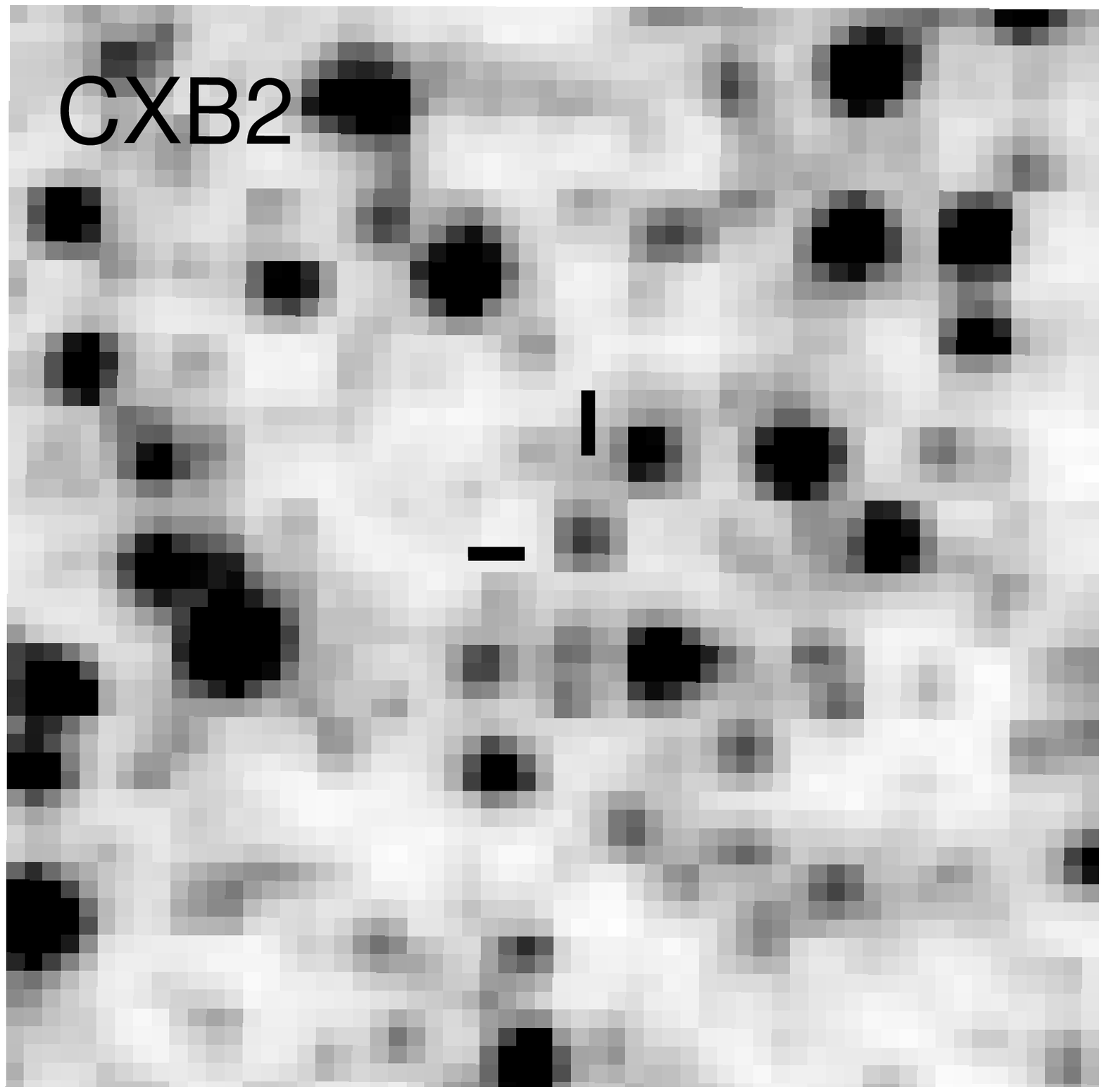}
    \includegraphics[width=1.5in]{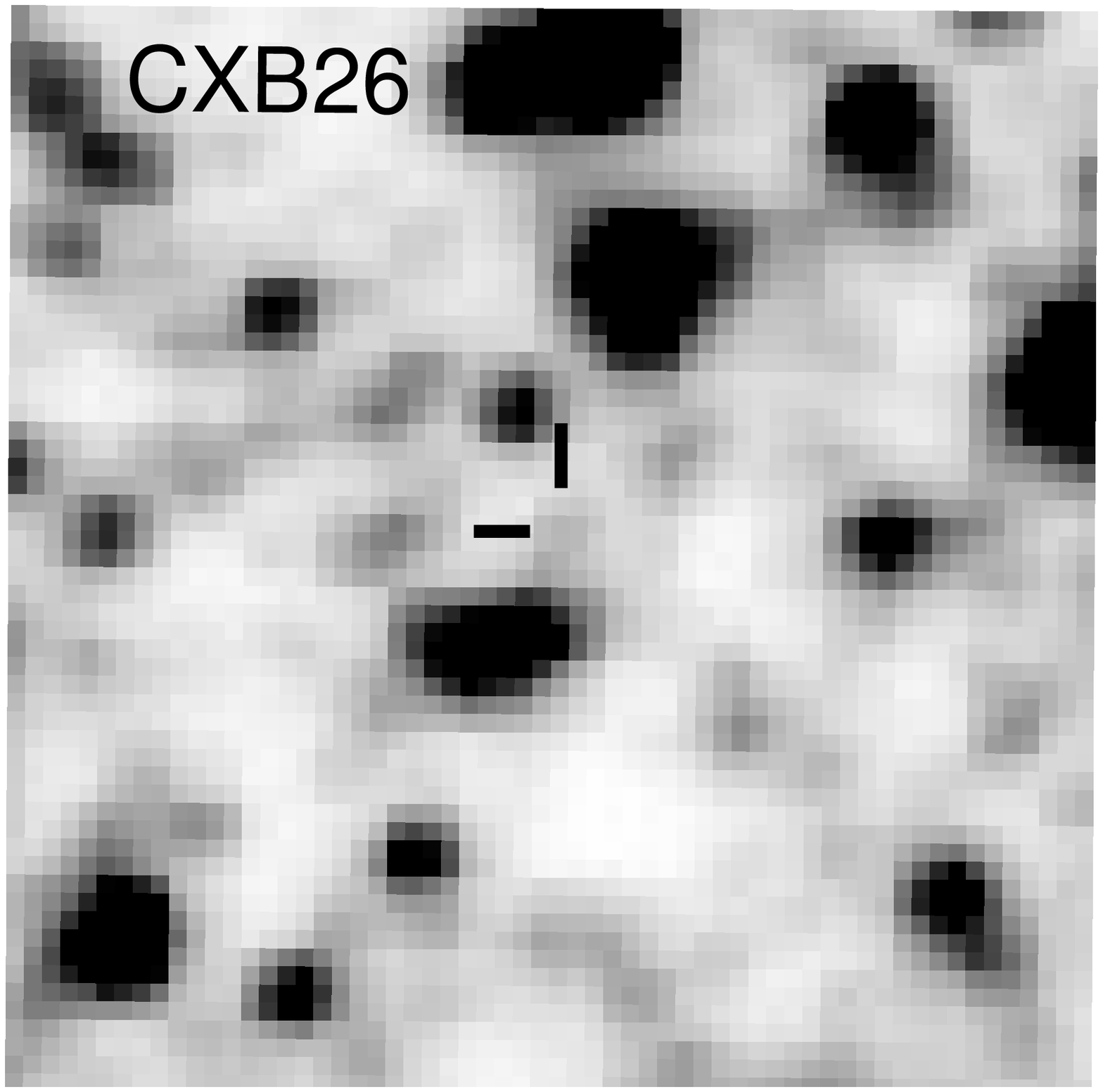}
    \includegraphics[width=1.5in]{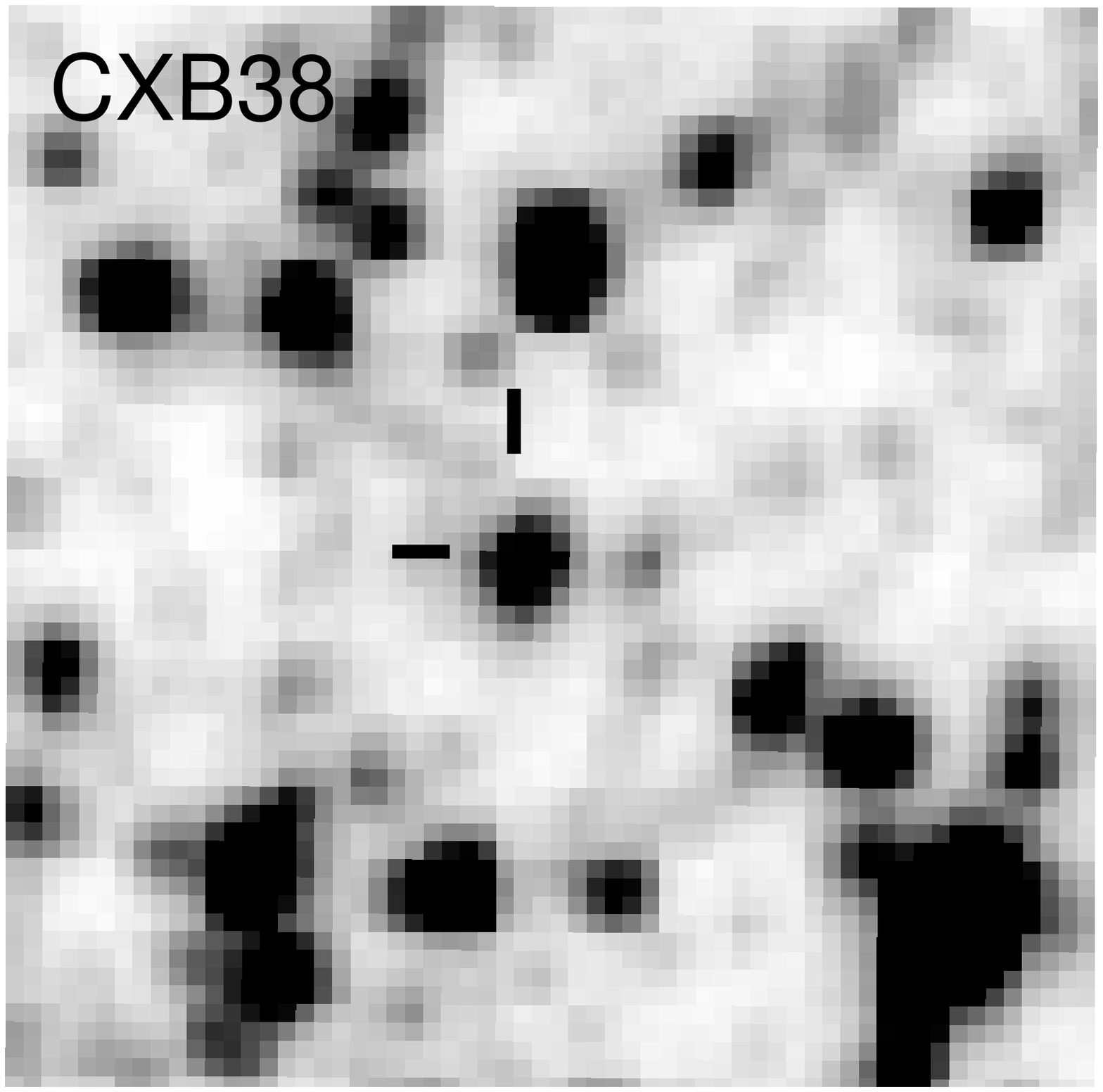}\\
    \includegraphics[width=1.5in]{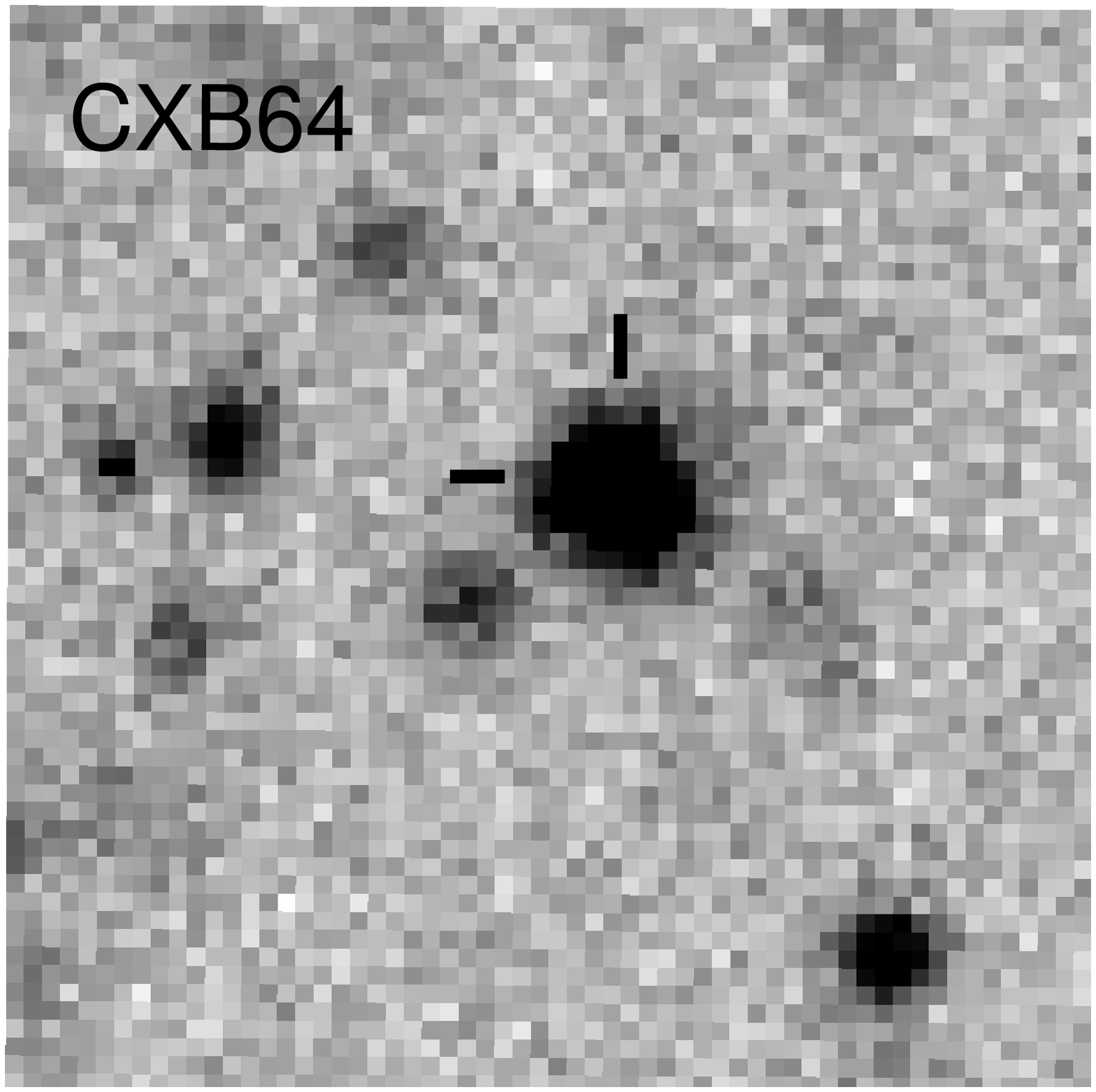}
    \includegraphics[width=1.5in]{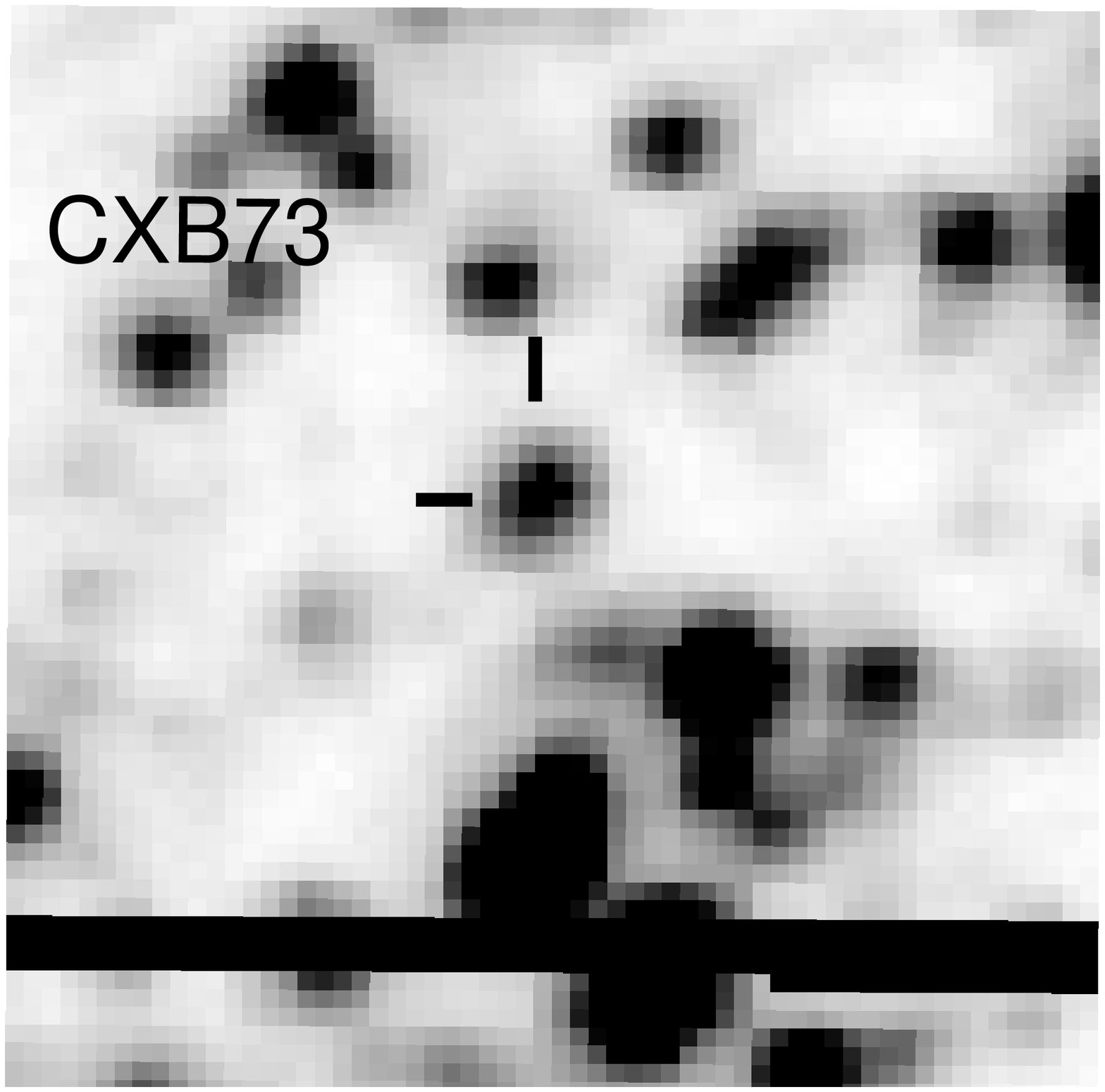}
    \includegraphics[width=1.5in]{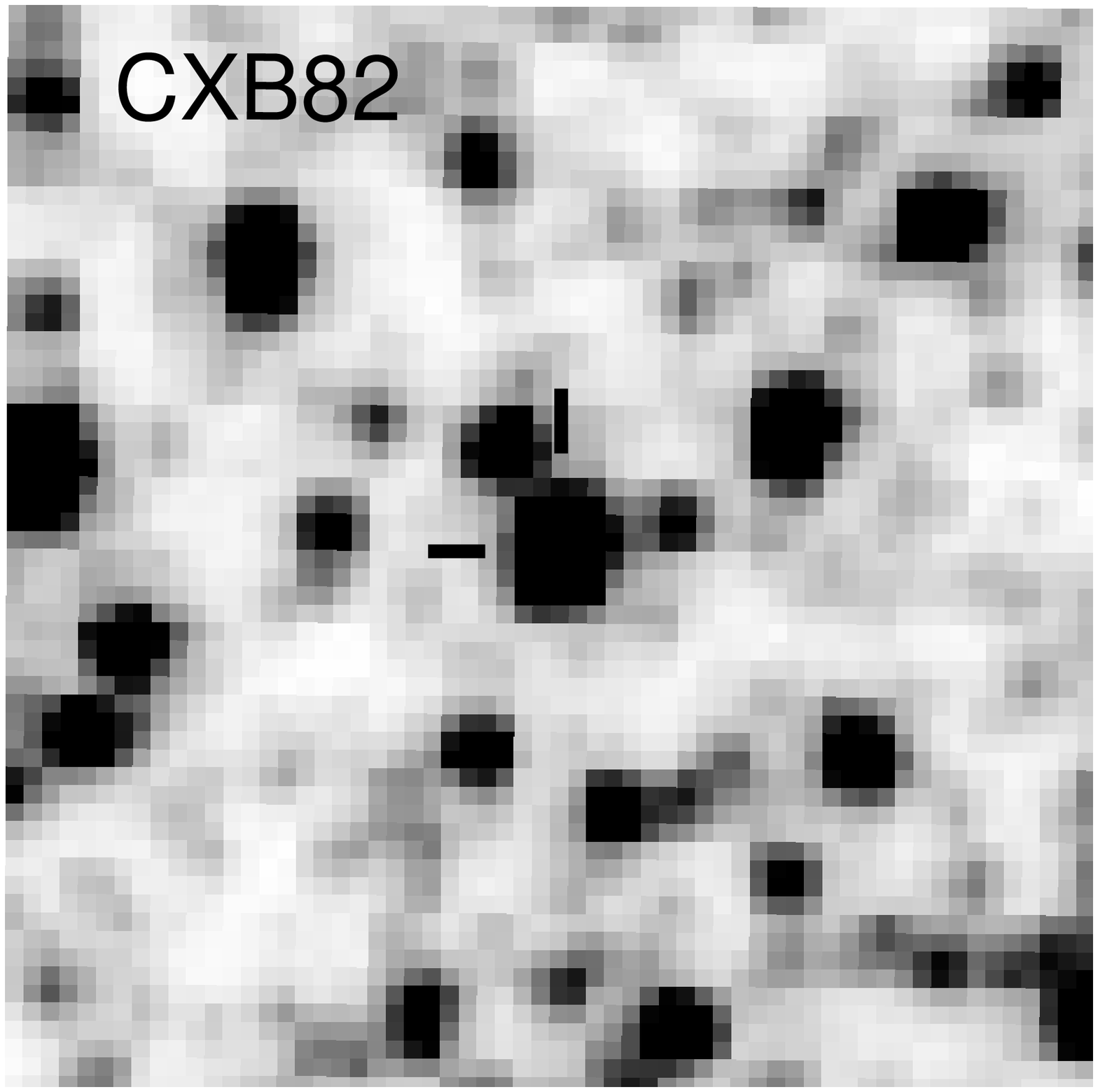}
    \caption{The $r^\prime$-band finding charts (generated from
    $r^\prime$-band images
    obtained by the Blanco 4-meter telescope) for optical counterparts of 
    18 of the 21 sources with Gemini/GMOS spectroscopy in this work
    (finding charts for the remaining three sources have been
    published in Torres et~al. 2014; also see Fig.~\ref{fc_cx377_fig}
    for the finding chart of CX377). The source 
    is indicated by the short horizontal and vertical bars in each chart. The sky 
    area of each chart is 20$^{\prime\prime}\times$20$^{\prime\prime}$, except for CXB113 
    which is 20$^{\prime\prime}\times$15$^{\prime\prime}$. North is up and 
    East is to the left.}
             \label{fc_fig}
\end{figure*}

\begin{figure*}
%    \figurenum{A1}
    \centering
    \includegraphics[width=1.5in]{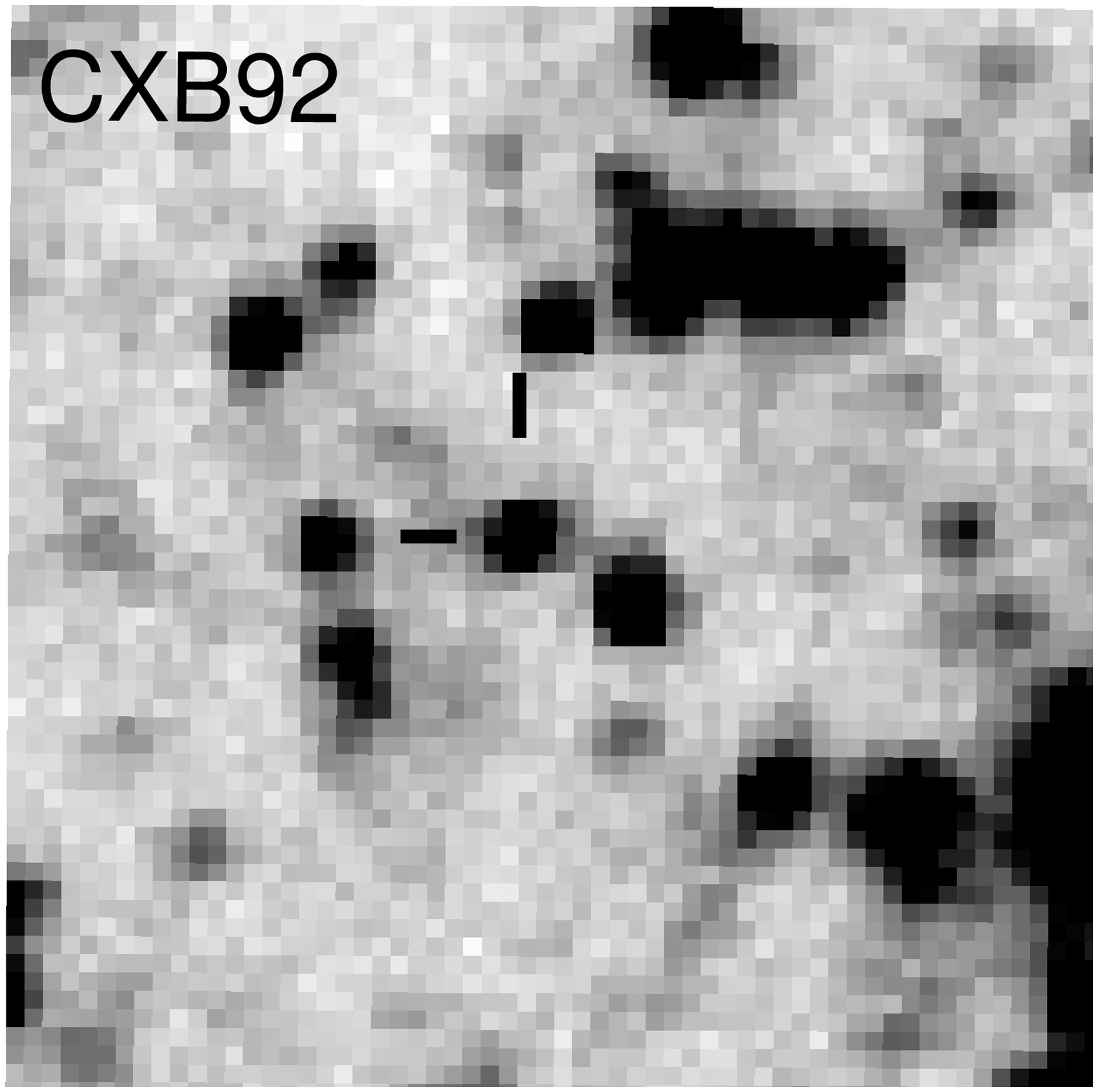}
    \includegraphics[width=1.5in]{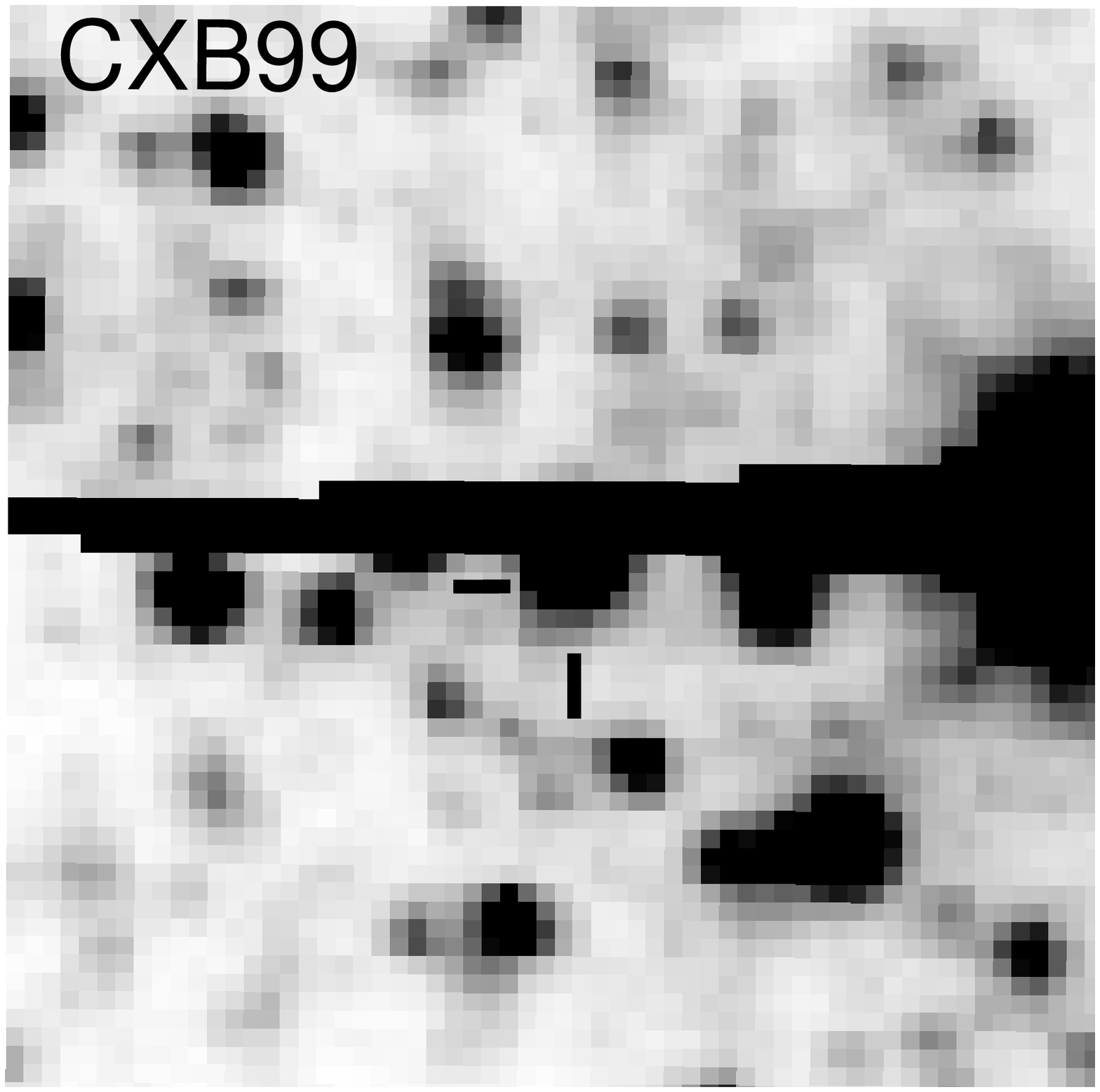}
    \includegraphics[width=1.5in]{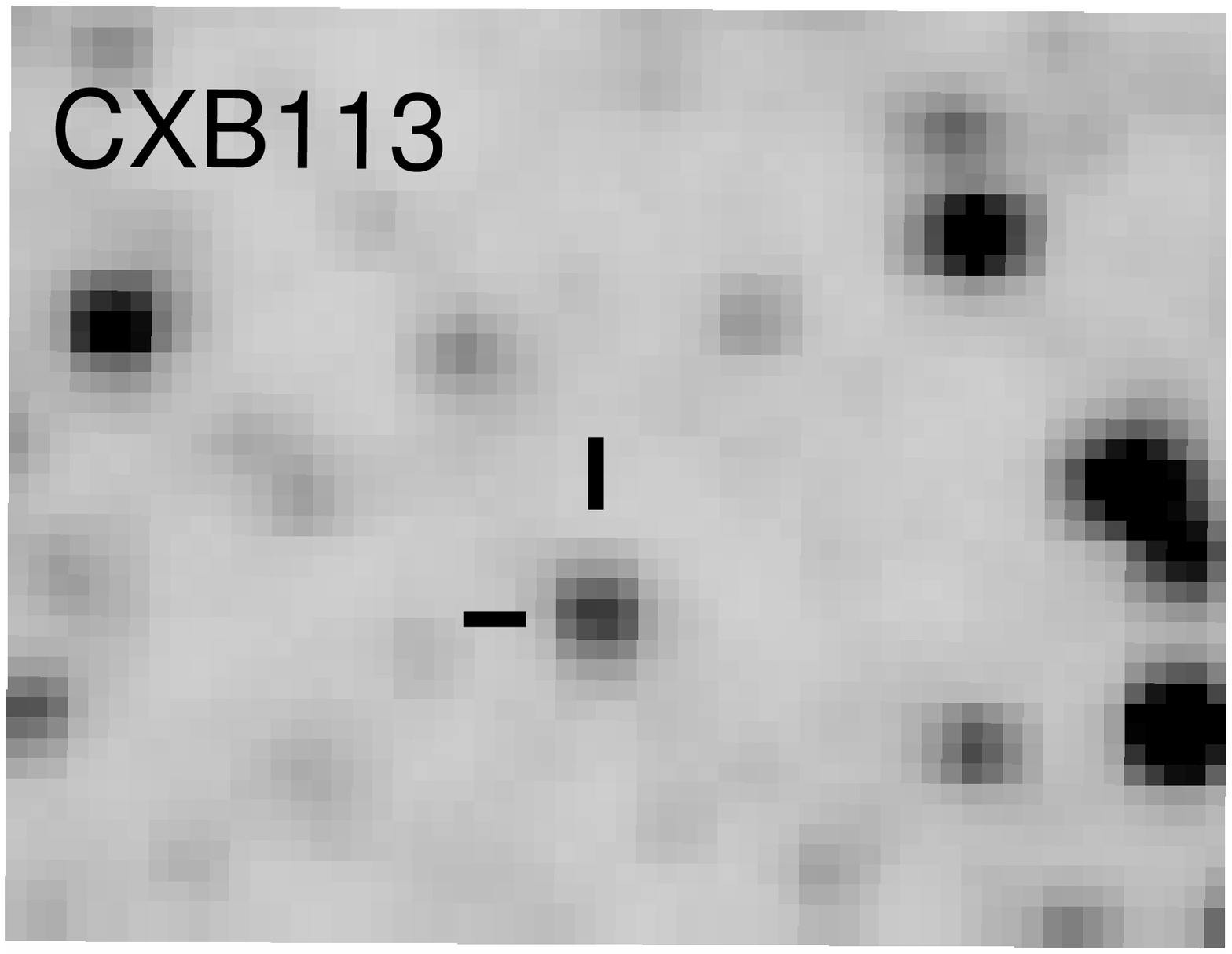}\\
    \includegraphics[width=1.6in]{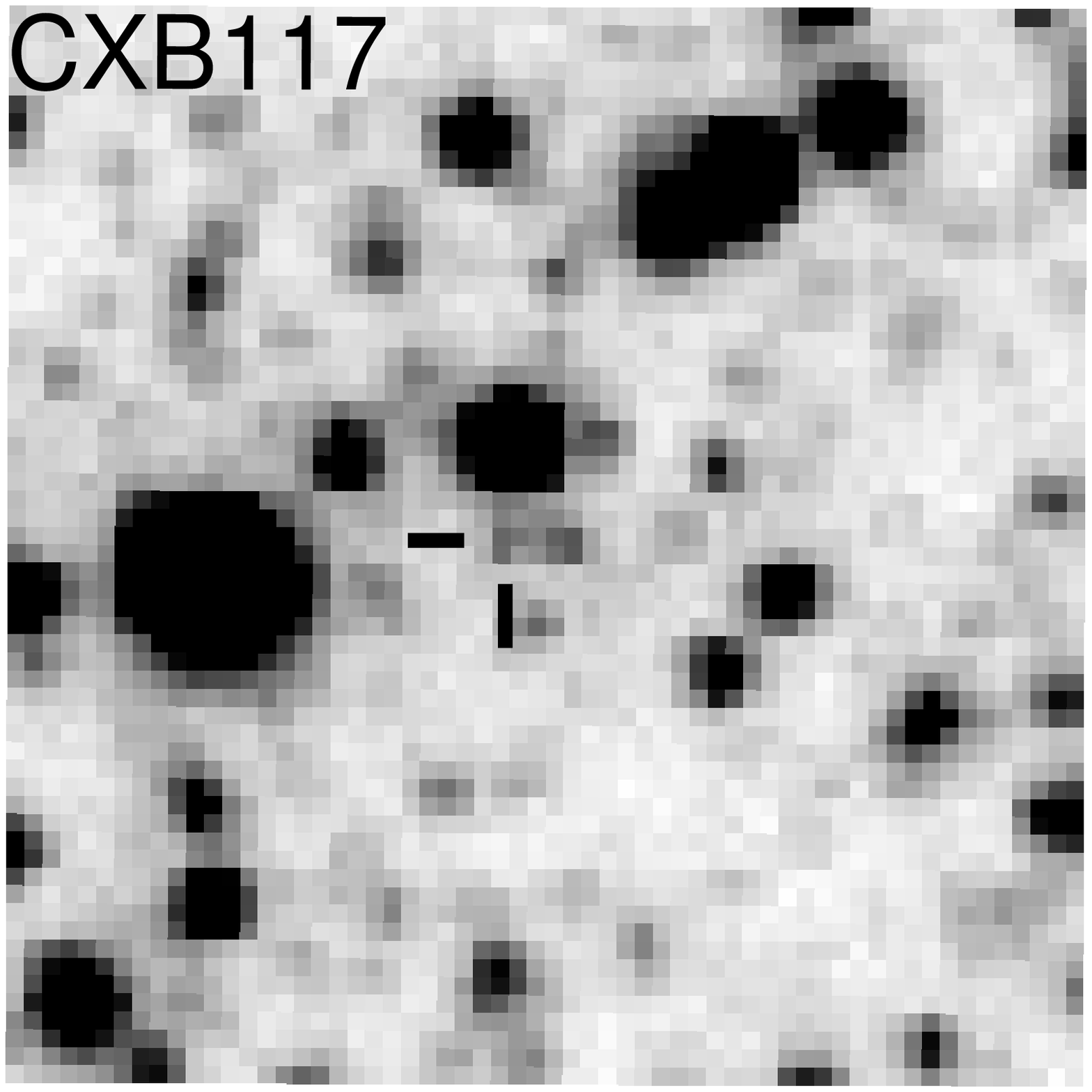}
    \includegraphics[width=1.5in]{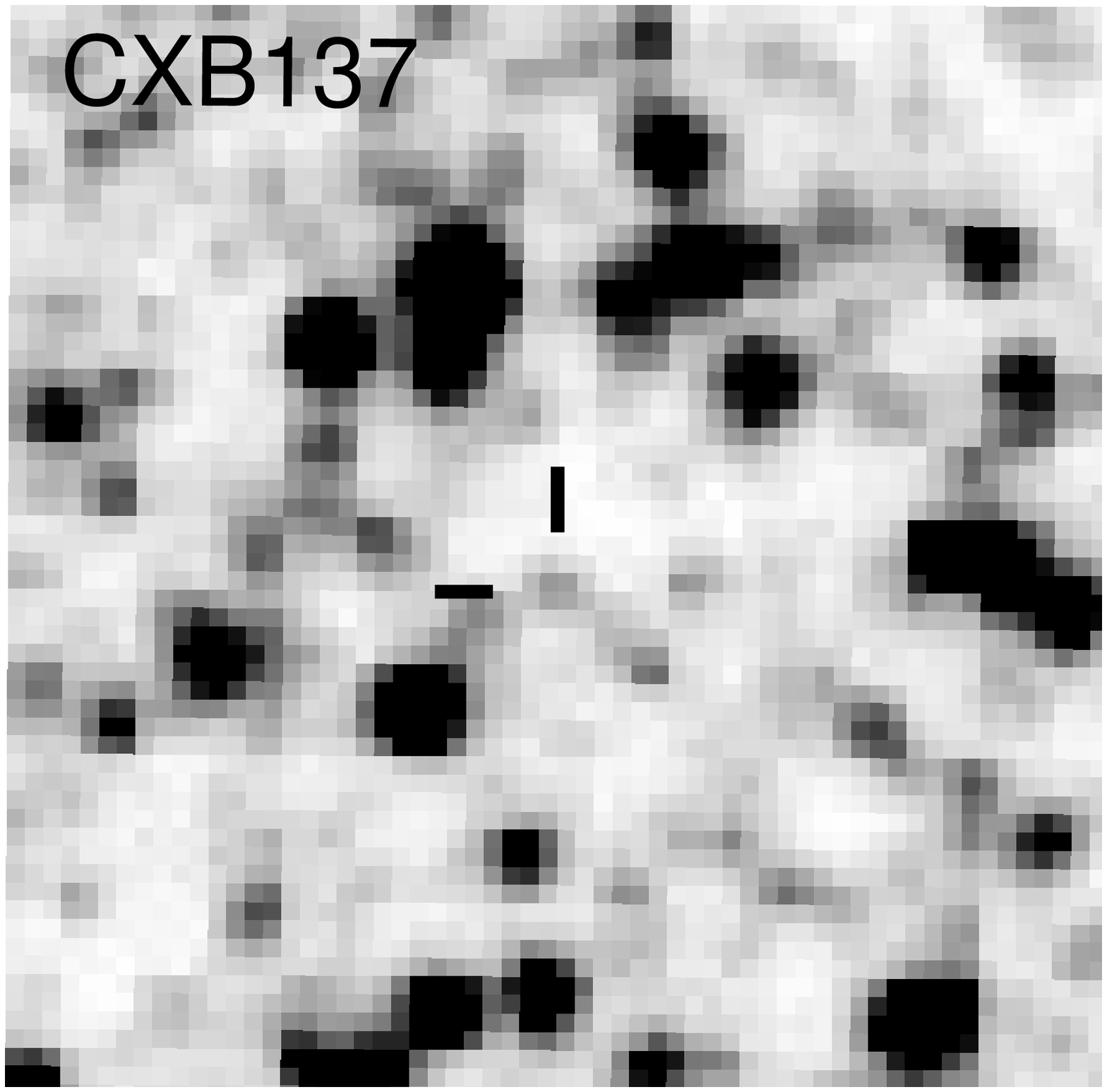}
    \includegraphics[width=1.5in]{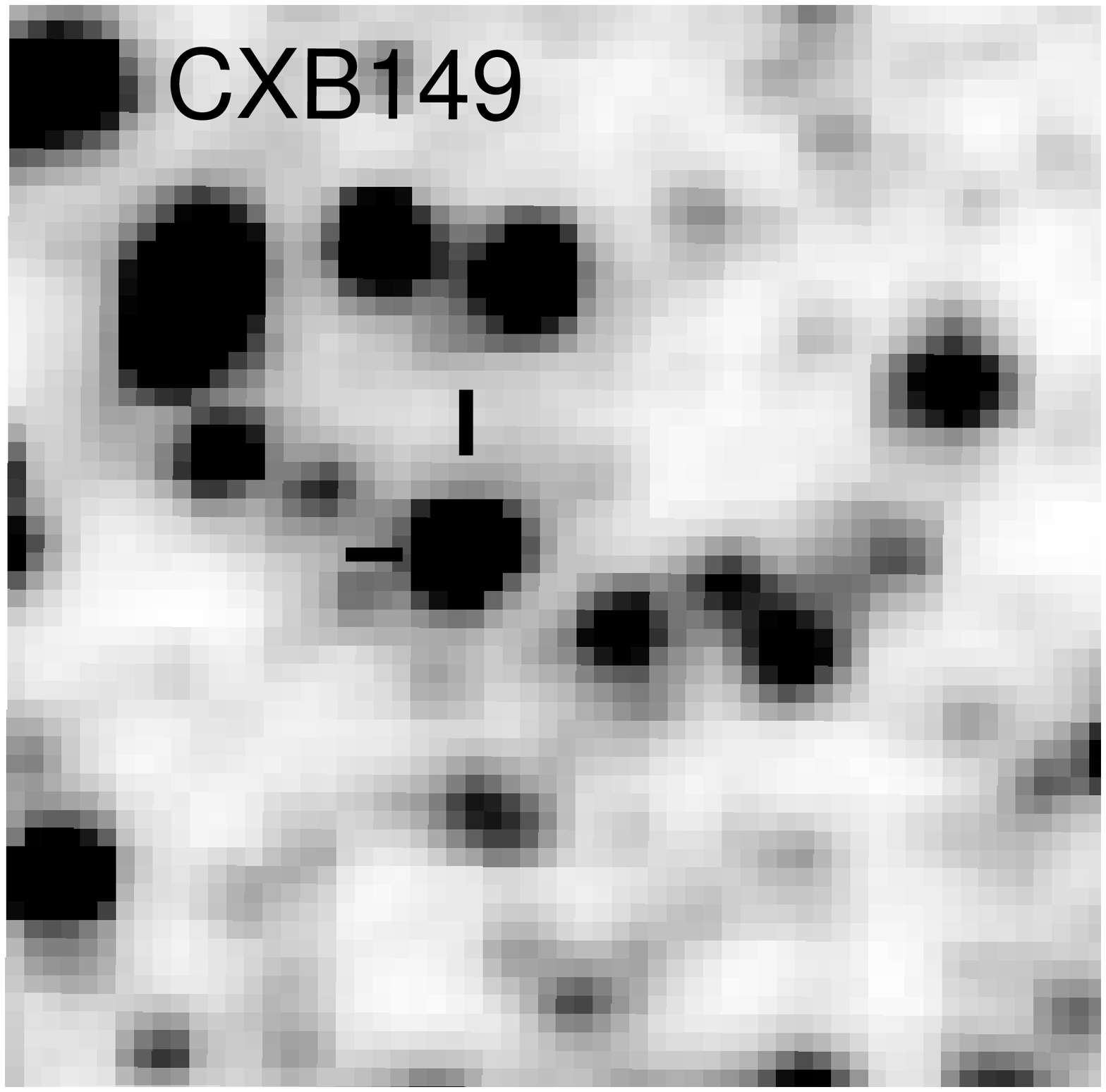}\\
    \includegraphics[width=1.5in]{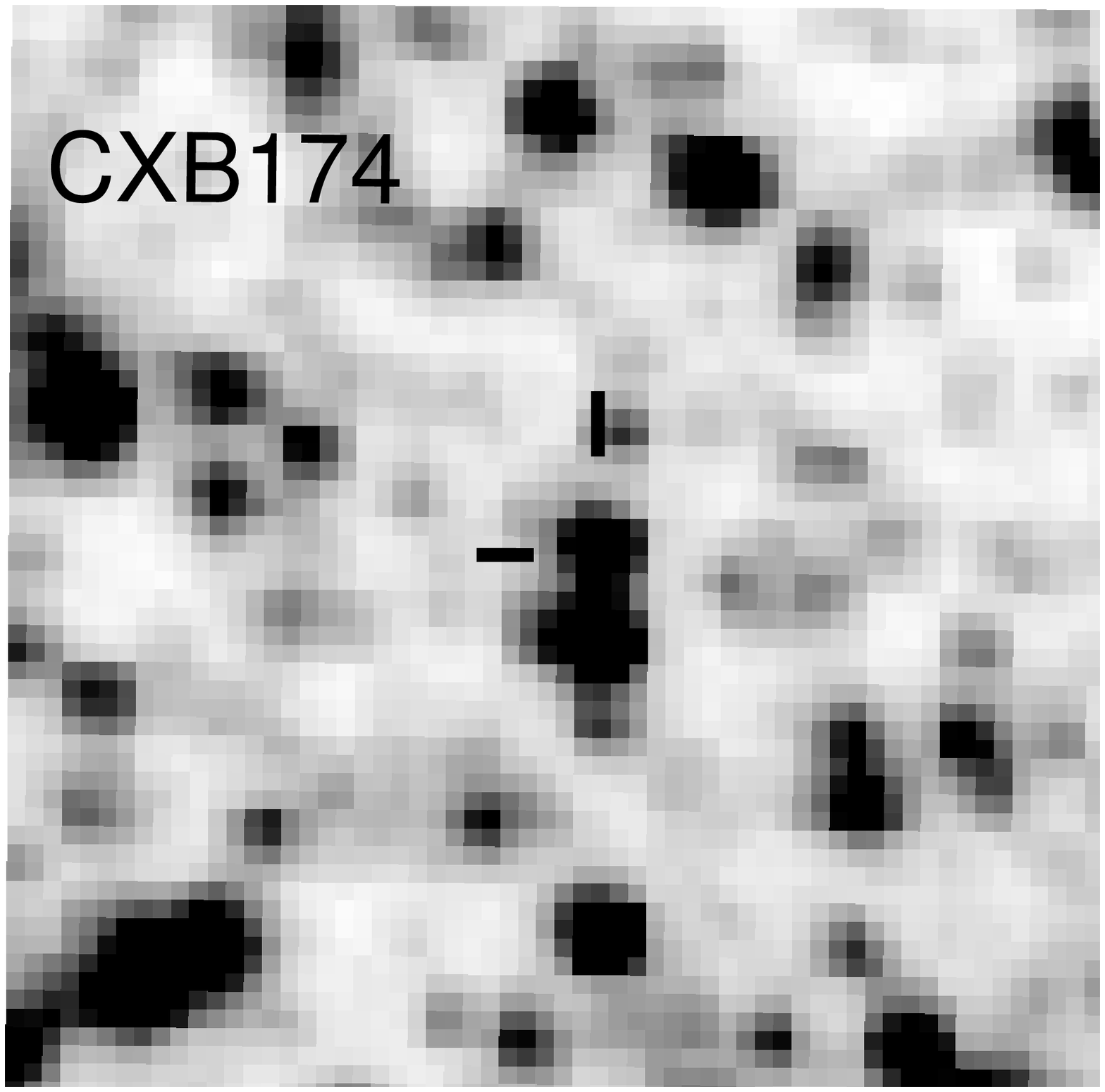}
    \includegraphics[width=1.5in]{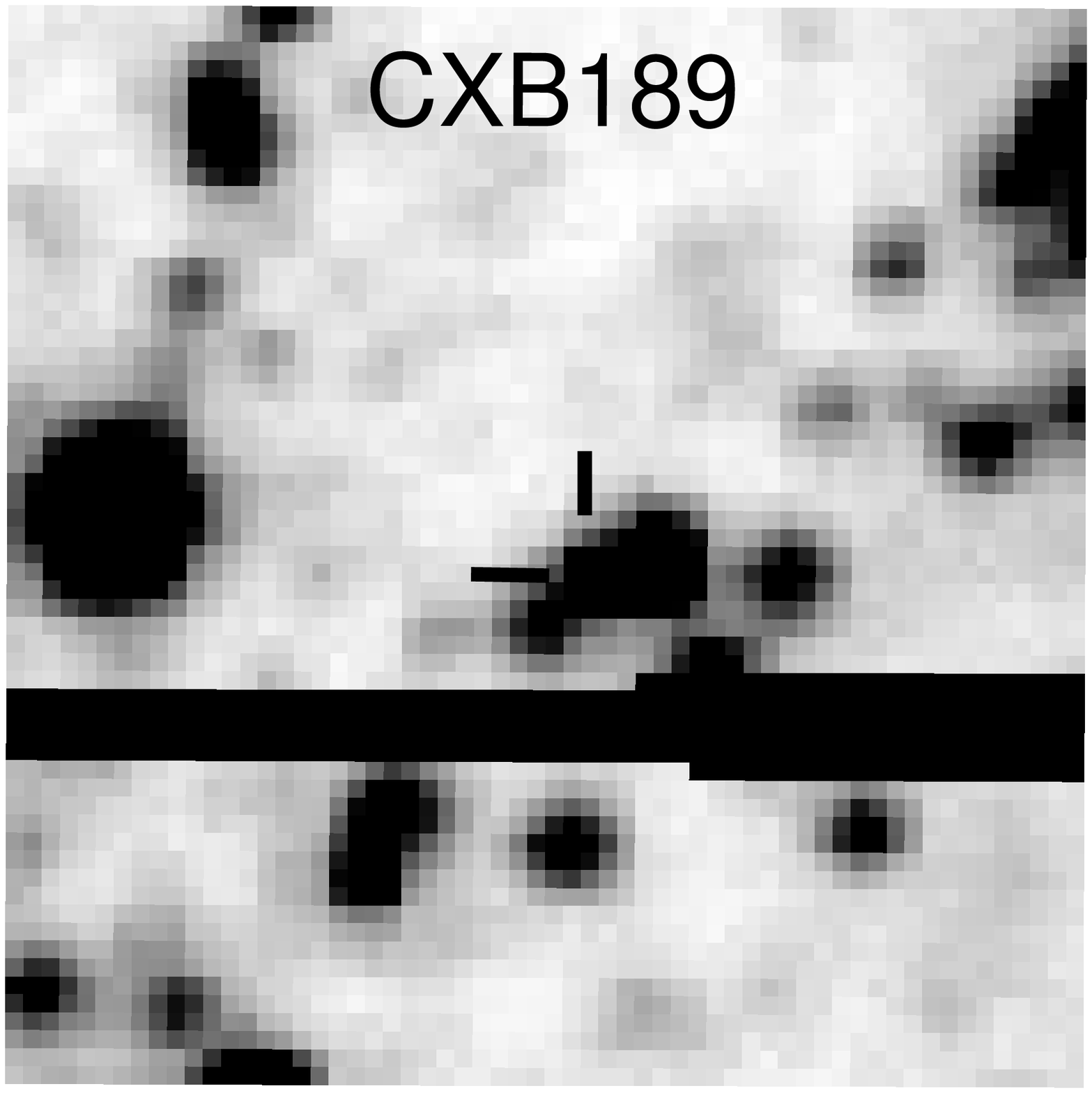}
    \includegraphics[width=1.5in]{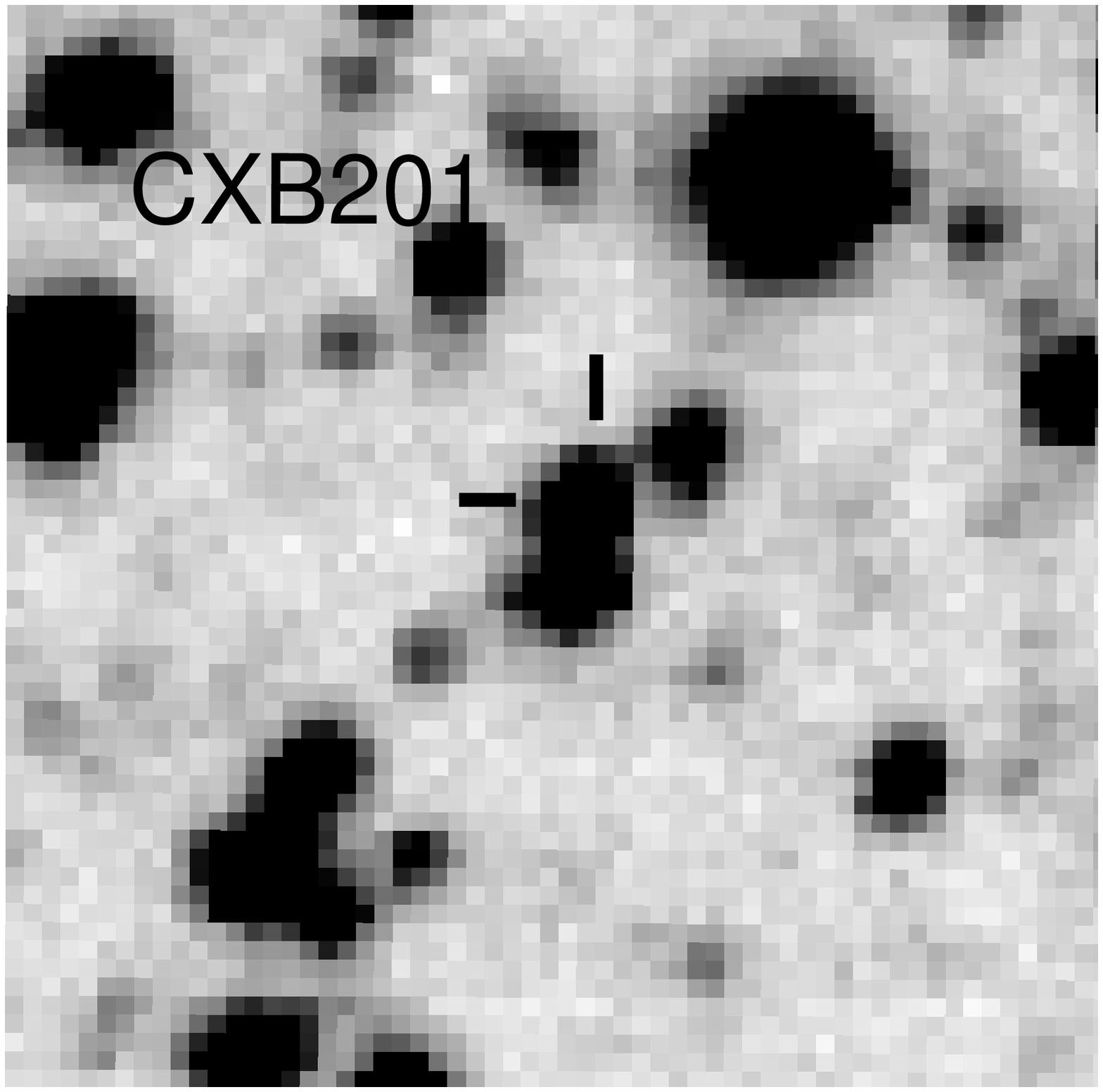}
    \caption{Finding charts -- \it Continued.}
             \label{fc_fig2}
\end{figure*}

The $r^\prime$-band finding charts for the optical counterparts of 18 of the
21 GBS sources presented  
in this paper are shown in Fig.~\ref{fc_fig} and Fig.~\ref{fc_fig2} . 
The sky area in each chart is 
20$^{\prime\prime}\times$20$^{\prime\prime}$ (except for CXB113 which is 
20$^{\prime\prime}\times$15$^{\prime\prime}$) with the source in the center and indicated by 
the thick horizontal and vertical short bars. North is up and East is to the left. 
CXB189 is blended with a nearby bright source because the field is crowded. Part of the 
emission from the nearby bright source may leak into the slit. The finding 
chart for CX377 is shown in Fig.~\ref{fc_cx377_fig}. The charts for CX446 and CX1004
can be found in Torres et~al. (2014). 

% -----------------------------------------------------------------------------
% Acknowledgments
% -----------------------------------------------------------------------------

\section*{Acknowledgments}
\noindent
We would like to thank the referee, Prof. P. Charles, for 
his comments which substantially improved the paper. The
software packages \verb+pamela+, and \verb+molly+  
utilized in this work were developed by T. Marsh. 
R.I.H., C.B.J., and C.T.B., acknowledge support from the National
Science Foundation under Grant No. AST-0908789 and from the National
Aeronautics and Space Administration through Chandra Award Number
AR3-14002X issued by the Chandra X-ray Observatory Center, which is
operated by the Smithsonian Astrophysical Observatory for and on
behalf of the National Aeronautics Space Administration under contract
NAS8-03060. D.S. acknowledges the support of the Science and
Technology Facilities Council, grant number ST/L000733/1. C.O.H. is 
supported by NSERC, an Ingenuity New Faculty Award, and an 
Alexander von Humboldt fellowship.
%\end{acknowledgments}

% -----------------------------------------------------------------------------
% Bibliography
% -----------------------------------------------------------------------------

% Figure A1
\clearpage

% Figure A1
\clearpage

% -----------------------------------------------------------------------------
% Tables
% -----------------------------------------------------------------------------

\clearpage
%Table 1: Observation log
\begin{center}
\begin{deluxetable}{lcccllc}
\tabletypesize{\footnotesize}
\tablecaption{The Observation Log of the
  Gemini/GMOS Spectroscopy \label{log_table}} 
\tablewidth{0pt}
\tablehead{ \colhead{GBS} & \colhead{R.A. (J2000)} & \colhead{Dec. (J2000)}
&           \colhead{Observation ID} & \colhead{Date (UT)} &
  \colhead{Integration} & \colhead{Seeing}\\
\colhead{ID} & \colhead{(HH:MM:SS.SS)} & \colhead{(DD:MM:SS.S)}
&           \colhead{(GS-2012A-Q)} & \colhead{} &
  \colhead{Time (s)} & \colhead{(arcsec)}
}
\startdata
CX84 & $17:38:12.84$ & $-29:06:12.4$ & 67-48 & 2012-05-31
&1200 & 0.5\\
& & & 67-9 & 2012-06-01 &1200 & 0.5\\
CX138 & $17:46:23.14$ & $-25:49:30.3$ & 67-52 &
2012-06-01 &1200 & 0.5\\
& & & 67-15 & 2012-06-14 &1200 & 1.1 \\
CX139 & $17:45:22.15$ & $-25:50:48.2$ & 67-3 &
2012-05-31 &900 & 0.6\\
& & & 67-44 & 2012-06-01 &900 & 0.6\\
CX377 & $17:43:16.54$ & $-27:45:37.0$ & 67-70 &
2012-06-01 &$900\times3$ & 0.6\\
& & & 67-72 & 2012-06-09 &$900\times3$ & 1.1\\
CX446 & $17:46:27.17$ & $-25:49:52.6$ & 67-96 & 2012-06-22
&$900\times3$ & 0.8 \\
& & & 67-109 & 2013-05-04 &$900\times3$ & 0.6 \\
CX1004 & $17:46:23.47$ & $-31:05:49.8$ & 44-63 &
2012-05-14 &$900\times3$ & 0.8\\
& & & 44-59 & 2012-05-17 &$900\times3$ & 0.8\\
& & & 44-3 & 2012-05-17 &$900\times3$ & 0.8\\
& & & 44-61 & 2012-05-17 &$900\times3$ & 0.7\\
CXB2 & $17:53:59.86$ & $-29:29:06.5$ & 44-53 &
2012-04-20 &$900\times3$ & 0.9\\
& & & 44-55 & 2012-04-20 &$900\times3$ & 0.9\\
& & & 44-57 & 2012-04-20 &$900\times3$ & 0.8\\
& & & 44-80 & 2012-05-19 &$900\times3$ & 0.9\\
CXB26 & $17:53:47.89$ & $-29:44:37.4$ & 44-49 &
2012-05-17 &$900\times3$ & 0.7\\
CXB38 & $17:58:33.81$ & $-27:30:22.9$ & 67-21 &
2012-06-09 &$900\times3$ & 1.3\\
CXB64 & $17:46:02.87$ & $-32:08:11.3$ & 67-74 &
2012-05-29 &$900\times4$ & 0.7\\
CXB73 & $17:52:29.07$ & $-30:03:21.7$ & 67-68 &
2012-06-22 &$900\times2$ & 1.1\\
CXB82 & $17:56:54.57$ & $-28:12:34.4$ & 67-66 &
2012-06-14 &$900\times2$ & 1.0\\
& & & 67-99 & 2012-06-22 &$900\times2$ & 1.0\\
CXB92 & $17:34:25.02$ & $-30:41:04.6$ & 44-47 &
2012-05-03 &$900\times3$ & 0.7\\
& & & 44-85 & 2012-05-19 &900 & 1.1\\
CXB99 & $17:54:59.36$ & $-29:10:21.1$ & 67-64 &
2012-06-09 &$900\times2$ & 1.2\\
CXB113 & $17:52:05.53$ & $-30:19:31.8$ & 67-62 &
2012-06-22 &900 & 1.1\\
CXB117 & $17:56:02.42$ & $-28:24:45.4$ & 67-60 &
2012-06-01 &$900\times2$ & 0.6\\
CXB137 & $17:57:10.01$ & $-27:57:54.9$ & 44-45 &
2012-05-17 &$900\times3$ & 0.8\\
CXB149 & $17:53:06.92$ & $-29:51:13.1$ & 67-58 &
2012-05-31 &$900\times2$ & 0.5\\
CXB174 & $17:56:18.52$ & $-28:45:40.5$ & 44-43 &
2012-04-19 &$900\times3$ & 0.6\\
CXB189 & $17:52:55.98$ & $-29:29:50.7$ & 44-41 &
2012-04-19 &$900\times3$ & 1.0\\
CXB201 & $17:33:26.08$ & $-30:40:24.4$ & 44-39 &
2012-04-19 &$900\times3$ & 0.9\\
\enddata\
%\tablenotetext{a}{The CXB IDs used in the Gemini observing program and
%  present in the FITS header files,
%  which are different from the finalized CXB IDs for most of our
%  sources. See Footnote~\ref{cxb_note}.}
\end{deluxetable}
\end{center}

\clearpage
%Table 2: Available Optical photometry for Gemini Sources
\begin{center}
\begin{deluxetable}{lcccc}
\tabletypesize{\footnotesize}
\tablecaption{Optical Photometry for the Gemini Counterparts\label{rmag_table}}
\tablewidth{0pt}
\tablehead{ \colhead{GBS ID} & \colhead{Mean($r^\prime$)} & \colhead{Err($r^\prime$)}\tablenotemark{a}
&  \colhead{$\sigma$($r^\prime$)}\tablenotemark{b} & \colhead{No. of Frames}
}
\startdata
CX84 & 19.018 & 0.007 & 0.029 & 35\\
CX377 & 18.885 & 0.009 & 0.060 & 36\\
CX446 & 21.162 & 0.062 & 0.166 & 37\\ 
CX1004 & 20.761 & 0.019 & 0.040 & 35\\
CXB2 & 20.350 & 0.003 & 0.14 & 110 \\
CXB64 & $\sim18.63$\tablenotemark{c} & 0.011 & 0.014 & 56 \\
CXB82 & 16.894 & 0.006 & 0.023 & 52 \\
CXB99 & 16.516 & 0.021 & 0.073 & 110 \\
CXB113 & $\sim19.13$\tablenotemark{c} & 0.018 & 0.041 & 55 \\
\enddata
\tablenotetext{a}{The mean error of the $r^\prime$ magnitudes for the
  object.}
\tablenotetext{b}{The standard deviation of the $r^\prime$ magnitudes for the
  object.}
\tablenotetext{c}{The magnitude here is approximate because of a lack of 
calibrator stars in the field of view.}
\end{deluxetable}
\end{center}

%\clearpage
%Table 3: Infrared photometry for Gemini Sources
\begin{center}
\begin{deluxetable}{lccccccc}
\tabletypesize{\footnotesize}
\tablecaption{Near-Infrared Photometry for the Gemini Counterparts\label{vvv_table}}
\tablewidth{0pt}
\tablehead{ \colhead{GBS ID} & \colhead{R.A.(VVV; J2000)} & \colhead{Dec.(VVV; J2000)}
&           \colhead{$\Delta\alpha$}\tablenotemark{a} &
  \colhead{$J$} & \colhead{$H$} & \colhead{$K$} \\
\colhead{} & \colhead{(HH:MM:SS.SS)} & \colhead{(DD:MM:SS.S)} 
& \colhead{(arcsec)} & \colhead{(mag)} & \colhead{(mag)} & \colhead{(mag)}
}
\startdata
CX84 & $17:38:12.84$ & $-29:06:12.5$ & $0.12$ & $14.566\pm0.018$ &
$13.570\pm0.019$ & $13.108\pm0.021$ \\ %& $2.68\times10^{-9}$ \\
CX138 & $17:46:23.12$ & $-25:49:30.5$ & $0.36$ & $13.122\pm0.004$ &
$12.134\pm0.004$ & $11.715\pm0.004$ \\ %& $3.53\times10^{-6}$ \\ 
CX139 & $17:45:22.15$ & $-25:50:48.3$ & $0.22$ & $12.555\pm0.002$ &
$11.914\pm0.003$ & $11.658\pm0.004$ \\ %& $1.67\times10^{-9}$\\ 
CX377 & $17:43:16.55$ & $-27:45:37.2$ & $0.20$ & $14.832\pm0.022$ &
$13.870\pm0.026$ & $13.443\pm0.027$ \\ %& $2.33\times10^{-6}$\\
%*CX446 & $266.61237$ & $-25.83159$ & $3.19$ & $15.473\pm0.036$ &
%$14.561\pm0.040$ & $14.039\pm0.039$ \\ %& $0.487$\\
%*CX1004 & $266.59824$ & $-31.09741$ & $1.59$ & $16.951\pm0.046$ &
%$15.199\pm0.036$ & $14.152\pm0.031$ \\ %& $2.90\times10^{-3}$\\ 
CXB2 & $17:53:59.85$ & $-29:29:06.4$ & $0.13$ & $17.251\pm0.344$ &
$16.408\pm0.314$ & $16.251\pm0.336$ \\ %& $1.58\times10^{-4}$\\
%*CXB26 & $268.44997$ & $-29.74437$ & $2.85$ & $13.618\pm0.012$ &
%$12.843\pm0.012$ & $12.547\pm0.012$ \\ %& $0.362$\\
CXB38 & $17:58:33.82$ & $-27:30:22.7$ & $0.25$ & $14.182\pm0.019$ &
$13.264\pm0.016$ & $12.958\pm0.016$ \\ %& $5.88\times10^{-4}$\\ 
%*CXB64 & $266.51183$ & $-32.13665$ & $0.67$ & $14.40\pm0.01$ &
%$13.31\pm0.01$ & $12.85\pm0.01$ \\ %& $5.36\times10^{-3}$\\
CXB73 & $17:52:29.08$ & $-30:03:21.5$ & $0.18$ & $13.978\pm0.017$ &
$13.143\pm0.016$ & $12.796\pm0.015$ \\ %& $2.38\times10^{-5}$\\ 
CXB82 & $17:56:54.57$ & $-28:12:34.3$ & $0.10$ & $13.514\pm0.011$ &
$12.732\pm0.010$ & $12.536\pm0.011$ \\ %& $1.82\times10^{-4}$\\ 
CXB92 & $17:34:25.02$ & $-30:41:04.5$ & $0.10$ & $14.866\pm0.016$ &
$13.967\pm0.018$ & $13.372\pm0.018$ \\ %& $7.38\times10^{-6}$\\
CXB99 & $17:54:59.34$ & $-29:10:21.1$ & $0.19$ & $14.069\pm0.022$ &
$13.356\pm0.022$ & $13.206\pm0.024$ \\ %& $6.13\times10^{-6}$\\ 
CXB113 & $17:52:05.52$ & $-30:19:31.8$ & $0.10$ & $14.108\pm0.020$ &
$13.599\pm0.023$ & $13.299\pm0.024$ \\ %& $5.46\times10^{-5}$\\ 
%*CXB117 & $269.00970$ & $-28.41260$ & $0.69$ & $15.711\pm0.095$ &
%$14.754\pm0.077$ & $14.399\pm0.069$ \\ %& $9.13\times10^{-5}$\\
%*CXB137 & $269.29242$ & $-27.96581$ & $3.63$ & $14.328\pm0.022$ &
%$13.406\pm0.019$ & $13.152\pm0.018$ \\ %& $0.774$\\ 
CXB149 & $17:53:06.93$ & $-29:51:13.0$ & $0.15$ & $14.242\pm0.022$ &
$13.720\pm0.027$ & $13.527\pm0.029$ \\ %& $4.46\times10^{-2}$\\ 
CXB174 & $17:56:18.53$ & $-28:45:40.5$ & $0.10$ & $14.388\pm0.029$ &
$13.552\pm0.026$ & $13.268\pm0.025$ \\ %& $1.46\times10^{-2}$\\
%*CXB189 & $268.23351$ & $-29.49769$ & $1.38$ & $14.839\pm0.043$ &
%*$13.912\pm0.035$ & $13.644\pm0.035$ \\ %& $1.07\times10^{-3}$\\ 
%*CXB201 & $263.35876$ & $-30.67383$ & $1.66$ & $13.811\pm0.010$ &
%$12.780\pm0.010$ & $12.328\pm0.010$ \\ %& $1.68\times10^{-2}$\\
\enddata
\tablenotetext{a}{The offset between near-infrared position and optical
position, in arcseconds.}
\end{deluxetable}
\end{center}

\clearpage
%Table 4: UVESPOP templates
\begin{center}
\begin{deluxetable}{cl}
\tabletypesize{\footnotesize}
\tablecaption{List of Stellar Templates from the UVES POP Library\label{uves_table}}
\tablewidth{0pt}
\tablehead{ \colhead{Spectral Type}  & \colhead{Star Name} 
}
\startdata
A0V & HD~162305  \\
A1V & HD~65810   \\
A2V & HD~60178   \\
A3V & HD~211998  \\
A4V & HD~145689  \\
A5V & HD~39060   \\
A7V & HD~187642  \\
A9V & HD~26612   \\
F0V & HD~109931  \\
F1V & HD~40136   \\
F2V & HD~33256   \\
F3V & HD~18692   \\
F4V & HD~37495   \\
F6V & HD~16673   \\
F8V & HD~45067   \\
F9V & HD~10647   \\
G0V & HD~105113  \\
G1V & HD~20807   \\
G2V & HD~14802   \\
G3V & HD~211415  \\
G4V & HD~59967   \\
G5V & HD~59468   \\
G6V & HD~140901  \\
G9V & HD~25069   \\
K2V & HD~22049   \\
K5V & HD~10361   \\
M0V & HD~156274  \\
M6V & HD~34055   \\
\enddata
\end{deluxetable}
\end{center}

%\clearpage
%Table 5: vsini and optsub 
\begin{center}
\begin{deluxetable}{lc}
\tabletypesize{\footnotesize}
\tablecaption{Spectral Classifications for the Gemini Counterparts\label{vsini_table}}
\tablewidth{0pt}
\tablehead{ \colhead{GBS ID} &
  \colhead{Spectral Type} %& \colhead{Uncertainty Range}
%            \colhead{} & \colhead{} & \colhead{} & \colhead{(\lyanv)} & \colhead{(C\,{\sc iv})} 
}
\startdata
CX84 & G9 \\
CX138 & G9 \\
CX139 & K2 \\
CX377 & F6 \\
CX1004\tablenotemark{a} & M0--M5 \\
CXB64\tablenotemark{a} & M0--M5 \\
CXB73 & G9 \\
CXB82 & G9 \\
CXB99 & K2 \\
CXB113\tablenotemark{a} & M0--M5 \\
CXB149 & G6 \\
CXB174 & F4 \\
\enddata
\tablenotetext{a}{These three sources cannot be spectrally classified
  via optimal subtraction. However, their spectra contain features (e.g., TiO
  band) of early-to-mid M-type stars.}
\end{deluxetable}
\end{center}

\clearpage
%Table 6: Radial velocity
\begin{center}
\begin{deluxetable}{lccc}
\tabletypesize{\footnotesize}
\tablecaption{Radial Velocity Results for the Gemini Counterparts\label{rvc_table}}
\tablewidth{0pt}
\tablehead{ \colhead{GBS ID} & \colhead{HJD} &
  \colhead{RV1 (km~s$^{-1}$)\tablenotemark{a}} & \colhead{RV2 (km~s$^{-1}$)\tablenotemark{b}}
%            \colhead{} & \colhead{} & \colhead{} & \colhead{(\lyanv)} & \colhead{(C\,{\sc iv})} 
}
\startdata
CX84 & $2456078.811$ & $0$ & $-104.3\pm3.3$ \\
& $2456079.677$ & $74.0\pm3.0$ & $-17.3\pm6.0$\\
CX138 & $2456079.760$ & $0$ & $-110.7\pm2.1$ \\
& $2456092.650$ & $75.6\pm6.5$ & $-44.2\pm5.4$\\
CX139 & $2456078.665$ & $0$ & $-96.2\pm10.6$ \\
& $2456079.783$ & $69.0\pm2.5$ & $-17.4\pm4.5$\\
CX377 & $2456079.714$ & $0$ & $-20.5\pm9.2$ \\
& $2456079.725$ & $-14.6\pm12.9$ & $-43.6\pm11.9$ \\
& $2456079.736$ & $19.0\pm8.8$ & $-17.2\pm6.6$ \\
%CXB26 & $2456064.839383728$ & $0$ & \nodata \\
%& $2456064.850318701$ & $4.5\pm3.4$ & \nodata \\
%& $2456064.861267759$ & $-4.7\pm4.1$ & \nodata \\
%CXB38 & $2456087.612078963$ & $0$ & \nodata \\
%& $2456087.623016412$ & $3.2\pm15.5$ & \nodata \\
%& $2456087.633965436$ & $12.5\pm9.7$ & \nodata \\
CXB73 & $2456100.751$ & $0$ & $-30.0\pm6.2$ \\
& $2456100.762$ & $-1.7\pm5.1$ & $-31.0\pm6.0$ \\
CXB82 & $2456092.798$ & $0$ & $-36.6\pm4.7$ \\
& $2456092.809$ & $-11.3\pm6.3$ & $-41.0\pm7.9$ \\
& $2456100.806$ & $-2.9\pm2.2$ & $-39.7\pm3.8$ \\
& $2456100.817$ & $-6.2\pm2.3$ & $-41.7\pm2.7$ \\
CXB99 & $2456087.659$ & $0$ & $21.5\pm7.0$ \\
& $2456087.670$ & $-11.0\pm3.8$ & $-1.7\pm10.2$ \\
%CXB117 & $2456079.801437841$ & $0$ & \nodata \\
%& $2456079.812377869$ & $0.3\pm2.0$ & \nodata \\
%CXB149 & $2456078.859577012$ & $0$ & \nodata \\
%& $2456078.870514698$ & $0.4\pm2.6$ & \nodata \\
%CXB174 & $2456036.814838650$ & $0$ & \nodata \\
%& $2456036.825778198$ & $-8.8\pm3.6$ & \nodata \\
%& $2456036.836724320$ & $-11.1\pm3.7$ & \nodata \\
%CXB189 & $2456036.741804772$ & $0$ & \nodata \\
%& $2456036.752739479$ & $1.7\pm9.2$ & \nodata \\
%& $2456036.763676686$ & $6.4\pm12.3$ & \nodata \\
%CXB201 & $2456036.698810112$ & $0$ & \nodata \\
%& $2456036.709745867$ & $-7.1\pm7.0$ & \nodata \\
%& $2456036.720684123$ & $-5.0\pm5.0$ & \nodata \\
\enddata
\tablecomments{Among the sources not listed in this table, seven of
  them (CX446, CX1004, CXB2, CXB64, CXB92, 
CXB113, and CXB137) do not show significant cross-correlations between their 
Gemini/GMOS spectra, while the other seven (CXB26, CXB38, CXB117,
CXB149, CXB174, CXB189, and CXB201) have relative radial velocities
consistent with zero.}% between their Gemini/GMOS spectra.}
\tablenotetext{a}{The radial velocity (RV) results when using the
  first source spectrum as a cross-correlation template; thus the RV1 value for the first
source spectra is always zero.}
\tablenotetext{b}{The radial velocity (RV) results when using the
  star template spectrum with the spectral type shown in
  Table~\ref{vsini_table} as a cross-correlation template.}
\end{deluxetable}
\end{center}

\clearpage
%Table 7: Halpha Emission/Absorption measurements
\begin{center}
\begin{deluxetable}{clccc}
\tabletypesize{\footnotesize}
\tablecaption{Measurements of the Broad H$\alpha$ Emission/Absorption Lines for Gemini Spectra\label{line_table}}
\tablewidth{0pt}
\tablehead{ & \colhead{GBS ID} & \colhead{EW (\AA)} & \colhead{FWHM(\AA)}\tablenotemark{a} & \colhead{FWHM(km s$^{-1}$)}\tablenotemark{b}
%&  \colhead{Epoch}
}
\startdata
%\multicolumn{2}{l}{Emission:} & & &\\
& CX377 & $3.1\pm0.6$ & $6.3\pm0.2$ & $233\pm9$ \\ %& 2012-06-01 \\
& CX446 & $-67.6\pm5.4$ & $27.5\pm1.2$ & $1250\pm50$ \\ %& 2012-06-22 \\
%& & $2.0\pm10.0$ & $29.2\pm2.8$ & 2013-05-04 \\
& CX1004\tablenotemark{c} & $-38.0\pm0.6$ & $55.1\pm1.9$ & $2500\pm100$ \\ %& 2012-05-17 \\
& CXB2 & $-4.4\pm0.7$ & $18.6\pm3.7$ & $800\pm160$ \\ %& 2012-04-20 \\
& CXB64 & $-2.0\pm0.1$ & $4.9\pm0.1$ & $63\pm5$ \\ %& 2012-05-29 \\
& CXB99 & $-1.3\pm0.1$ & $6.3\pm0.6$ & $185\pm25$ \\ %& 2012-06-09 \\
& CXB113 & $-5.3\pm0.1$ & $5.3\pm0.1$ & $110\pm5$ \\ %& 2012-06-22 \\
\hline
& CX84 & $-2.8\pm0.1$ & $6.0\pm0.3$ & $225\pm15$ \\ %& \nodata\\ 
& CX138 & $-2.6\pm0.1$ & $30.0\pm1.5$ & $1350\pm75$ \\ %& \nodata\\ 
& CX139 & $-2.2\pm0.1$ & $6.4\pm0.6$ & $250\pm25$ \\ %& \nodata\\
& CX377 & $-2.8\pm0.1$ & $15.0\pm0.7$ & $660\pm30$ \\ %& \nodata\\
& CXB73 & $-1.3\pm0.1$ & $9.0\pm1.0$ & $350\pm50$ \\ %& \nodata\\
& CXB82 & $-1.3\pm0.1$ & $6.1\pm0.2$ & $180\pm10$ \\ %& \nodata\\
& CXB99 & $-4.1\pm0.1$ & $6.1\pm0.3$ & $180\pm15$ \\ %& \nodata\\
%\multicolumn{2}{l}{Absorption:} & & &\\
\enddata
\tablecomments{The upper part of the table shows the measurements of
  H$\alpha$ emission/absorption lines in the Gemini/GMOS spectra of
  the GBS sources. The lower part of the tables lists the measurements
  of the H$\alpha$ emission lines in the residual spectra. The residual 
  spectra are obtained by optimally subtracting the best-fit stellar 
  templates (see Fig.~\ref{resid_fig} for residual spectra).} 
%\tablenotetext{a}{The equivalent width (EW) of the H$\alpha$
%  emission/absorption lines. Emission/absorption lines have
%  negative/positive values, respectively.} 
\tablenotetext{a}{The FWHM of the H$\alpha$
  emission/absorption lines before correcting for instrumental
  broadening, in unit of \AA. } 
\tablenotetext{b}{The FWHM of the H$\alpha$
  emission/absorption lines after correcting for instrumental
  broadening, in unit of km~s$^{-1}$. The FWHM values quoted in the
  text (in unit of km~s$^{-1}$) are also corrected for instrumental
  broadening. }
\tablenotetext{c}{The measurements listed here are for the full
  H$\alpha$ emission profile. For the measurements of each individual
  peak, see the text in \S\ref{discuss:cx446}.} 
\end{deluxetable}
\end{center}

% -----------------------------------------------------------
% Figures
% --------------------------------------------------------------------

\end{document}